\documentclass[oneside,11pt]{article}

\usepackage{url,mathrsfs}
\usepackage{amsmath,wrapfig,color,algorithm,algorithmic}
\usepackage{amsfonts}
\usepackage{graphicx}
\usepackage{subfigure}
 \usepackage{ifthen}
 \usepackage[round]{natbib}

\def\pa{\parallel}
\def\pe{\perp}

\def\sym{\mathbb{S}}
\def\ds{\wt{\beta}_{\lambda}}
\def\dss{\wt{\beta}_{\lambda}^{\ell_2}}
\def\rer{\wh{\beta}_{\lambda}^{\ell_2}}

\newboolean{isjournal}
\setboolean{isjournal}{true}
\newcommand{\inner}[1]{\left\langle#1\right\rangle}

\usepackage{amsfonts}
\usepackage{amssymb}
\usepackage{amsmath}
\usepackage{amsthm}
\usepackage{latexsym}
\usepackage{verbatim}
\usepackage{epsfig}
\usepackage{color}
\usepackage{url}
										
\usepackage{graphicx}
\usepackage{amsbsy}
\usepackage{amscd}
\usepackage{bm}

\def\pa{\parallel}
\def\pe{\perp}

\def\sym{\mathbb{S}}
\def\ds{\wt{\beta}_{\lambda}}
\def\dss{\wt{\beta}_{\lambda}^{\ell_2}}
\def\rer{\wh{\beta}_{\lambda}^{\ell_2}}

\newcommand{\TT}{{\mathbb T}}
\newcommand{\HH}{{\mathbb H}}
\newcommand{\B}{{\mathbb B}}
\newcommand{\R}{{\mathbb R}}

\newcommand{\C}{{\mathbb C}}

\DeclareMathOperator*{\tr}{tr}

\DeclareMathOperator*{\E}{\mathbf{E}}
\DeclareMathOperator*{\argmin}{argmin}

\DeclareMathOperator*{\conv}{conv}

\DeclareMathOperator*{\p}{\mathbf{P}}
\providecommand{\wt}[1]{\widetilde{#1}}
\providecommand{\wh}[1]{\widehat{#1}}

\providecommand{\norm}[1]{\left \lVert#1 \right \rVert}
\providecommand{\nnorm}[1]{ \lVert#1 \rVert}
\newcommand{\scp}[2]{\left\langle#1, #2\right\rangle}

\providecommand{\mc}[1]{\mathcal#1}
\providecommand{\T}{\top}

\newcommand{\blanco}[1]{  }

\newcommand{\deriv}[3]{%
\ifthenelse{#1 = 1}{\frac{d\,#2}{d\,#3}}{\frac{d^{{#1}} #2}{d{#3}^{{#1}}}}
}

\newcommand{\partials}[3]{%
\ifthenelse{#1 = 1}{\frac{\partial\,#2}{\partial\,#3}}{\frac{\partial^{#1}
    #2}{\partial#3^{#1}}}
}

\def\su{\sum_{i=1}^n}

\def \coloneq{\mathrel{\mathop:}=}

\def \eps{\varepsilon}

\def \gec{\succeq}

\newtheorem{propo}{Theorem}

\newtheorem{conditio}{Theorem}

\newtheorem{definitioA}{Theorem}[section]

  \newtheorem{prop}[propo]{Proposition}

\newtheorem{cond}[conditio]{Condition}

\newtheorem{defnApp}[definitioA]{Definition}

\newenvironment{bew}{\begin{proof}[Proof]}{\end{proof}}

\begin{document}
\title{\Huge Methods for Sparse and Low-Rank Recovery
under Simplex Constraints\vspace{0.2in}}

\author{ \bf{Ping Li} \\
         Department of Statistics and Biostatistics\\\hspace{0.1in}
         Department of Computer Science\\
       Rutgers University\\
          Piscataway, NJ 08854, USA\\
       \texttt{pingli@stat.rutgers.edu}\\\\
       \and
\bf{Syama Sundar Rangapuram} \\
         Department of Computer Science\\
       Saarland University\\
          Saarbr\"ucken, Germany\\
       \texttt{srangapu@mmci.uni-saarland.de}
       \and
\bf{Martin Slawski} \\
         Department of Statistics and Biostatistics\\\hspace{0.1in}
         Department of Computer Science\\
       Rutgers University\\
          Piscataway, NJ 08854, USA\\
       \texttt{martin.slawski@rutgers.edu}
}

\date{}

\maketitle

\begin{abstract}
\noindent The de-facto standard approach of promoting sparsity by means of $\ell_1$-regularization
becomes ineffective in the presence of simplex constraints, i.e.,~the target is known
to have non-negative entries summing up to a given constant. The situation is analogous
for the use of nuclear norm regularization for low-rank recovery of Hermitian  positive semidefinite
matrices with given trace. In the present paper, we discuss several strategies to deal with this situation, from simple to more complex. As a starting point, we consider empirical risk minimization (ERM). It follows from existing theory that ERM enjoys  better theoretical properties w.r.t.~prediction and $\ell_2$-estimation error than $\ell_1$-regularization. In light of this, we argue that ERM combined with a subsequent sparsification step like thresholding is superior to the heuristic of using $\ell_1$-regularization after dropping the sum constraint and subsequent normalization.\\

\noindent At the next level, we show that any sparsity-promoting regularizer under simplex constraints
cannot be convex. A novel sparsity-promoting regularization scheme based on the inverse or
negative of the squared $\ell_2$-norm is proposed, which avoids shortcomings of various alternative methods from the literature. Our approach naturally extends to Hermitian positive semidefinite matrices with given trace. Numerical studies concerning compressed sensing, sparse mixture density estimation, portfolio optimization and quantum state tomography are used to illustrate the key points of the paper.
\end{abstract}

\newpage

\section{Introduction}\label{sec:intro}
In this paper, we study the case in which the parameter of interest $\beta^*$ is
sparse and non-negative with known sum, i.e.,~$\beta^* \in c \Delta^p \cap \B_0^p(s)$, where
for $c > 0$ and $1 \leq s \leq p$, $c \Delta^p = \{\beta \in \R_+^p:\, \bm{1}^{\T} \beta = c \}$ is the
(scaled) canonical simplex in $\R^p$, $\B_0^p(s) = \{\beta \in \R^p:\, \norm{\beta}_0 \leq s \}$ and
$\norm{\beta}_0 = |S(\beta)| = |\{j:\,\beta_j \neq 0 \}|$ is referred to as $\ell_0$-norm (the cardinality of the support $S(\beta)$).
Unlike the constant $c$, the sparsity level $s$ is regarded as unknown. The specific value of the constant
$c$ is not essential; in the sequel, we shall work with $c = 1$ as for all problem instances studied
herein, the data can be re-scaled accordingly. The elements of $\Delta^p$ can represent probability
distributions over $p$ items, proportions or normalized weights, which
are quantities frequently arising in instances of contemporary data analysis. A selection is listed
below.
\begin{itemize}
\item \emph{Estimation of mixture proportions.}\hspace{0.2in}
Specific examples are the determination of the proportions of chemical constituents in a given sample or the endmember composition of pixels in hyperspectral imaging \citep{Keshava2003}.

\item \emph{Probability density estimation}, cf.~\citet{Bunea2010}\textbf{.}\hspace{0.2in}
Let $(\mc{Z}, \mc{A}, P)$ be a probability space with $P$ having a density $f$ w.r.t.~some
dominating measure $\nu$. Given a sample $\{ Z_i \}_{i=1}^n \overset{\text{i.i.d.}}{\sim} P$ and a dictionary
$\{ \phi_j \}_{j = 1}^p$ of densities (w.r.t.~$\nu$), the goal is to find
a mixture density $\phi_{\beta} = \sum_{j = 1}^p \beta_j \phi_j$ well approximating $f$, where $\beta \in \Delta^p$.
\item\emph{Convex aggregation/ensemble learning.}\hspace{0.2in}
The following problem has attracted much interest in non-parametric estimation, cf.~\citet{Nemirovski2000}.
Let $f$ be the target in a regression/classification problem and let $\{ \phi_{j} \}_{j=1}^p$ be an ensemble of regressors/classifiers.  The goal is to approximate $f$ by a convex combination of $\{ \phi_{j}\}_{j = 1}^p$.
\item\emph{Markowitz portfolios\ifthenelse{\boolean{isjournal}}{ \citep{Markowitz1952}}{} without short positions.}\hspace{0.2in}
Given $p$ assets whose returns have expectation $r = (r_1,\ldots,r_p)^{\T}$ and covariance $\Sigma$, the goal
is to distribute the total investment according to proportions $\beta \in \Delta^p$ such
that the variance $\beta^{\T} \Delta \beta$ is minimized subject to a lower bound on the
expected return $\beta^{\T} r$.
\end{itemize}
Sparsity is often prevalent or desired in these applications.
\begin{itemize}
\item In hyperspectral imaging, a single pixel usually contains few endmembers.
\item  In density estimation, the underlying density may be concentrated in certain regions of the sample space.
\item In aggregation, it is common to work with a large
      ensemble to improve approximation capacity, while
      specific functions may be well approximated by few
        members of the ensemble.
\item A portfolio involving a small number of assets incurs less transaction costs and is much easier to manage.
\end{itemize}
At the same time, promoting sparsity in the presence of the constraint
$\beta \in \Delta^p$ appears to be more difficult as $\ell_1$-regularization
does not serve this purpose anymore. As clarified in $\S$\ref{sec:basicperformance}, the naive approach of
employing $\ell_1$-regularization and dropping the sum constraint
is statistically not sound. The situation is similar for nuclear norm regularization and low-rank matrices that are Hermitian positive semidefinite with
fixed trace as arising (e.g.,) in quantum state tomography \citep{Gross2010},
where the constraint set is given by $\bm{\Delta}^{m} =
\{B \in \C^{m \times m}: B = B^H,\, B \gec 0,\, \tr(B) = 1\}$, where $^{H}$ denotes
conjugate transposition. Both the presence of simplex constraints and its
matrix counterpart thus demand different strategies to deal with sparsity and
low-rankedness, respectively. The present paper proposes such strategies that are statistically sound, straightforward to implement, adaptive in the sense that the sparsity level $s$ (resp.~the rank in the matrix case) is not required to be known, and that
work with a minimum amount of hyperparameter tuning.\\

\noindent{\bf Related work.} The problem outlined above is nicely discussed in \citet{Kyrilidis2013}. In that
paper, the authors consider the sparsity level $s$ to be known as well and suggest to deal
with the constraint set $\Delta_0^p(s) = \Delta^p \cap \B_0^p(s)$ by projected gradient descent based
on a near-linear time algorithm for computing the projection. This approach can be seen as a natural
extension of iterative hard thresholding (IHT, \ifthenelse{\boolean{isjournal}}{\citet{Blumensath2009}}{\citet{Blumensath2009, Garg2011}}). It has clear merits
in the matrix case, where significant computational savings are achieved in the projection step. On
the other hand, it is a serious restriction to assume that $s$ is known. Moreover, the method has
theoretical guarantees only under the restricted isometry property (RIP, \citet{CandesTao2007}) and a certain choice of the
step size depending on RIP constants, which are not accessible in practice.

\citet{Pilanci2012} suggest the regularizer $\beta \mapsto 1/\norm{\beta}_{\infty}$ to promote sparsity
on $\Delta^p$ and show that in the case of squared loss, the resulting non-convex optimization
problem can be reduced to $p$ second-order cone programs. In practice, however, in particular when combined
with tuning of the regularization parameter, the computational cost quickly becomes prohibitive.

Relevant prior work also includes \citet{Ugander2011} and \citet{Shashanka07} who discuss the problem
of this paper in the context of latent variable representations (PLSA, \citet{Hofmann99}) for image and bag-of-words data.
\citet{Ugander2011} proposes a so-called pseudo-Dirichlet prior correponding to the log-penalty in \citet{Candes2007}.
\citet{Shashanka07} suggest to use the Shannon entropy as regularizer. Both approaches are cumbersome regarding
optimization due to the singularity of the logarithm at the origin which is commonly handled by adding a
constant $\eps > 0$ to the argument, thereby introducing one more hyperparameter; we refer to $\S$\ref{subsec:comparison} below for a more detailed discussion.

A conceptually different approach is pursued in \citet{Jojic2011}. Instead of the
usual loss + $\ell_1$-norm formulation with the $\ell_1$-norm arising as the convex envelope\ifthenelse{\boolean{isjournal}}{ (also known
  as bi-conjugate, cf.~\citet{Rockafellar1970}, \S12)}{} of the $\ell_0$-norm on the unit $\ell_{\infty}$-ball, the authors suggest to work with the
convex envelope of loss + $\ell_0$-norm. A major drawback is computational cost since
evaluation of the convex envelope already entails solving a convex optimization problem.

\paragraph{Outline and contributions of the paper.} As a preliminary step, we provide a brief
analysis of high-dimensionsal estimation under simplex constraints in $\S$\ref{sec:basicperformance}.
Such analysis provides valuable insights when designing suitable sparsity-promoting schemes. An important
observation is that empirical risk minimization (ERM) and elements of $\Delta^p$ contained in a
``high confidence set'' for $\beta^*$ (a construction inspired by the Dantzig selector of \citet{CandesTao2007})
already enjoy nice statistical guarantees, among them adaptation to sparsity under a restricted strong convexity condition
weaker than that in \citet{Negahban2009}. This situation does not seem to be properly acknowledged in the work cited
in the previous paragraph. As a result, methods have been devised that may not even achieve the guarantees of ERM. By
contrast, we here focus on schemes designed to improve over ERM, particularly with respect to sparsity
of the solution and support recovery. As a basic strategy, we consider simple two-stage procedures, thresholding
and re-weighted $\ell_1$ regularization on top of ERM in \S\ref{sec:twostage}.

Alternatively, we propose a novel regularization-based scheme in $\S$\ref{sec:regularization} in which $\beta \mapsto 1/\norm{\beta}_2^2$ serves
as a relaxation of the $\ell_0$-norm on $\Delta^p$. This approach naturally extends to the matrix
case (positive semidefinite Hermitian matrices of trace one) as discussed in $\S$\ref{sec:tomography}. On the optimization side, the approach can be implemented easily by DC (difference of convex) programming \citep{DCprogramming}. Unlike other popular forms of
concave regularization such as the SCAD, capped $\ell_1$ or MCP penalties \citep{ZhangZhang2013} no extra parameter besides
the regularization parameter needs to be specified. For this purpose, we consider a generic BIC-type criterion (\citet{Schwarz1978}; \citet{Kim2012}) with the aim to achieve correct model selection respectively rank selection in the matrix case. The effectiveness of both the two-stage procedures and the regularization-based approach is demonstrated by several sets of numerical experiments
covering applications in compressed sensing, density estimation, portfolio optimization and quantum state tomography. All proofs can be found in the appendix.

\if\boolean{isjournal}{\paragraph{Notation.}} For the convenience of readers, we here gather essential notation. We denote by $\nnorm{\cdot}_q$, $0 \leq q \leq \infty$, the
usual $\ell_q$-norm or the Schatten $\ell_q$-norm depending on the context, and by $\scp{\cdot}{\cdot}$ the usual Euclidean inner product. We use $|\cdot|$ for the cardinality of a set. We write
$\mathbf{I}(\cdot)$ for the indicator function. We denote by $\{e_1,\ldots,e_d\}$ the canonical basis of $\R^d$. For $A \subseteq \R^d$, $\Pi_A: \R^d \rightarrow A$ denotes the Euclidean projection
on $A$. For functions $f(n)$ and $g(n)$, we write $f(n) \gtrsim g(n)$ and $f(n) \lesssim g(n)$ to express that $f(n) \geq C g(n)$ respectively
$f(n) \leq C g(n)$ for some constant $C > 0$. We write $f(n) \asymp g(n)$ if both $f(n) \gtrsim g(n)$ and $f(n) \lesssim g(n)$. We also use the Landau symbols $O(\cdot)$ and $o(\cdot)$.

\section{Simplex constraints in high-dimensional statistical inference: basic analysis}\label{sec:basicperformance}

Before designing schemes promoting sparsity under the constraint $\beta \in \Delta^p$, it is
worthwhile to derive basic performance bounds and to establish adaptivity to underlying sparsity when
only simplex constraints are used for estimation, without explicitly enforcing sparse solutions. Note
that the constraint $\beta \in \Delta^p$ is stronger than the $\ell_1$-ball constraint $\nnorm{\beta}_1 \leq 1$.
As a consequence, it turns out that ERM enjoys properties known from the analysis of (unconstrained) $\ell_1$-regularized
estimation, including adaptivity to sparsity under certain conditions. This already sets a substantial limit on what
can be achieved in addition by sparsity-promoting schemes.

We first fix some notation. Let $\{Z_1,\ldots,Z_n \}$ be i.i.d.~copies of a random variable $Z$ following
a distribution $P$ on some sample space $\mc{Z} \subseteq \R^d$. Let further $L:\R^p \times \mc{Z} \rightarrow \R$
a loss function such that $\forall z \in \mc{Z}$, $L(\cdot, z)$ is convex and differentiable. For $\beta \in \R^p$,
$R(\beta) = \E[L(\beta, Z)]$ denotes the expected risk and $R_n(\beta) = n^{-1} \su L(\beta, Z_i)$ its empirical
counterpart. The goal is to recover $\beta^* = \argmin_{\beta \in \Delta^p} \E[L(\beta, Z)]$.

Besides ERM which yields $\wh{\beta} \in \argmin_{\beta \in \Delta^p} R_n(\beta)$, our analysis simultaneously
covers all elements of the set
\begin{equation}\label{eq:fcs}
\mc{D}(\lambda) = \{\beta \in \Delta^p:\, \nnorm{\nabla R_n(\beta)}_{\infty} \leq \lambda \},
\end{equation}
for suitably chosen $\lambda \geq 0$ as precised below. The construction of $\mc{D}(\lambda)$ is inspired
by the constraint set of the Dantzig selector \citep{CandesTao2007}, which is extended to general loss
functions in \citet{Lounici2008, James2009, Fan2013}. Elements of $\mc{D}(\lambda)$ will be shown to have
performance comparable to $\wh{\beta}$. The set $\mc{D}(\lambda)$ need not be convex in general. For
squared loss it becomes a convex polyhedron. It is non-empty as long as $\lambda \geq \lambda_*$, where
$\lambda_* = \nnorm{\nabla R_n(\beta^*)}_{\infty}$. In many settings of interest (cf.~\cite{Lounici2008, Negahban2009}),
it can be shown that
\begin{equation}\label{eq:gradstar}
\vspace{-1ex}
\p(\lambda_* \gtrsim \sqrt{\log(p)/n}) = o(1) \; \; \, \text{as} \; n \rightarrow \infty.
\end{equation}\if\not{\boolean{isjournal}}{Here and the following, for functions $f(n)$ and $g(n)$, we write $f(n) \gtrsim g(n)$ and $f(n) \lesssim g(n)$ to express that $f(n) \geq C g(n)$ respectively
$f(n) \leq C g(n)$ for some constant $C > 0$. We write $f(n) \asymp g(n)$ if both $f(n) \gtrsim g(n)$ and $f(n) \lesssim g(n)$.}\fi
\subsection{Excess risk}
The first result bounds the excess risk of $\wh{\beta}$ and $\ds$, where in the sequel $\ds$
represents an arbitrary element of $\mc{D}(\lambda)$.
\begin{prop}\label{prop:excessrisk} For $\beta \in \R^p$, let $\psi_n(\beta) = R_n(\beta) - R(\beta)$ and
$\overline{\psi}_n(\beta) = \psi_n(\beta) - \psi_n(\beta^*)$. For $r > 0$, let $\B_1^p(r;\beta^*) = \{\beta \in \R^p: \norm{\beta - \beta^*}_1 \leq r \}$
denote the $\ell_1$-ball of radius $r$ centered at $\beta^*$ and $\overline{\Psi}_n(r) = \sup\{|\overline{\psi}_n(\beta)|:\; \beta \in \B_1^p(r;\beta^*) \}$. We then have
\begin{align*}
R(\wh{\beta}) - R(\beta^*) &\leq \overline{\Psi}_n(\nnorm{\wh{\beta} - \beta^*}_1) \leq \overline{\Psi}_n(2),\\
R(\ds) - R(\beta^*) &\leq \overline{\Psi}_n(\nnorm{\ds - \beta^*}_1) +  \lambda \nnorm{\ds - \beta^*}_1 \\
                    &\leq \overline{\Psi}_n(2) + 2 \lambda.
\end{align*}
\end{prop}

The excess risk of ERM and points in $\mc{D}(\lambda)$ can thus be bounded by controlling $\overline{\Psi}_n(r)$,
the supremum of the empirical process $\overline{\psi}_n(\beta)$ over all $\beta$ with $\ell_1$-distance at most $r$
from $\beta^*$. This supremum is well-studied in the literature on $\ell_1$-regularization. For example,
for linear regression with fixed or random sub-Gaussian design and sub-Gaussian errors as well as for
Lipschitz loss (e.g.~logistic loss), it can be shown that \citep{Geer2008}
\begin{equation}\label{eq:eprocess}
\p( \overline{\Psi}_n(r) \gtrsim r \sqrt{\log(p)/n} ) = o(1) \; \; \, \text{as} \; n \rightarrow \infty.
\end{equation}
Using that $\wh{\beta} \in \Delta^p$ and $\ds \in \Delta^p$, choosing $\lambda \asymp \lambda_*$ and
invoking \eqref{eq:gradstar}, Proposition \ref{prop:excessrisk} yields that the excess risk of ERM
and points in $\mc{D}(\lambda)$ scales as $O(\sqrt{\log(p)/n})$. As a result, ERM and finding a
point in $\mc{D}(\lambda)$ constitute persistent procedures in the sense of \citet{GreenshteinRitov2004}.
\subsection{Adaptation to sparsity}
Proposition \ref{prop:excessrisk} does not entail further assumptions on $\beta^*$ or $R_n$. In
the present subsection, however, we suppose that $\beta^* \in \Delta_0^p(s)$  and that $R_n$ obeys
a restricted strong convexity (\textsf{RSC}) condition defined as follows. Consider the set
\ifthenelse{\boolean{isjournal}}{\begin{align}\label{eq:union_cones}
\mc{C}^{\Delta}(s) =
\{\delta \in \R^p:\; \exists J \subseteq \{1,\ldots,p \}, |J| \leq s \;\,\,\,
\text{s.t.} \; \bm{1}^{\T} \delta_{J^c} = -\bm{1}^{\T} \delta_J
, \; \; \delta_{J^c} \gec 0 \}.
\end{align}}{\begin{align}\label{eq:union_cones}
\begin{split}
\mc{C}^{\Delta}(s) =
\{\delta \in \R^p:&\; \exists J \subseteq \{1,\ldots,p \}, |J| \leq s \\
&\text{s.t.} \; \bm{1}^{\T} \delta_{J^c} = -\bm{1}^{\T} \delta_J
, \; \; \delta_{J^c} \gec 0 \}.
\end{split}
\end{align}}
Observe that $\{\beta - \beta^*:\;\beta \in \Delta^p \} \subseteq \mc{C}^{\Delta}(s)$. For the
next result, we require $R_n$ to be strongly convex over $\mc{C}^{\Delta}(s)$.
\begin{cond}\label{cond:RSC} We say that the $\Delta$-\emph{\textsf{RSC}} condition
is satified for sparsity level $1 \leq s \leq p$ and constant $\kappa > 0$  if for all
$\beta \in \Delta_0^p(s)$ and all $\delta \in \mc{C}^{\Delta}(s)$, one has
\begin{equation*}
\vspace{-1ex}
R_n(\beta + \delta) - R_n(\beta) - \nabla R_n(\beta)^{\T} \delta  \geq \kappa \nnorm{\delta}_2^2.
\end{equation*}
\end{cond}
Condition \ref{cond:RSC} is an adaptation of a corresponding condition employed in \citet{Negahban2009}
for the analysis of (unconstrained) $\ell_1$-regularized ERM. Our condition here is weaker, since the
\textsf{RSC} condition in \citet{Negahban2009} is over the larger set
\ifthenelse{\boolean{isjournal}}{\begin{align*}
\mc{C}(\alpha, s) = \{\delta \in \R^p:\; \exists J \subseteq \{1,\ldots,p \}, |J| \leq s
\;\,\,\, \text{s.t.} \; \nnorm{\delta_{J^c}}_1 \leq
\alpha \nnorm{\delta_{J}}_1  \}, \; \, \text{for} \; \alpha \geq 1.
\end{align*}}{
\begin{align*}
\vspace{-1ex}
\mc{C}(\alpha, s) = \{\delta \in \R^p:&\; \exists J \subseteq \{1,\ldots,p \}, |J| \leq s \\
& \text{s.t.} \; \nnorm{\delta_{J^c}}_1 \leq
\alpha \nnorm{\delta_{J}}_1  \}, \; \, \alpha \geq 1.
\end{align*}}
We are now in position to state a second set of bounds.
\begin{prop}\label{prop:adaptation} Suppose that the $\Delta$-\emph{\textsf{RSC}} condition
is satisfied for sparsity level $s$ and $\kappa > 0$. We then have
\begin{align*}
\vspace{-1ex}
&\nnorm{\wh{\beta} - \beta^*}_2^2 \leq \frac{4 s \lambda_*^2}{\kappa^2}, \quad \nnorm{\ds - \beta^*}_2^2 \leq \frac{4 s (\lambda + \lambda_*)^2}{\kappa^2},\\
&\nnorm{\wh{\beta} - \beta^*}_1 \leq \frac{4s \lambda_*}{\kappa}, \quad \nnorm{\ds - \beta^*}_1 \leq \frac{4s (\lambda  +  \lambda_*)}{\kappa}
\end{align*}
\end{prop}
\vspace{-1ex}
Invoking \eqref{eq:gradstar} and choosing $\lambda \asymp \lambda_*$, we recover the rates
$O(s \log(p)/n)$ for squared $\ell_2$-error and $O(s \sqrt{\log(p)/n})$ for $\ell_1$-error, respectively. Combining
the bounds on $\ell_1$-error with \eqref{eq:eprocess} and Proposition \ref{prop:excessrisk}, we obtain
\begin{equation*}
R(\wh{\beta}) - R(\beta^*) \lesssim \frac{s \log p}{n}, \quad R(\ds) - R(\beta^*) \lesssim \frac{s \log p}{n}.
\end{equation*}
The above rates are known to be minimax optimal for the parameter set $\B_0^p(s)$ and squared loss \citep{Ye2010}.
Under the $\Delta$-\textsf{RSC} condition, there hence does not seem to be much room for improving over $\wh{\beta}$ and
$\ds$ as far as the $\ell_1$-error, $\ell_2$-error and excess risk are concerned. An additional plus of $\wh{\beta}$ is
that it does not depend on any hyperparameter.
\section{Two-stage procedures}\label{sec:twostage}
While $\wh{\beta}$ has appealing adaptation properties with regard to underlying sparsity, $\nnorm{\wh{\beta}}_0$
may be significantly larger than the sparsity level $s$. Note that the $\ell_2$-bound of Proposition \ref{prop:adaptation}
yields that $S(\wh{\beta}) \supseteq S(\beta^*)$ as long as $b_{\min}^* \gtrsim \lambda^* \sqrt{s}$, where
$b_{\min}^* = \min\{\beta_j^*:\, j \in S(\beta^*) \}$. If the aim is to construct an
estimator $\wh{\theta} $ achieving support recovery, i.e.,~$S(\wh{\theta}) = S(\beta^*)$, $\wh{\beta}$ needs to
be further sparsified by a suitable form of post-processing. We here consider two common schemes, thresholding
and weighted $\ell_1$-regularization. The latter is often referred to as ``adaptive lasso''~\citep{Zou2006}. Specifically, we
define
\begin{align}
\vspace{-1ex}
\text{thresholding:}&\quad\wh{\beta}_{\tau} = (\wh{\beta}_j \cdot \textbf{I}(\wh{\beta}_j \geq \tau))_{1 \leq j \leq p}  \label{eq:thresholding_refitting} \\
\text{weighted $\ell_1$:}&\quad\wh{\beta}_{\lambda}^{w} \in \argmin_{\beta \in \Delta^p} R_n(\beta)  + \lambda \scp{w}{\beta} \label{eq:adalasso},
\end{align}
where $\textbf{I}(\cdot)$ denotes the indicator function and $w = (w_j)_{j=1}^p$ is a vector of non-negative
weights. We here restrict ourselves to the common choice $w_j = 1/\wh{\beta}_j$ if $\wh{\beta}_j > 0$ and
$w_j = +\infty$ otherwise (in which case $(\wh{\beta}_{\lambda}^{w})_j = 0$), $j=1,\ldots,p$.

A third approach is to ignore the unit sum constraint first, so that $\ell_1$-regularization has a sparsity-promoting
effect, and then divide the output by its sum as a simple way to satisfy the constraint. Altogether, the two stages
are as follows.
\begin{align}\label{eq:naiveapproach}
1.\;\text{Compute} \; \wh{\beta}_{\lambda}^{\ell_1} \in \argmin_{\beta \in \R_+^p} R_n(\beta) + \lambda \bm{1}^{\T}\beta \qquad
2.\; \text{Return}\; \wh{\beta}_{\lambda}^{\ell_1}/(\bm{1}^{\T}\wh{\beta}_{\lambda}^{\ell_1}).
\end{align}

An alternative to the second step is to compute the Euclidean projection of $\wh{\beta}_{\lambda}^{\ell_1}$ on $\Delta^p$.
From the point of view of optimization, \eqref{eq:naiveapproach} has some advantages. Non-negativity constraints alone
tend to be easier to handle than simplex constraints. For projected gradient-type algorithms the projection on the constraint
set becomes trivial. Moreover, coordinate descent is applicable as non-negativity constraints do not couple several variables
(whereas simplex constraints do). Coordinate descent is one of the fastest algorithms for sparse estimation \citep{Friedman2010, Mazumder2011}, in particular for large values of $\lambda$. On the other hand, from a statistical perspective,
it is advisable to prefer $\wh{\beta}$ since it incorporates all given constraints into the optimization problem,
which leads to a weaker $\textsf{RSC}$ condition and eliminates the need to specify $\lambda$ appropriately. Indeed,
taking a large value of $\lambda$ in \eqref{eq:naiveapproach} in order to obtain a highly sparse solution increases the bias and may lead to false negatives. In addition, \eqref{eq:naiveapproach} may also lead to false positives if the ``irrepresentable condition''
\citep{ZhaoYu2006} is violated. Our experimental results (cf.~\S\ref{sec:simulations}) confirm that \eqref{eq:naiveapproach} has
considerably larger estimation error than ERM.

\paragraph{Model selection.} In this paragraph, we briefly discuss a data driven-approach for selecting the parameters $\tau$ and $\lambda$ in \eqref{eq:thresholding_refitting} and \eqref{eq:adalasso} when the aim is support recovery. It suffices to pick $\tau$ from $T = \{\wh{\beta}_j\}_{j=1}^p$, whereas for \eqref{eq:adalasso} we consider a finite set $\Lambda \subset \R^+$. We first obtain $\{\wh{\beta}_{\tau}, \; \tau \in T \}$ resp. $\{\wh{\beta}_{\lambda}^w, \; \lambda \in \Lambda \}$ and then perform model selection on the candiate models induced by the support sets $\{S(\wh{\beta}_{\tau}), \; \tau \in T \}$ resp. $\{ S(\wh{\beta}_{\lambda}^w), \; \lambda \in \Lambda \}$. Model selection can either be done by means of a hold-out
dataset or an appropriate model selection criterion like the RIC in the case of squared loss \citep{Foster1994}. To
be specific, let $Z_i = (X_i, Y_i)$, $i=1,\ldots,n$, and suppose that
\begin{equation}\label{eq:linearmodel}
Y_i = X_i^{\T} \beta^* + \eps_i, \quad \eps_i \sim N(0, \sigma^2), i=1,\ldots,n.
\end{equation}
Then for $S \subseteq \{1,\ldots,p\}$, the RIC is defined as
\begin{equation*}
\vspace{-1ex}\text{RIC}(S) = \min_{\beta \in \R^p: \beta_{S^c} = 0} \frac{1}{n} \su (Y_i - X_i^{\T} \beta)^2 + \frac{2 \sigma^2 \log p}{n} |S|.
\end{equation*}
Note that minimizing $\text{RIC}$ with respect to $S$ is equivalent to solving an $\ell_0$-norm-regularized least
squares problems with regularization parameter $2 \sigma^2 \log(p)/n$. This approach is known to be model selection
consistent in a high-dimensional setting \citep{Kim2012, ZhangZhang2013}. This implies that as long as $\ell_0$-norm-regularized least
squares with above choice of the regularization parameter achieves support recovery, the same is achieved by thresholding as long as the indices of the $s$ largest coefficients of $\wh{\beta}$ equal $S(\beta^*)$ (and thus
$S(\beta^*) \in \{S(\wh{\beta}_{\tau})\}_{\tau \in T}$). The same is true for weighted $\ell_1$-regularization provided $S(\beta^*) \in \{ S(\wh{\beta}_{\lambda}^w) \}_{\lambda \in \Lambda}$. In fact, the approach is not specific to thresholding/weighted $\ell_1$-regularization and applies to any other method yielding candidate support sets indexed by a tuning parameter.

If $\sigma$ is unknown, it has to be
replaced by an estimator $\wh{\sigma}$ at least obeying $\wh{\sigma} \asymp \sigma$ \citep{Kim2012}. We refer to
\citet{SunZhang2012, Fan2012} for suggestions of specific estimators $\wh{\sigma}$.

\section{Regularization with the negative $\ell_2$-norm}\label{sec:regularization}
A natural concern about ERM (optionally followed by a sparsification step) is that possible
prior knowledge about sparsity is not incorporated into estimation. The hope is that if
such knowledge is taken into account, the guarantees of $\S$\ref{sec:basicperformance} can even be improved upon.
In particular, one may be in position to weaken $\Delta$-\textsf{RSC} significantly.

It turns out that any sparsity-promoting regularizer $\Omega$ on $\Delta^p$ cannot be convex. To see this,
note that if $\Omega$ is sparsity-promoting, it should assign strictly smaller values to the vertices of $\Delta^p$
(which are maximally sparse) than to its barycentre (which is maximally dense), i.e.,
\begin{equation}\label{eq:regularizer_requirement}
\Omega(e_j) < \Omega( \{ e_1 + \ldots + e_p \}/p), \; \; j=1,\ldots,p,
\end{equation}
where $\{ e_j \}_{j = 1}^p$ is the standard basis of $\R^p$. However, \eqref{eq:regularizer_requirement}
contradicts convexity of $\Omega$, since by Jensen's inequality
\begin{equation*}
\Omega(\{ e_1 + \ldots + e_p \}/p) \leq \{\Omega(e_1) + \ldots \Omega(e_p) \}/p.
\end{equation*}

\subsection{Approach}\label{subsec:approach}
For $0 \neq \beta \in \R^p$, consider $\Omega(\beta) = \nnorm{\beta}_1^2 / \nnorm{\beta}_2^2$. $\Omega$ can
be seen as a   ``robust'' measure of sparsity. One has $\nnorm{\beta}_0 \geq \Omega(\beta)$
with equality holding iff $\{|\beta_j|, \; j \in S(\beta) \}$ is constant. By ``robustness'' we here
mean that $\Omega$ is small for vectors that have few entries of large magnitude while the number
of non-zero elements may be as large as $p$. Using that $\nnorm{\beta}_2^2 \leq \nnorm{\beta}_{\infty} \nnorm{\beta}_1$ yields
the alternative $\overline{\Omega}(\beta) = \nnorm{\beta}_1 / \nnorm{\beta}_{\infty}$. As $\nnorm{\beta}_1 = 1 \; \forall\beta \in \Delta^p$, we have
\begin{equation}\label{eq:hierarchy}
\frac{1}{\nnorm{\beta}_{\infty}} \leq \frac{1}{\nnorm{\beta}_{2}^2} \leq \nnorm{\beta}_0 \; \; \forall\beta \in \Delta^p.
\end{equation}
The map $\beta \mapsto 1/\nnorm{\beta}_{\infty}$ is proposed as a sparsity-promoting regularizer on $\Delta^p$ in \citet{Pilanci2012}. It
yields a looser lower bound on $\beta \mapsto \nnorm{\beta}_0$ than the map $\beta \mapsto 1/\nnorm{\beta}_{2}^2$ advocated
in the present work. Both these lower bounds are sparsity-promoting on $\Delta^p$ as indicated by Figure \ref{fig:contours}.\\

\begin{figure}[h!]
\begin{center}
\begin{tabular}{lll}
\includegraphics[width=0.24\textwidth]{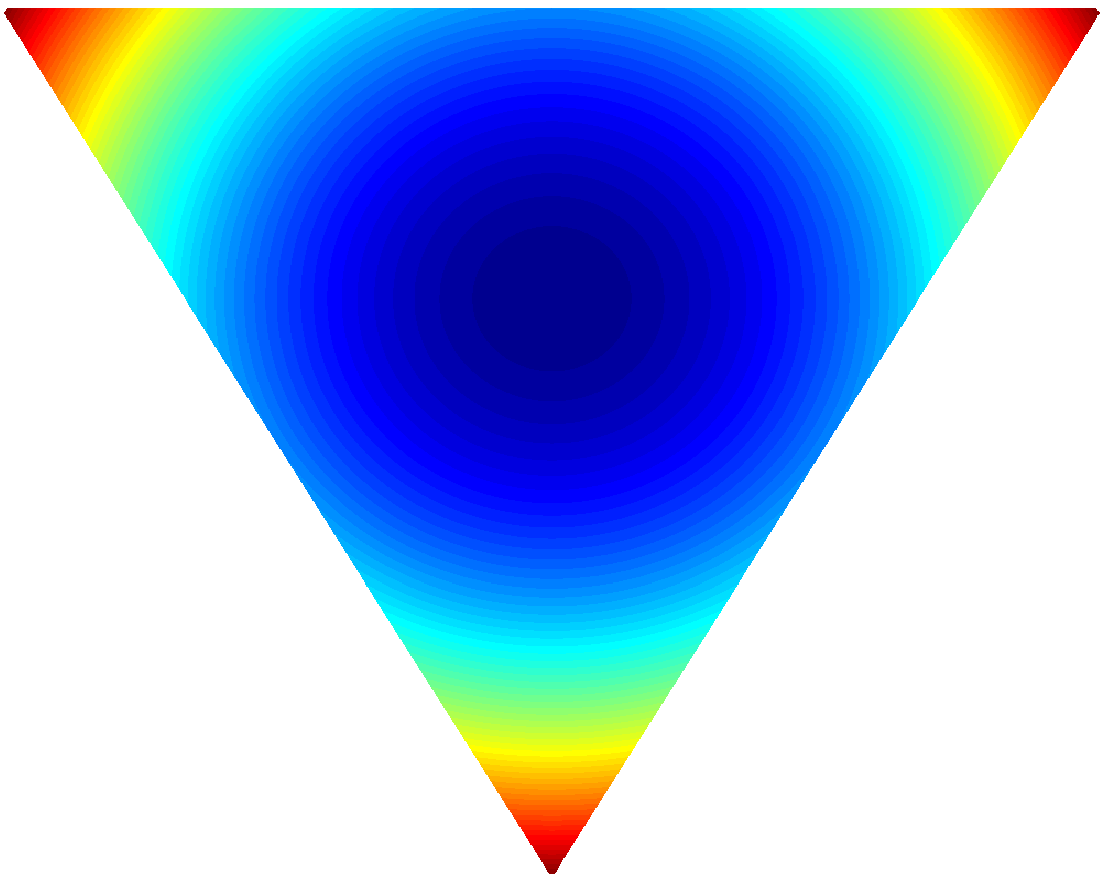} & \includegraphics[width=0.24\textwidth]{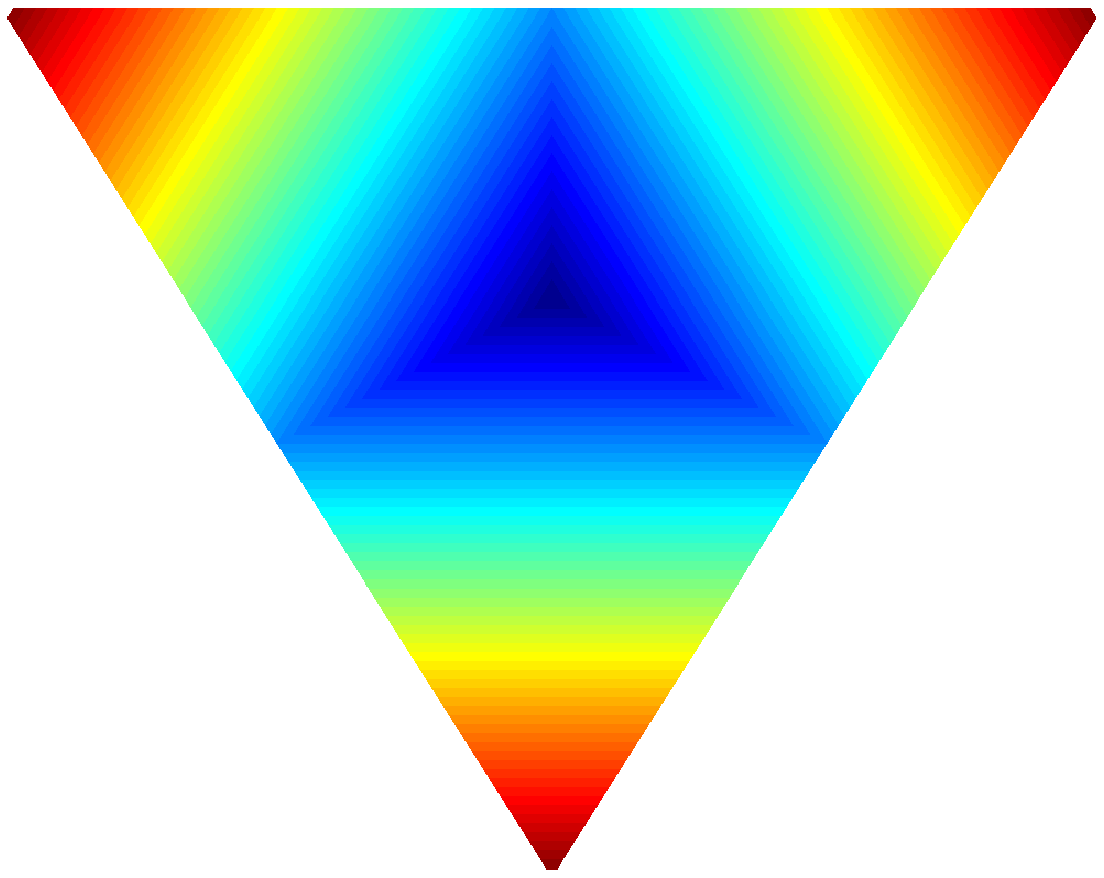} &
\includegraphics[height=0.12\textheight]{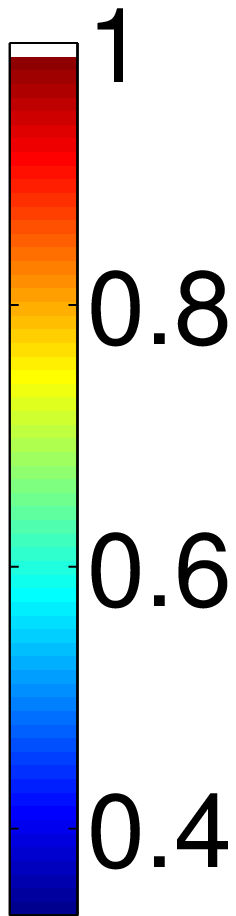}
\end{tabular}
\end{center}
\caption{Contours of $\beta \mapsto \nnorm{\beta}_2^2$ (left) and $\beta \mapsto \nnorm{\beta}_{\infty}$ (right) on $\Delta^3$.}
\label{fig:contours}
\end{figure}

From the above discussion, we conclude that finding points of large $\ell_2$-norm is a reasonable surrogate
for finding sparse solutions. This makes us propose the following modifications of $\wh{\beta}$ and $\wt{\beta}_{\lambda}$,
respectively.
\begin{align}
\vspace{-1ex}
&\rer \in \argmin_{\beta \in \Delta^p} R_n(\beta) - \lambda \nnorm{\beta}_2^2, \label{eq:rem} \\
&\dss \in \argmin_{\beta \in \mc{D}(\lambda)} -\nnorm{\beta}_2^2, \; \;  \text{with $\mc{D}(\lambda)$ as in \eqref{eq:fcs}}  \label{eq:dss}
\end{align}
Note the correspondence of \eqref{eq:rem}$\,$/$\,$\eqref{eq:dss} on the one hand, and the lasso respectively
Dantzig selector on the other hand.

Regarding \eqref{eq:rem}, it appears more consequent to use $1/\nnorm{\beta}_2^2$ in place of $-\nnorm{\beta}_2^2$ in
light of \eqref{eq:hierarchy}. Eventually, this is a matter of parameterization. While $\beta \mapsto \nnorm{\beta}_0$ is
the canonical measure of sparsity, $\beta \mapsto -1/\nnorm{\beta}_0$ is another one. It is lower bounded by $\beta \mapsto -\nnorm{\beta}_2^2$. We prefer the negative over the inverse for computational reasons: the optimization problem
in \eqref{eq:rem} is a ``difference of convex'' (DC) program \citep{DCprogramming} and hence more amenable to
optimization; see the next subsection for details.

The problem in \eqref{eq:dss} is also a DC program if $\mc{D}(\lambda)$ is convex. Note that
for \eqref{eq:dss} minimizing the negative $\ell_2$-norm is equivalent to minimizing the inverse $\ell_2$-norm.

\subsection{Least squares denoising}
In order to show that the negative $\ell_2$-norm, when combined with simplex constraints, promotes exactly sparse solutions, we elaborate on \eqref{eq:rem} in the simple setup of denoising. Let
$Z_i = \beta_i^* + \eps_i$, $i=1,\ldots,n=p$, where $\beta^* \in \Delta_0^n(s)$ and
the $\{\eps_i\}_{i = 1}^n$ represent random noise. We consider squared loss, i.e.,~$L(\beta, Z_i) = (Z_i - \beta)^2$,
$i=1,\ldots,n$. This yields the optimization problem
\begin{equation}\label{eq:denoising}
\vspace{-1ex}
\min_{\beta \in \Delta^n} \frac{1}{n}\nnorm{\mathbf{Z} - \beta}_2^2 - \lambda \nnorm{\beta}_2^2, \; \; \mathbf{Z}=(Z_i)_{i=1}^n.
\end{equation}
It turns out that \eqref{eq:denoising} can be recast as Euclidean projection of $\mathbf{Z}/\gamma$ on $\Delta^n$, where $\gamma$ is
a function of $\lambda$. Based on this re-formulation, one can derive conditions on $\beta^*$ and $\lambda$ such that $\rer$ achieves support
recovery.
\begin{prop}\label{prop:denoising}
Consider \eqref{eq:denoising} and suppose that $z_{(1)} > \ldots > z_{(n)}$, where the $\{ z_{(i)} \}_{i = 1}^n$
denote the ordered realizations of the $\{ Z_i \}_{i = 1}^n$. For all $\lambda \geq 1/n$, we have
$\rer = \big(\mathbf{I}(Z_i = z_{(1)})\big)_{i = 1}^n$. For all $0 \leq \lambda < 1/n$, we have
$\rer = \argmin_{\beta \in \Delta^n} \nnorm{\mathbf{Z}/\gamma - \beta}_2^2$, where $\gamma = 1 - n \lambda$. Moreover, if $2 s \max_{1 \leq i \leq n}|\eps_i|/n < \lambda < 1/n$ and
$b_{\min}^* > (n \lambda)/s + 2 \max_{1 \leq i \leq n}|\eps_i|$, we have $S(\rer) = S(\beta^*)$.
\end{prop}
In particular, for $\lambda = (1 + \delta) 2s  \max_{1 \leq i \leq n}|\eps_i|/n$ for any $\delta > 0$, the required lower bound on $b_{\min}^*$ becomes
$4 (1 + \delta)  \max_{1 \leq i \leq n}|\eps_i|$. For the sake of reference, we note that in the Gaussian sequence model with $\eps_i \sim N(0, \sigma^2/n)$ (cf.~\citet{monographJohnstone}), we have $\max_{1 \leq i \leq n}|\eps_i| \asymp \sqrt{\log(n)/n}$.

The denoising problem \eqref{eq:denoising} can be seen as least squares regression problem in which
the design matrix equals the identity. For general design matrices, analysis becomes harder, in particular
because the optimization problem may be neither convex nor concave. In the latter case, the minimum
is attained at one of the vertices of $\Delta^p$.

\subsection{Optimization}
Both \eqref{eq:rem} and \eqref{eq:dss} are non-convex in general. Even more, maximizing the Euclidean norm over a convex
set is NP-hard in general \citep{Pardalos1991}. To get a handle on these two problems, we exploit  the fact that both objectives are in
DC form, i.e.,~can be represented as $f(\beta) = g(\beta) - h(\beta)$ with $g$ and $h$ both being convex. Linearizing
$-h$ at a given point yields a convex majorant of $f$ that is tight at that point. Minimizing the majorant
and repeating yields an iterative procedure known as concave-convex procedure (CCCP, \citet{Yuille2003}) that falls into the
more general framework of majorization-minimization (MM) algorithms \citep{Lange2000}. When applied to
\eqref{eq:rem} and \eqref{eq:dss}, this approach yields Algorithm \ref{alg:dc}.
\begin{algorithm}\caption{}\label{alg:dc}
\eqref{eq:rem}: $\min_{\beta \in \Delta^p} R_n(\beta) - \lambda \nnorm{\beta}_2^2$
\begin{algorithmic}
\STATE {\bfseries Initialization:} $\beta^0 \in \Delta^{p}$
\STATE \textbf{repeat} $\;\beta^{k+1} \in \argmin_{\beta \in \Delta^p} R_n(\beta) - 2 \scp{\beta^k}{\beta - \beta^k}$
\STATE \textbf{until}  $\;\;\,R_n(\beta^{k+1}) -2\scp{\beta^k}{\beta^{k+1}}=R_n(\beta^k)$ 
\end{algorithmic}
\hrule
\vspace*{0.6ex}
\eqref{eq:dss}: $\min_{\beta \in \mc{D}(\lambda)} -\nnorm{\beta}_2^2$
\begin{algorithmic}
\STATE {\bfseries Initialization:} $\beta^0 \in \mc{D}(\lambda)$
\STATE \textbf{repeat} $\beta^{k+1} \in \argmin_{\beta \in \mc{D}(\lambda)} -2\scp{\beta^k}{\beta-\beta^k}$
\STATE \textbf{until} $\scp{\beta^k}{\beta^{k+1}-\beta^k}=0$
\end{algorithmic}
\end{algorithm}\hfill

For the second part of Algorithm \ref{alg:dc} to be practical, it is assumed that $\mc{D}(\lambda)$ is convex. The
above algorithms can be shown to yield strict descent until convergence to a limit point satisfying
the first-order optimality condition of the problems \eqref{eq:rem}/\eqref{eq:dss}. This is the content
of the next proposition.
\begin{prop}\label{prop:convergence}
Let $f$ denote the objective in \eqref{eq:rem} or \eqref{eq:dss}. The elements of the sequence $\{ \beta^k \}_{k \geq 0}$
produced by the above algorithms are feasible for problems \eqref{eq:rem} resp.~\eqref{eq:dss} and
satisfy $f(\beta^{k+1}) < f(\beta^k)$ until convergence. Moreover, the limit
satisfies the first-order optimality condition of the respective problem.
\end{prop}
An appealing feature of Algorithm \ref{alg:dc} is that solving each sub-problem in the
repeat step only involves minor modifications of the computational approaches used for ERM
resp.~finding a feasible point in $\mc{D}(\lambda)$.

When selecting the parameter $\lambda$ by means of a grid search, we suggest solving the associated
instances of \eqref{eq:rem}/\eqref{eq:dss} from the smallest to the largest value of $\lambda$, using
the solution from the current instance as initial iterate for the next one. For the smallest value of $\lambda$, we recommend
using $\wh{\beta}$ and any point from $\mc{D}(\lambda)$ as initial iterate for \eqref{eq:rem} respectively \eqref{eq:dss}.
Running Algorithm \ref{alg:dc} for formulation \eqref{eq:dss} has the advantage that all iterates are contained in $\mc{D}(\lambda)$
and thus enjoy at least the statistical guarantees of $\ds$ derived in $\S$\ref{sec:basicperformance}. According to our
numerical results, it is formulation \eqref{eq:rem} that achieves better performance (cf.~$\S$\ref{sec:simulations}).

\subsection{Comparison to other regularization schemes}\label{subsec:comparison}

There exist a variety of other regularization schemes one may want to consider for the given problem, in particular those already mentioned in the introduction. The present subsection provides a more detailed overview on such alternatives, and we justify why we think that our proposal offers advantages.\\

\noindent\textbf{Iterative Hard Thresholding} \citep{Kyrilidis2013}\textbf{.}\hspace{0.2in} This is an iterative algorithm whose iterates are given by the relation
\begin{equation*}
\beta^{k+1} \leftarrow \Pi_{\Delta_0^p(s)}(\beta^k - \alpha^k \nabla R_n(\beta^k)),
\end{equation*}
where the projection operator $\Pi_{\Delta_0^p(s)}$ can be evaluated in near-linear time \citep{Kyrilidis2013}. As pointed out above, $s$ is typically not known. Numerical results suggest that the approach is sensitive to over-specification of $s$ (cf.~). A second issue is the choice of the step-sizes $\{\alpha^k\}$. \citet{Kyrilidis2013} consider a constant step-size based on the Lipschitz constant of the gradient as normally used for projected gradient descent with
convex constraint sets \citep{Bertsekas1999}, as well as strategy developed in \citet{Kyrilidis2011}. Neither of these has been supported by theoretical analysis. For squared loss, convergence guarantees can be established under the restricted isometry property (RIP) and a constant step-size depending on RIP constants, which are not accessible in practice.\\

\noindent\textbf{Inverse $\ell_{\infty}$-regularization} \citep{Pilanci2012}\textbf{.}\hspace{0.2in} \citet{Pilanci2012} consider regularized risk estimation with the inverse of the $\ell_{\infty}$-norm as regularizer, that is
\begin{equation}\label{eq:pilanci}
\argmin_{\beta \in \Delta^p} R_n(\beta) + \lambda (1/\nnorm{\beta}_{\infty}).
\end{equation}
As discussed above and illustrated by Figure \ref{fig:contours}, the function $\beta \mapsto 1/\nnorm{\beta}_{\infty}$ is sparsity-promoting on $\Delta^p$. While this function is non-convex, problem \eqref{eq:pilanci} can be reduced to $p$ convex optimization problems (one for each coordinate, \citet{Pilanci2012}). This is an appealing property, but the computational effort that is required to solve these $p$ convex optimization problem is out of proportion compared to most other approaches considered herein. In noiseless settings, computation is more manageable since $\lambda$ does not need to be tuned. In our numerical studies, the performance is inferior to \eqref{eq:rem}/\eqref{eq:dss}.\\

\noindent\textbf{Entropy regularization} \citep{Shashanka07}\textbf{.}\hspace{0.2in}
For $\beta \in \Delta^p$, consider the Shannon entropy $\beta \mapsto \mc{H}(\beta) \coloneq -\sum_{j = 1}^p \log(\beta_j) \beta_j$ with the convention $\log(0) \cdot 0 = 0$. While $\mc{H}$ is a natural measure of sparsity on $\Delta^p$, there is a computational issue that makes it less attractive. It is not hard to show that the subdifferential \citep{Rockafellar1970} of $\mc{H}$ is empty at any point in
$\partial \Delta^p \coloneq \{x \in \Delta^p:\,\exists j \; \text{s.t.}\; x_j = 0 \}$. Because of this, approaches that make use of (sub-)gradients of the regularizer such as projected (sub-)gradient descent or DC programming) cannot be employed. As already mentioned, coordinate descent is not an option given the simplex constraints. A workaround would be the use of $\mc{H}_{\eps}(\beta) = -\sum_{j = 1}^p \log(\beta_j + \eps) \beta_j$ for some $\eps > 0$. This leads to an extra parameter in addition to the regularization parameter, which we consider as a serious drawback. In fact, $\eps$ needs to be tuned as values very close to zero may negatively affect convergence of iterative algorithms used for optimization.\\

\noindent\textbf{$\ell_q$-regularization for $0 < q < 1$.}\hspace{0.2in}
The regularizer $\beta \mapsto \nnorm{\beta}_q^q = \sum_{j = 1}^p \beta_j^q$ for $\beta \in \R_+^p$suffers from the same issue as the entropy regularizer: the subdifferential is empty for all points in $\partial \Delta^p$. Apart from that, it is not clear how close $q$ needs to be to zero so that appropriately sparse solutions are obtained.\\

\noindent\textbf{Log-sum regularization} \citep{Candes2007, Ugander2011}\textbf{.}\hspace{0.2in}
The regularizer $\beta \mapsto \sum_{j = 1}^p \log(\beta_j + \eps)$ for $\beta \in \R_+^p$ and $\eps > 0$ is suggested in
\citet{Candes2007}. When restricted to $\Delta^p$, the above regularizer can be motivated from a Bayesian viewpoint as
in terms of a specific prior on $\beta$ called pseudo-Dirichlet in \citet{Ugander2011}. The above regularizer is concave and
continuously differentiable and can hence be tackled numerically via DC programming in the same manner as in Algorithm \ref{alg:dc}.
Again, the choice of $\eps$ may have a significant effect on the result/convergence  and thus needs to be done with care. Small values of $\eps$ lead to a close approximation of the $\ell_0$-norm. While this is desired in principle, optimization (e.g.~via Algorithm \ref{alg:dc}) becomes more challenging as the regularizer approaches discontinuity at zero.\\

\noindent\textbf{Minimax concave penalty} \citep{Zhang2010} \textbf{and related regularizers.}\hspace{0.2in}
The minimax concave penalty (MCP) can be written as $P(\beta)= \lambda \nnorm{\beta}_1 + Q(\beta)$ with
\begin{equation*}
Q(\beta) = \sum_{j = 1}^p \left\{ -\frac{\beta_j^2}{2b}  \textbf{I}(|\beta_j| \leq b \lambda) + \left( \frac{b \lambda^2}{2} - \lambda |\beta_j| \right) \textbf{I}(|\beta_j| > b \lambda) \right\},
\end{equation*}
where $\lambda, b > 0$ are parameters. Since $\nnorm{\beta}_1 = 1 \, \forall \beta \in \Delta^p$, the MCP reduces to $Q$ under simplex constraints. Using the re-parameterization $(\lambda, b) \mapsto (\lambda, \theta \coloneq b \lambda)$, we have
\begin{equation*}
Q(\beta) = \sum_{j = 1}^p \left\{ -\frac{\lambda}{2 \theta}  \beta_j^2 \textbf{I}(|\beta_j| \leq \theta)
 +  \left(\frac{\lambda \theta}{2} - \lambda \beta_j)  \textbf{I}(|\beta_j| > \theta \right) \right\}
\end{equation*}
Under simplex constraints, $\theta$ can be restricted to $(0,1]$. Note that for $\theta = 1$,
$Q(\beta) = -(\lambda/2) \nnorm{\beta}_2^2$, in which case the MCP boils down to the negative $\ell_2$-regularizer
suggested in the present work. On the other hand, as $\theta \rightarrow 0$, the MCP behaves like the $\ell_0$-norm. The MCP is concave
and continuously differentiable so that the approach of Algorithm \ref{alg:dc} can be applied. Clearly, by optimizing the choice of the
parameter $\theta$, MCP has the potential to improve over the negative $\ell_2$-regularizer in \eqref{eq:rem}, which results as special case for $\theta = 1$. Substantial improvement is not guaranteed though and entails additional effort for tuning $\theta$.

The SCAD \citep{FanLi2001} and capped $\ell_1$-regularization \citep{ZhangZhang2013} follow a design principle similar to
that of MCP. All three are coordinate-wise separable and resemble the $\ell_1$-norm for values close to zero and flatten out for larger values so as to serve as better proxy of the $\ell_0$-norm than the $\ell_1$-norm. A parameter (like $\theta$ for MCP) controls what
exactly is meant by ``close to zero'' respectively ``large''. Note the conceptual difference between this class of regularizers
and negative $\ell_2$-regularization: members of the former have been designed as concave approximations to the $\ell_0$-norm at the level
of a single coordinate, whereas the latter has been motivated as a proxy for the $\ell_0$-norm on the simplex.\\

\noindent\textbf{Convex envelope} \citep{Jojic2011}\textbf{.}\hspace{0.2in}
An approach completely different from those discussed so far can be found in \citet{Jojic2011}. A common way
of motivating $\ell_1$-regularization is that the $\ell_1$-norm is the convex envelope (also known as bi-conjugate, \citet{Rockafellar1970}), i.e.,~the tightest convex minorant, of the $\ell_0$-norm on the unit $\ell_{\infty}$-ball. Instead of using the convex envelope of only the regularizer, \citet{Jojic2011} suggest that one may also use convex envelope of the entire regularized risk $R_n(\beta) + \lambda \nnorm{\beta}_0$. The authors argue that this approach is suitable for promoting sparsity under simplex constraints. Moreover, the resulting optimization problem is convex so that computing a global optimum appears to be feasible. However, it is in general not possible to derive a closed form expression for the convex envelope of the entire regularized risk. In this case, alreay evaluating the convex envelope at a single point requires solving a convex optimization problem via an iterative algorithm. It is thus not clear whether there exist efficient algorithms for minimizing the resulting convex envelope.

\section{Extension to the matrix case}\label{sec:tomography}
As pointed out in the introduction, there is a matrix counterpart to the problem discussed in the previous settings in
which the object of interest is a low-rank Hermitian positive semidefinite matrix of unit trace. This set of matrices
covers density matrices of quantum systems \citep{Nielsen2000}. The task of reconstructing such density matrices from so-called
observables (as e.g.~noisy linear measurements) is termed quantum state tomography \citep{Paris2004}. In the past few years, quantum state tomography based on Pauli measurements has attracted considerable interest in mathematical signal processing and statistics \citep{Gross2010, Gross2011, Koltchinskii2011, Wang2013, Cai2015}.\\

Specifically, the setup that we have in mind throughout this section is as follows. Let $\HH^m = \{B \in \C^{m \times m}: B = B^H \}$ be the Hilbert space of complex Hermitian matrices with innner product $\scp{F}{G} = \tr(FG)$, $(F,G) \in \HH \times \HH$, and let henceforth $\nnorm{\cdot}_q$, $0 \leq q \leq \infty$, denote the Schatten $q$-'norm' of a Hermitian matrix, defined as the $\ell_q$-norm of its eigenvalues. In particular, $\nnorm{\cdot}_0$ denotes the number of non-zero eigenvalues, or equivalently, the rank. We suppose that the target $B^*$ is contained in the set $\bm{\Delta}_0^{m}(r) \coloneq \bm{B}_0^{m}(r) \cap \bm{\Delta}^{m}$, where
\ifthenelse{\boolean{isjournal}}{
\begin{equation*}
\bm{B}_0^{m}(r) \coloneq \{B \in \HH^{m}:\;\nnorm{B}_0 \leq r \}, \quad\;\, \text{and} \;\;\bm{\Delta}^m \coloneq \{B \in \HH^{m}: B \gec 0,\, \tr(B) = 1\}.
\end{equation*}}
{\begin{align*}
\vspace{-1ex}
\bm{B}_0^{m}(r) &= \{B \in \HH^{m}:\;\nnorm{B}_0 \leq r \},\\
\bm{\Delta}^m &= \{B \in \HH^{m}: B \gec 0,\, \tr(B) = 1\},
\end{align*}}
i.e.,~$B^*$ is additionally positive semidefinite, of unit trace and has rank at most
$r$. In low-rank matrix recovery, the Schatten 1-norm (typically referred
to as the nuclear norm) is commonly used as convex surrogate for the rank of a matrix \citep{Recht2010}. Since
the nuclear norm is constant over $\bm{\Delta}^m$, one needs a different strategy to promote low-rankendess under
that constraint. In the sequel, we carry over our treatment of the vector case to the matrix case. The analogies
are rather direct; at certain points, however, an extension to the
matrix case may yield additional complications as detailed below. To keep matters simple, we restrict ourselves to the setup in which the $Z_i = (X_i, Y_i)$ are such that
\begin{equation}\label{eq:tracereg}
\vspace{-1ex}
Y_i = \scp{X_i}{B^*} + \eps_i, \;\; \eps_i \sim N(0, \sigma^2), \; i=1,\ldots,n,
\end{equation}
with $\{X_i \}_{i=1}^n \subset \HH^m$. Equivalently,
\begin{equation*}
\mathbf{Y} = \mc{X}(B^*) + \eps, \;\; \mathbf{Y} = (Y_i)_{i=1}^n, \; \eps = (\eps_i)_{i=1}^n,
\end{equation*}
where $\mc{X}:\HH^m \rightarrow \R^n$ is a linear operator defined by $(\mc{X}(B))_i = \scp{X_i}{B}$, $B \in \HH^m$, $i=1,\ldots,n$. We consider squared loss, i.e.,~for $B \in \bm{\Delta}^m$ the empirical risk reads
\begin{equation*}
\vspace{-1ex}
R_n(B) = \nnorm{\mathbf{Y} - \mc{X}(B)}_2^2/n.
\end{equation*}
\paragraph{Basic estimators.} As basic estimators, we consider empirical risk minimization given by $\wh{B} \in \argmin_{B \in \bm{\Delta}^m} R_n(B)$, as
well as $\wt{B}_{\lambda}$, where $\wt{B}_{\lambda}$ is any point in the set
\begin{equation}\label{eq:fcs_matrix}
\bm{D}(\lambda) = \{B \in \bm{\Delta}^m: \nnorm{\nabla R_n(B)}_{\infty} \leq \lambda \} = \Big\{B \in \bm{\Delta}^m: \frac{2}{n} \nnorm{\mc{X}^{\star}(\mc{X}(B)-y)}_{\infty} \leq \lambda \Big\},
\end{equation}
where $\mc{X}^{\star}: \R^n \rightarrow \HH^m$ is the adjoint of $\mc{X}$. Both $\wh{B}$ and $\wt{B}_{\lambda}$ show adaptation to the rank of $B^*$ under the following
restricted strong convexity condition. For $B \in \bm{B}_0^m(r) \subset \HH^m$, let $\TT(B)$ be the tangent space of $\bm{B}_0^m(r) \subset \HH^m$ at $B$ \ifthenelse{\boolean{isjournal}}{(see Definition \ref{defn:tangentspace} in the appendix)}{(cf.~supplement)}. and let $\Pi_V$ denote the projection on a subspace $V$ of $\HH^m$.
\begin{cond}\label{cond:RSC_m} We say that the $\bm{\Delta}$-\emph{\textsf{RSC}} condition
is satified for rank $r$ and constant $\kappa > 0$  if $\forall \Phi \in \mc{K}^{\bm{\Delta}}(r)$
$\nnorm{\mc{X}(\Phi)}_2^2/n \geq \kappa \nnorm{\Phi}_2^2$, where
\begin{align*}
\vspace{-1ex}
&\mc{K}^{\bm{\Delta}}(r) = \{\Phi \in  \mathbb{H}^m: \,\exists B \in \bm{B}_0^m(r) \, \text{s.t.} \\
&\tr(\Pi_{\mathbb{T}(B)^{\perp}}(\Phi)) = -\tr( \Pi_{\mathbb{T}(B)} \Phi) \; \text{and} \;  \Pi_{\mathbb{T}(B)^{\perp}}(\Phi) \gec 0 \}.
\end{align*}
\end{cond}
\vspace{-1ex}
The $\bm{\Delta}$-{\textsf{RSC}} condition is weaker than the corresponding
condition employed in \citet{Negahban2011}, which in turn is weaker than the matrix RIP condition
\cite{Recht2010}. The next statement parallels Proposition \ref{prop:adaptation},
asserting that the constraint $B \in \bm{\Delta}^m$ alone is strong enough to take advantage of low-rankedness.
\begin{prop}\label{prop:adaptation_m}
Suppose that the $\bm{\Delta}$-\emph{\textsf{RSC}} condition
is satisfied for rank $r$ and $\kappa > 0$. Set $\lambda_* = 2 \nnorm{\mc{X}^{\star}(\eps)}_{\infty}/n$, where
$\mc{X}^{\star}: \R^n \rightarrow \HH^m$ is the adjoint of $\mc{X}$. We then have
\begin{align*}
&\nnorm{\wh{B} - B^*}_2^2 \leq \frac{4 s \lambda_*^2}{\kappa^2}, \quad \nnorm{\wt{B}_{\lambda} - B^*}_2^2 \leq \frac{4 s (\lambda + \lambda_*)^2}{\kappa^2},\\
&\nnorm{\wh{B} - B^*}_1 \leq \frac{4s \lambda_*}{\kappa}, \quad \nnorm{\wt{B}_{\lambda} - B^*}_1 \leq \frac{4s (\lambda + \lambda_*)}{\kappa}.
\end{align*}
\end{prop}
\paragraph{Obtaining solutions of low rank.} While $\wh{B}$ may have low estimation error its rank can by far
exceed the rank of $B^*$, even though the extra non-zero eigenvalues of $\wh{B}$ may be small. The simplest approach
to obtain solutions of low rank is to apply thresholding to the spectrum of $\wh{B} = \wh{U} \wh{\Phi} \wh{U}^{\T}$ (the r.h.s.~representing the usual spectral decomposition), that is $\wh{B}_{\tau} = \wh{U} \wh{\Phi}_{\tau} \wh{U}^{\T}$, where $\wh{\Phi}_{\tau} = \text{diag}(\{\mathbf{I}(\wh{\phi}_j \geq \tau) \}_{j = 1}^m)$ for a threshold $\tau > 0$. Likewise, one may consider the following analog to weighted
$\ell_1$-regularization:
\begin{align}\label{eq:weightedL1_m}
\begin{split}
&\wh{B}_w  = \wh{U} \text{diag}(\{\wh{\phi}_{w,j} \}_{j = 1}^m) \wh{U}^{\T}, \\%
\text{where} \; \, &\wh{\phi}_w
\in \argmin_{\phi \in \Delta^p} \frac{1}{n} \nnorm{\mathbf{Y} - \mc{X}(\wh{U} \text{diag}(\{\phi_j \}_{j=1}^m) \wh{U}^{\T})}_2^2 + \lambda \scp{w}{\phi}
\end{split}
\end{align}
for non-negative weights $\{ w_j \}_{j = 1}^m$ as in the vector case. Note that the matrix of eigenvectors $\wh{U}$ is kept fixed at the second stage; the optimization is only over the eigenvalues. Alternatively, one may think of optimization over $\bm{\Delta}^m$ with
regularizer $B \mapsto \nnorm{B}_w = \sum_{j = 1}^m w_j \phi_j(B)$ for eigenvalues $\phi_1(B) \geq \ldots \geq \phi_m(B) \geq 0$ of $B$ in
decreasing order. However, from the point of view of optimization $\nnorm{\cdot}_w$ poses difficulties, possible non-convexity (depending on $w$) in particular.
\paragraph{Regularization with the negative $\ell_2$-norm.} One more positive aspect about the regularization scheme proposed in $\S$\ref{sec:regularization} is
that it allows a straightforward extension to the matrix case, including the algorithm used for optimization (Algorithm \ref{alg:dc}). By contrast, for regularization with the inverse $\ell_{\infty}$-norm, which can be reduced to $p$ convex optimization problems in the vector case, no such reduction seems to be possible in the matrix case. The analogs of
\eqref{eq:rem}/\eqref{eq:dss} are given by
\begin{align}
\vspace{-1ex}
&\wh{B}_{\lambda}^{\ell_2} \in \argmin_{B \in \bm{\Delta}^m} R_n(\beta) - \lambda \nnorm{B}_2^2, \label{eq:rem_m} \\
&\wt{B}_{\lambda}^{\ell_2} \in \argmin_{B \in \bm{D}(\lambda)} -\nnorm{B}_2^2 \label{eq:dss_m}.
\end{align}
Algorithm \ref{alg:dc} can be employed for optimization mutatis mutandis. In the vector case and for squared loss, formulations \eqref{eq:rem} and \eqref{eq:dss}are
comparable in terms of computational effort: each minimization problem inside the 'repeat'-loop becomes a quadratic respectively
a linear program with a comparable number of variables/constraints. In the matrix case, however, \eqref{eq:rem_m} appears to be preferable as the
sub-problems are directly amenable to a proximal gradient method. By contrast, the constraint set in \eqref{eq:dss_m} requires a more sophisticated approach.
\paragraph{Denoising.} Negative $\ell_2$-regularization in combination with the constraint set $\bm{\Delta}^m$ enforces solution
of low rank as exemplified here in the special case of denoising of a real-valued matrix (i.e.,~$B^* \in \HH^m \cap \R^{m \times m}$) contaminated by Gaussian noise. Specifically, the sampling operator $\mc{X}(\cdot) = (\scp{X_i}{\cdot})_{i = 1}^n$, $n = m(m+1)/2$, here equals the symmetric vectorization operator, that is
\begin{align}\label{eq:Xi_svec}
\begin{split}
&X_1 = e_1 e_1^{\T},\;\; X_2 = \frac{e_1 e_2^{\T} + e_2 e_1^{\T}}{\sqrt{2}},\ldots,X_{m}=\frac{e_1 e_m^{\T} + e_m e_1^{\T}}{\sqrt{2}},\;\;
X_{m+1} = e_2 e_2^{\T},\ldots,\\
&X_{2m-1} = \frac{e_2 e_m^{\T} + e_m e_2^{\T}}{\sqrt{2}},\ldots,X_{m(m+1)/2}= \frac{e_{m-1} e_m^{\T} + e_m e_{m-1}^{\T}}{\sqrt{2}}
\end{split}
\end{align}
The following proposition makes use of a result in random matrix theory due to \cite{Peng2012}.
\begin{prop}\label{prop:denoising_matrix}
Let $B^* \in \bm{\Delta}_0^{m}(r) \cap \R^{m \times m}$ with eigenvalues $\phi_1^* \geq \ldots \geq \phi_r^* > 0$ and
$\phi_{r+1}^* = \ldots = \phi_m^* = 0$, let $\mc{X}$ be defined according to \eqref{eq:Xi_svec}, and let futher
$\eps \sim N(0, \sigma^2 I_m/m)$, $\mathbf{Y} = \mc{X}(B^*) + \eps$. Consider the optimization problem
\begin{equation*}
\min_{B \in \bm{\Delta}^m} \frac{1}{n} \nnorm{\mathbf{Y} - \mc{X}(B)}_2^2 - \lambda \nnorm{B}_2^2
\end{equation*}
with minimizer $\wh{B}_{\lambda}^{\ell_2}$ and define $\Upsilon = B^* + \mc{X}^{\star}(\eps)$. Then, for all $\lambda \geq 1/n$, we have
$\wh{B}_{\lambda}^{\ell_2} = u_1 u_1^{\T}$, where $u_1$ is the eigenvector of $\Upsilon$ corresponding to its largest eigenvalue. For all $0 \leq \lambda < 1/n$, we have $\wh{B}_{\lambda}^{\ell_2} = \argmin_{B \in \bm{\Delta}^m} \nnorm{\Upsilon/\gamma - B}_2^2$, where $\gamma = 1 - n \lambda$. Moreover, there exists constants $c_0,c,C > 0$ so that if $r < c_0 m$, $\lambda \geq 6 \sigma r/n$ and $\phi_r^* \geq 5 \sigma + n \lambda / r$, we have $\nnorm{\wh{B}_{\lambda}^{\ell_2}}_0 = r$ with probability at least $1 - C \exp(-c m)$.
\end{prop}
In particular, for $\lambda = (1 + \delta) 6 \sigma  r / n$ for some $\delta > 0$, the required lower bound on $\phi_r^*$ becomes
$11(1+ \delta) \sigma$, which is proportional to the noise level of the problem as follows from the proof of the proposition.

\paragraph{Alternative regularization schemes.} The approaches of $\S$\ref{subsec:comparison} can in principle all be
extended to the matrix case by defining corresponding regularizers in terms of the spectrum of a positive semidefinite
matrix. For example, the Shannon entropy becomes the von Neumann entropy \citep{Nielsen2000}. It is important to note that
in \cite{Koltchinskii2011}, the \emph{negative} of the von Neumann entropy is employed, without constraining
the target to be contained in $\bm{\Delta}^m$. The negative von Neumann entropy still ensures positive semidefiniteness
and boundedness of the solution. Schatten $\ell_q$-regularization with $0 < q < 1$ for low rank matrix recovery is discussed in
\citet{Mohan2012, Rohde2011}. In a recent paper, \cite{Gui2015} analyze statistical properties of the counterparts of
MCP and SCAD in the matrix case. For details, we refer to the cited references and the references therein; our reasoning
for prefering negative $\ell_2$-regularization persists.

\section{Empirical results}\label{sec:simulations}
We have conducted a series of simulations to compare the different methods considered herein and to provide
additional support for several key aspects of the present work. Specifically, we study compressed
sensing, least squares regression, mixture density estimation, and quantum state tomography based on
Pauli measurements in the matrix case. The first two of these only differ by the presence respectively absence
of noise. We also present a real data analysis example concerning portfolio optimization for NASDAQ stocks
based on weekly price data from 03/2003 to 04/2008.

\subsection{Compressed sensing}\label{subsec:CS}
We consider the problem of recovering $\beta^* \in \Delta_0^p(s)$ from a small
number of random linear measurements $Y_i = \scp{X_i}{\beta^*}$, where $X_i$ is standard
Gaussian, $i=1,\ldots,n$. In short, $\mathbf{Y} = \bm{X} \beta^*$ with
$\mathbf{Y} = (Y_i)_{i=1}^n$ and $\bm{X}$ having the $\{X_i\}_{i = 1}^n$ as its rows.
Identifying $\beta^*$ with a probability distribution $\pi$ on $\{1,\ldots,p\}$, we may think
of the problem as recovering such distribution from expectations of the form
$Y_i = \sum_{j = 1}^p (X_i)_j \pi(\{ j \})$. We here show the results for $p = 500$,
$s = 50$ and $n = c s \log(p/s)$ with $c \in [0.8,2]$ (cf.~Figure \ref{fig:noiseless}). The
target $\beta^*$ is generated by selecting its support uniformly at random, drawing the non-zero
entries randomly from $[0,1]$ and normalizing subsequently. This is replicated $50$ times for
each value of $n$.

The following approaches are compared for the given task, assuming squared loss
$R_n(\beta) = \nnorm{\mathbf{Y} - \bm{X}\beta}_2^2/n$.\\

\noindent\textsf{'Feasible set'}: Note that ERM here amounts to finding a point in
$\mc{D}(0)$. The output is used as initial iterate for \textsf{'L2'}, \textsf{'weighted L1'}, and \textsf{'IHT'} below.\\[1ex]
\textsf{'L2}': $\ell_2$-norm maximization \eqref{eq:dss} with $\lambda = 0$, i.e.,~over
\begin{align}\label{eq:D0}
\begin{split}
\mc{D}(0) &= \{\beta \in \Delta^p:\; \bm{X}^{\T}(\bm{X} \beta - \mathbf{Y}) = 0 \} \\
          &= \{\beta \in \Delta^p:\; \bm{X} \beta  =  \mathbf{Y} \}  \;\;\text{with probability 1}.
\end{split}
\end{align}

\noindent\textsf{'Pilanci'}: The method of \cite{Pilanci2012} that maximizes the $\ell_{\infty}$-norm over \eqref{eq:D0}.\\

\noindent\textsf{'weighted L1'}: Weighted $\ell_1$-norm minimization (cf.~$\S$\ref{sec:twostage}) over \eqref{eq:D0}.\\

\noindent\textsf{'IHT'}: Iterative hard threshold under simplex constraints \citep{Kyrilidis2013}. Regarding the
step size used for gradient projection, we use the method in \citet{Kyrilidis2011} which empirically turned
out to be superior compared to a constant step size. \textsf{'IHT'} is run with the correct value of $s$ and is hence
given an advantage.\\

\noindent\textbf{Results.}\hspace{0.1in} Figure \ref{fig:noiseless} visualizes the fractions of recovery out of $50$ replications. A general
observation is that the constraint $\beta \in \Delta^p$ is powerful enough to reduce the required number of measurements considerably
compared to $2s \log(p/s)$ when using standard $\ell_1$-minimization without constraints.
At this point, we refer to \cite{DonohoTanner2005b} who gave a precise asymptotic characterization of this phenomenon
in a ``proportional growth'' regime, i.e.,~$n/p \rightarrow c \in (0,1)$ and $s/n \rightarrow c' \in (0,1)$. When solving
the feasibility problem, one does not explicitly exploit sparsity of the solution (even though the constraint implicitly
does). Enforcing sparsity via \textsf{'Pilanci', 'IHT', 'L2'} further improves performance. The improvements achieved by \textsf{'L2'}
are most substantial and persist throughout all sparsity levels. \textsf{'weighted L1'} does not consistenly improve over
the solution of the feasibility problem.

\begin{figure}
\begin{center}
\includegraphics[width = 0.4\textwidth]{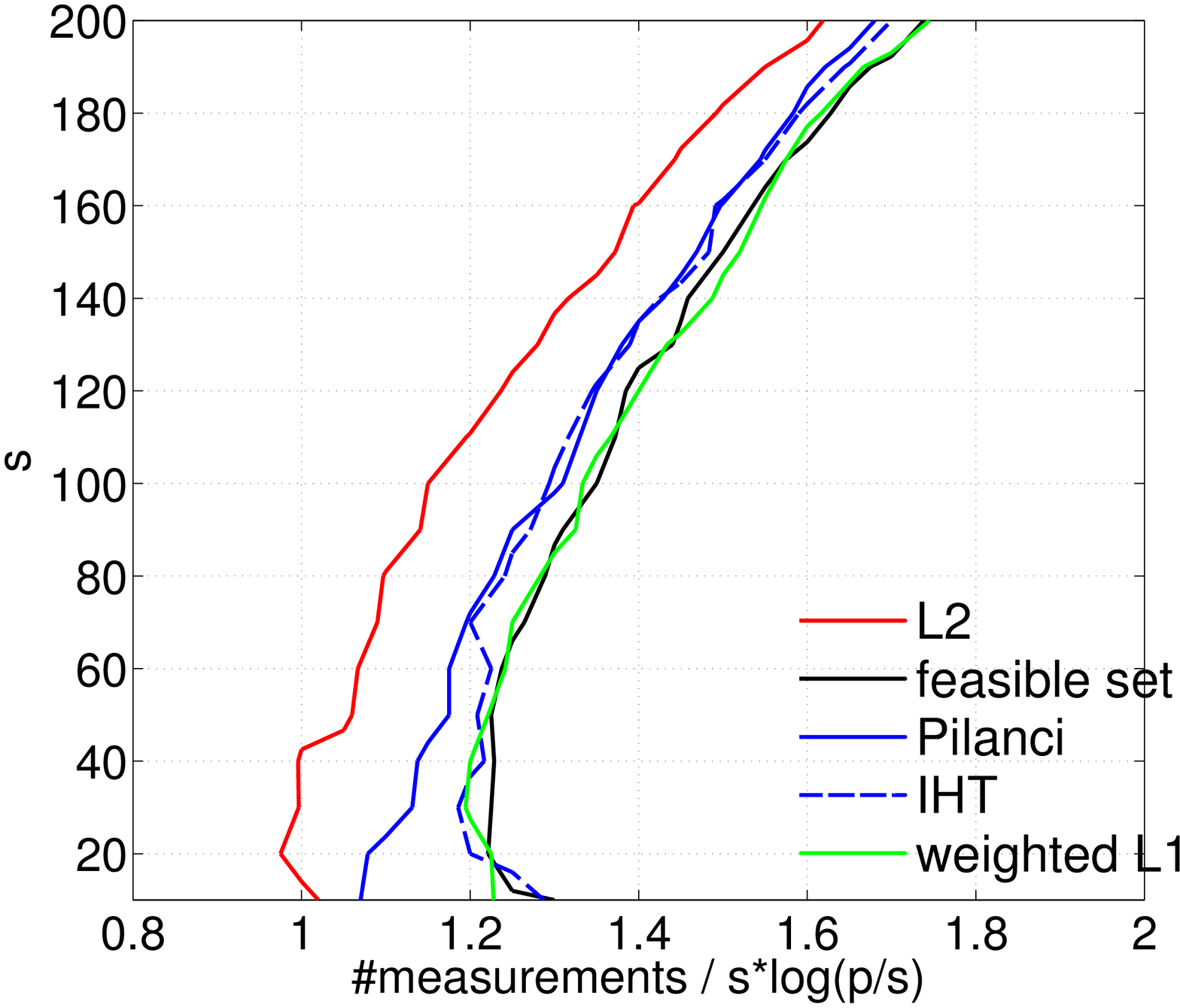} \hspace{0.3in}
\includegraphics[width = 0.4\textwidth]{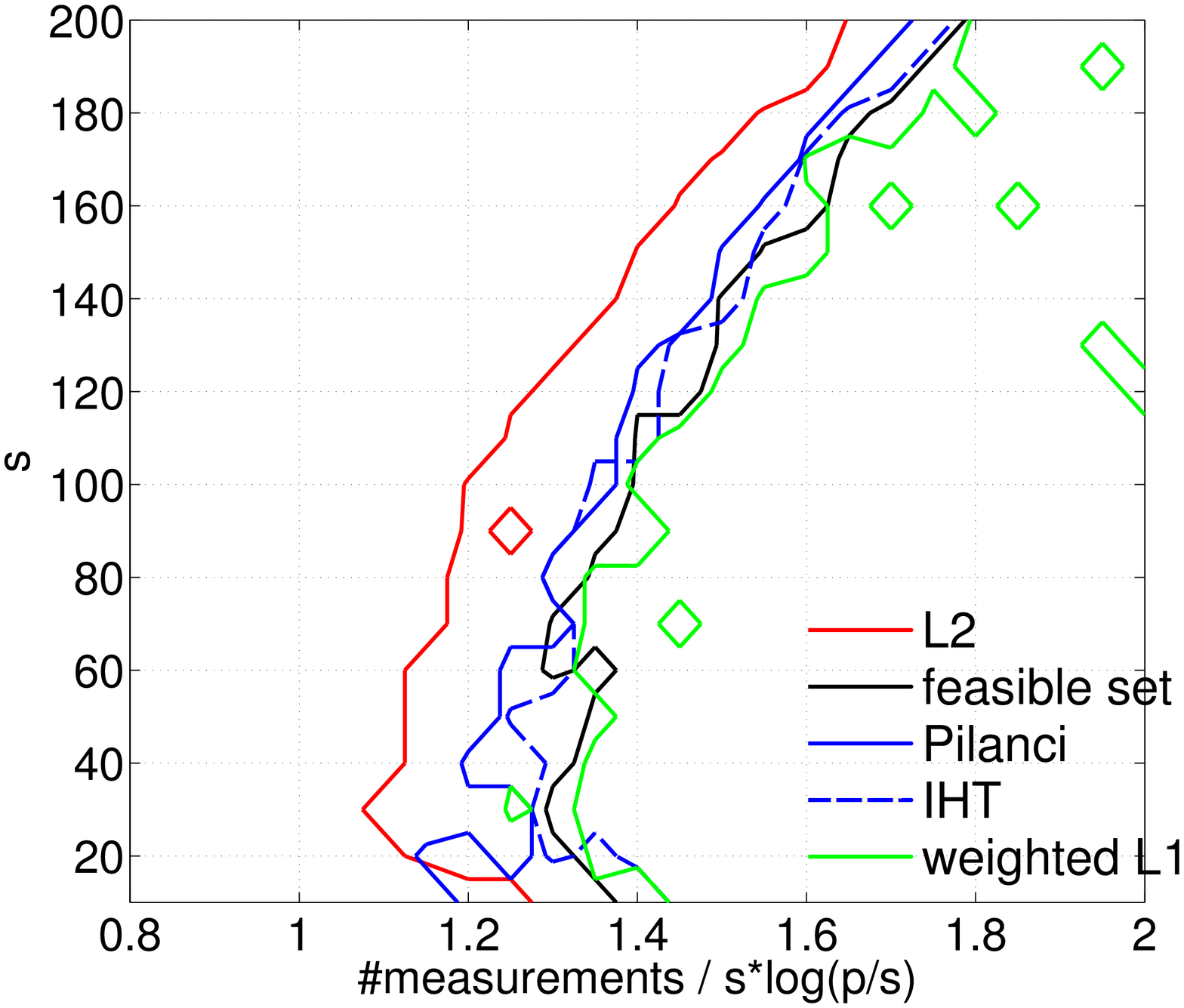}
\end{center}
\vspace{-0.2in}
\caption{Contour plots of the empirical relative frequencies of exact recovery in dependency of the number of
measurements (horizontal axis) and $s$ (vertical axis). The left and right plot show the contour levels $.75$ and $.99$, respectively. Note that the smaller the area ``left'' to and ``above'' the curve, the better the performance.}\label{fig:noiseless}
\end{figure}

\subsection{Least squares regression}\label{subsec:regression}

We next consider the Gaussian linear regression model \eqref{eq:linearmodel} with the $\{X_i \}_{i = 1}^n$
as in the previous subsection. Put differently, the previous data-generating model is changed by an
additive noise component. The target $\beta^*$ is generated as before, with the change that
the subvector $\beta_{S(\beta^*)}^*$ corresponding to $S(\beta^*)$ is projected
on $[b_{\min}^*, 1]^s \cap \Delta^s$ to ensure sufficiently strong signal, where $b_{\min}^* = \varrho \sigma \sqrt{2 \log(p)/n}$ with $\sigma = s^{-1}$
and $\varrho = 1.7$ controlling the signal strength relative
to the noise level $\lambda_0 = \sigma \sqrt{2 \log(p)/n}$. The following approaches are compared.\\

\noindent\textsf{'ERM'}: Empirical risk minimization.\\

\noindent\textsf{'Thres'}: \textsf{'ERM'} followed by hard thresholding (cf.~$\S$\ref{sec:twostage}).\\

\noindent\textsf{'L2-ERM'}: Regularized ERM with negative $\ell_2$-regularization \eqref{eq:rem}. For
the parameter $\lambda$, we consider a grid $\Lambda$ of 100 logarithmically spaced
points from $0.01$ to $\phi_{\max}(\bm{X}^{\T} \bm{X}/n)$, the maximum eigenvalue of $\bm{X}^{\T} \bm{X}/n$.
Note that for $\lambda \geq \phi_{\max}(\bm{X}^{\T} \bm{X}/n)$, the optimization problem \eqref{eq:rem} becomes
concave and the minimizer must consequently be a vertex of $\Delta^p$, i.e.,~the solution is maximally sparse
at this point, and it hence does not make sense to consider even larger values of $\lambda$. When computing
the solutions $\{\rer, \, \lambda \in \Lambda \}$, we use a homotopy-type scheme in which for each
$\lambda \in \Lambda$, Algorithm \ref{alg:dc} is initialized with the solution for the previous $\lambda$, using
the output $\wh{\beta}$ of \textsf{'ERM'} as initialization for the smallest value of $\lambda$.\\

\noindent\textsf{'L2-D'}: $\ell_2$-norm maximization \eqref{eq:dss} over $\mc{D}(C \lambda_0)$ with $\lambda_0$ being the
noise level defined above and $C \in \{0.5,0.55,\ldots,2\}$. Algorithm \ref{alg:dc} is initialized with $\wh{\beta}$
provided it is feasible. Otherwise, a feasible point is computed by linear programming.\\[1ex]
\textsf{'weighted L1'}: The approach in \eqref{eq:adalasso}. Regarding the regularization parameter,
we follow \citet{Geer2013} who let $\lambda = C \lambda_0^2$. We try 100 logarithmically spaced
values between 0.1 and 10 for $C$.\\

\noindent\textsf{'IHT'}: As above, again with the correct value of $s$. We perform a second sets of experiments though in which $s$ is over-specified by different factors ($1.2$, $1.5$, $2$)
in order to investigate the sensitivity of the method w.r.t.~the choice of the sparsitity level.\\

\noindent\textsf{'L1'}: The approach \eqref{eq:naiveapproach}, i.e.,~dropping
the unit sum constraint and normalizing the output of the non-negative $\ell_1$-regularized estimator
$\wh{\beta}_{\lambda}^{\ell_1}$. We use $\lambda = \lambda_0$ as recommended in the literature, cf.~e.g.~\cite{Negahban2009}.\\

\noindent\textsf{'oracle'}: ERM given knowledge of the support $S(\beta^*)$.\\

For \textsf{'Thres'},\textsf{'L2-ERM'} and other methods for which multiple values of a hyperparameter are
considered, hyperparameter selection is done by minimizing the RIC as defined in $\S$\ref{sec:twostage} after
evaluating each support set returned for a specific value of the hyperparameter.\\

\begin{figure}[ht!]
\begin{tabular}{lll}
\hspace*{-0.017\textwidth}\includegraphics[width = 0.333\textwidth]{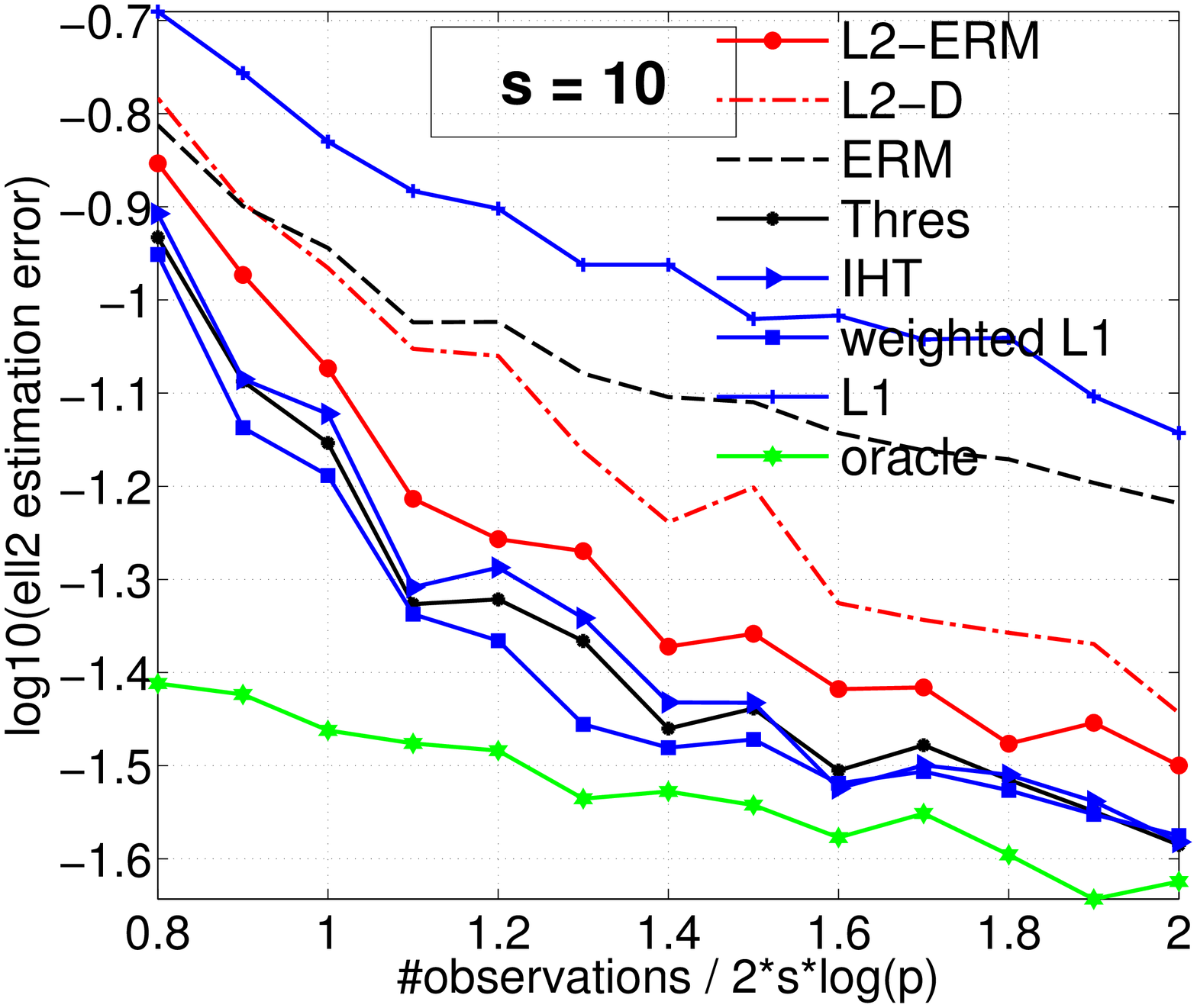} & \hspace*{-0.015\textwidth}\includegraphics[width = 0.333\textwidth]{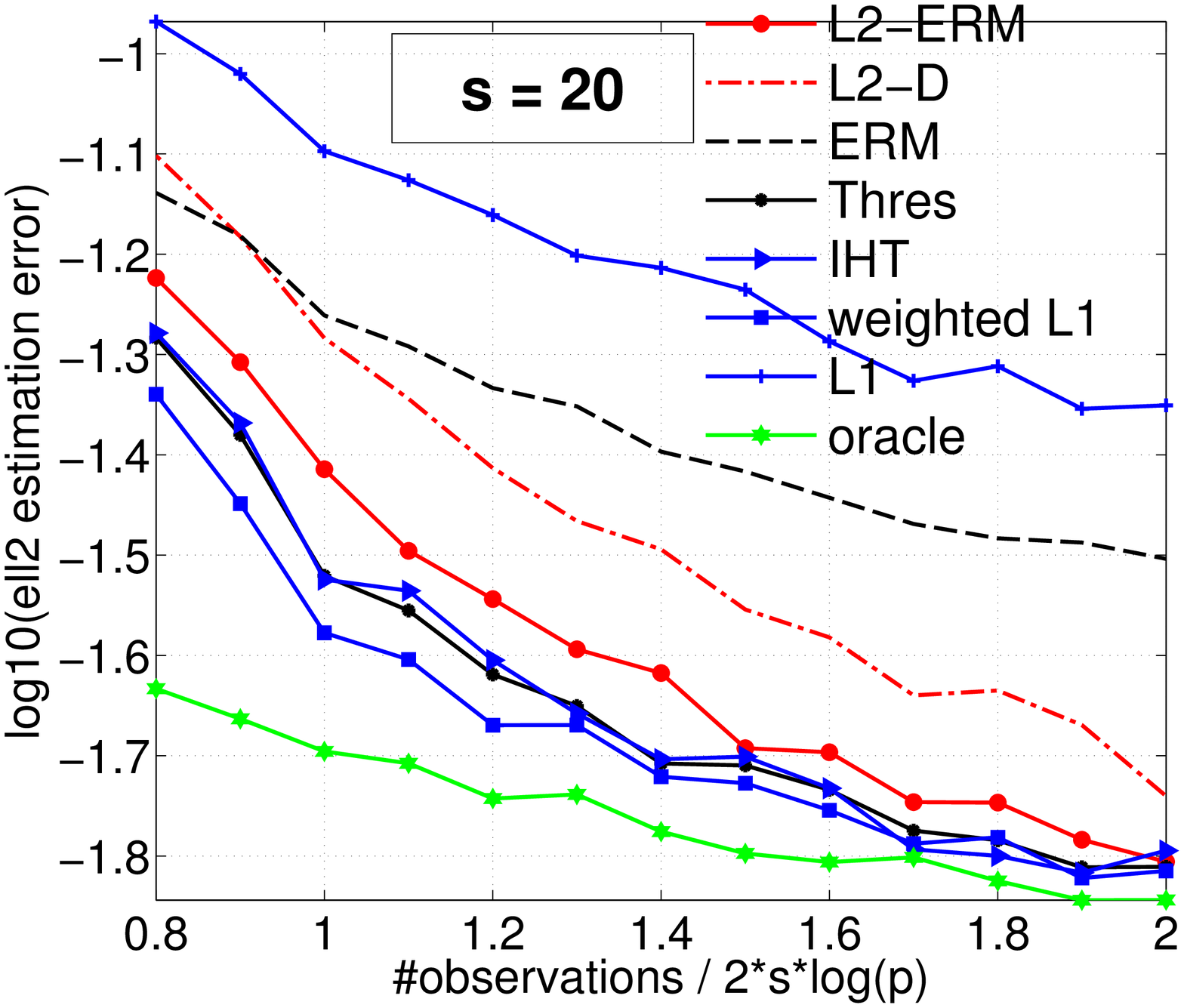} & \hspace*{-0.015\textwidth}\includegraphics[width = 0.333\textwidth]{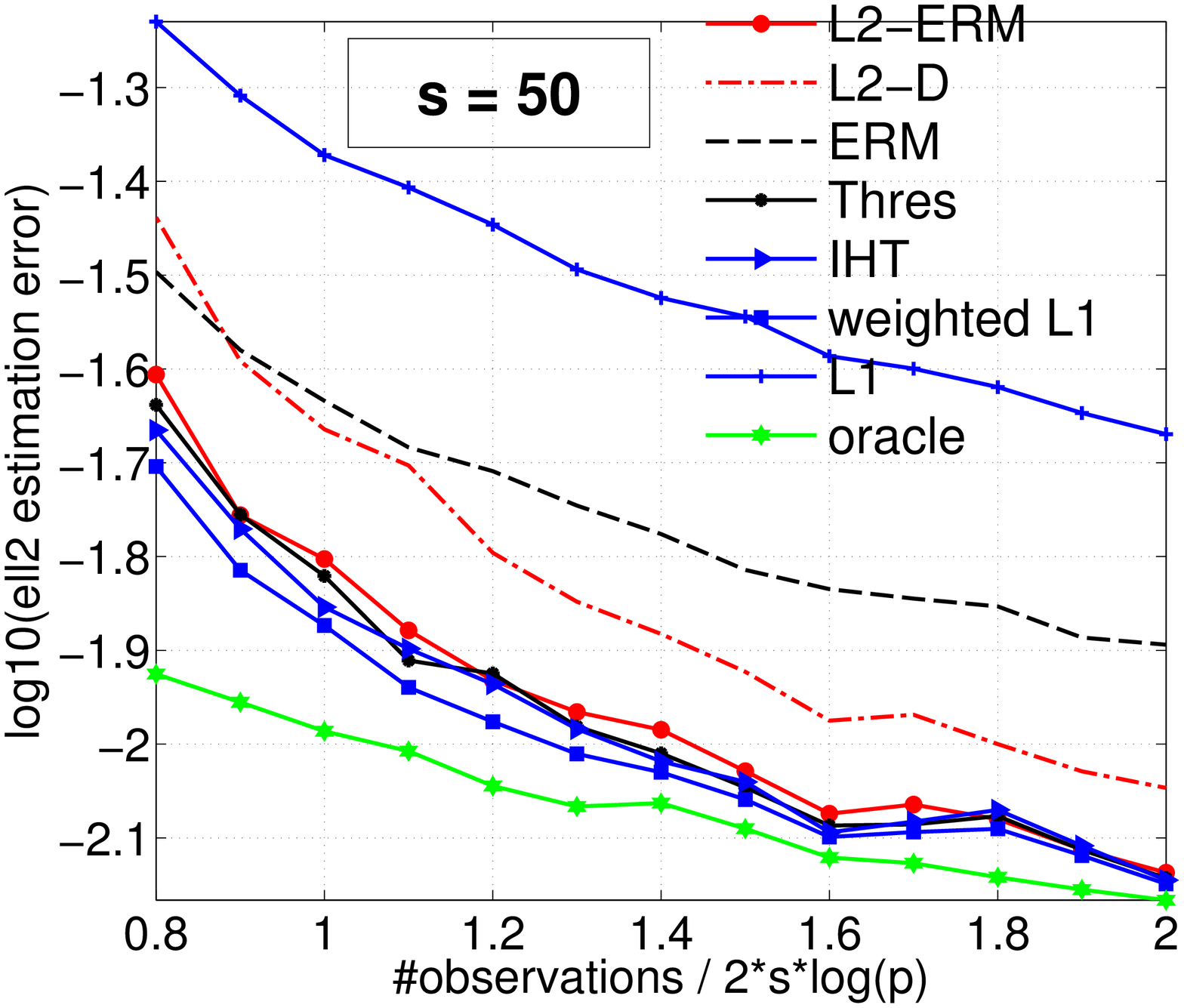} \\
{\tiny 
 $\begin{array}{l} \text{\textbf{Standard errors}(rounded):} \\
\begin{array}{|c|c|c|c|}\hline
\textbf{\textsf{L2-ERM}}  &  \textbf{\textsf{L2-D}} & \textbf{\textsf{ERM}} &  \textbf{\textsf{Thres}} \\
.04  & .04 & .02 & 0.04 \\
\hline
\textbf{\textsf{weightedL1}} & \textbf{\textsf{IHT}} & \textbf{\textsf{L1}} & \textbf{\textsf{oracle}} \\
.04 & .04 & .02 & .02 \\
\hline
 \end{array}
\end{array}
$} & {\tiny 
 $\begin{array}{l} \text{\textbf{Standard errors}(rounded):} \\
\begin{array}{|c|c|c|c|}\hline
\textbf{\textsf{L2-ERM}}  &  \textbf{\textsf{L2-D}} & \textbf{\textsf{ERM}} &  \textbf{\textsf{Thres}} \\
.03  & .03 & .01 & 0.03 \\
\hline
\textbf{\textsf{weightedL1}} & \textbf{\textsf{IHT}} & \textbf{\textsf{L1}} & \textbf{\textsf{oracle}} \\
.03 & .03 & .01 & .01 \\
\hline
 \end{array}
\end{array}
$} &  {\tiny 
 $\begin{array}{l} \text{\textbf{Standard errors}(rounded):} \\
\begin{array}{|c|c|c|c|}\hline
\textbf{\textsf{L2-ERM}}  &  \textbf{\textsf{L2-D}} & \textbf{\textsf{ERM}} &  \textbf{\textsf{Thres}} \\
.02  & .02 & .01 & 0.02 \\
\hline
\textbf{\textsf{weightedL1}} & \textbf{\textsf{IHT}} & \textbf{\textsf{L1}} & \textbf{\textsf{oracle}} \\
.02 & .02 & .01 & .01 \\
\hline
 \end{array}
\end{array}
$} \\
\\
\hspace*{-0.033\textwidth} \includegraphics[width=0.333\textwidth]{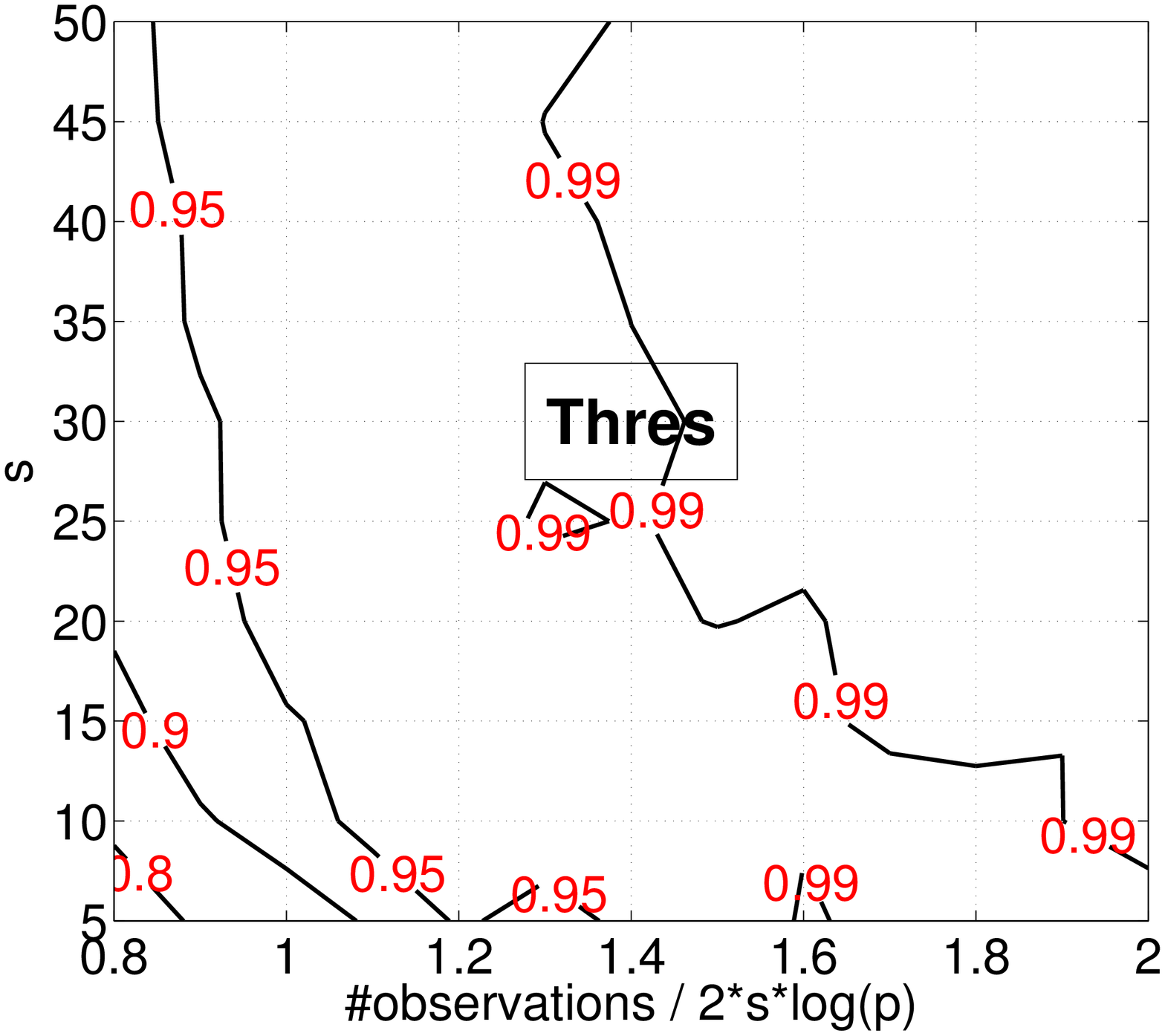}
        & \hspace*{-0.028\textwidth}      \includegraphics[width=0.333\textwidth]{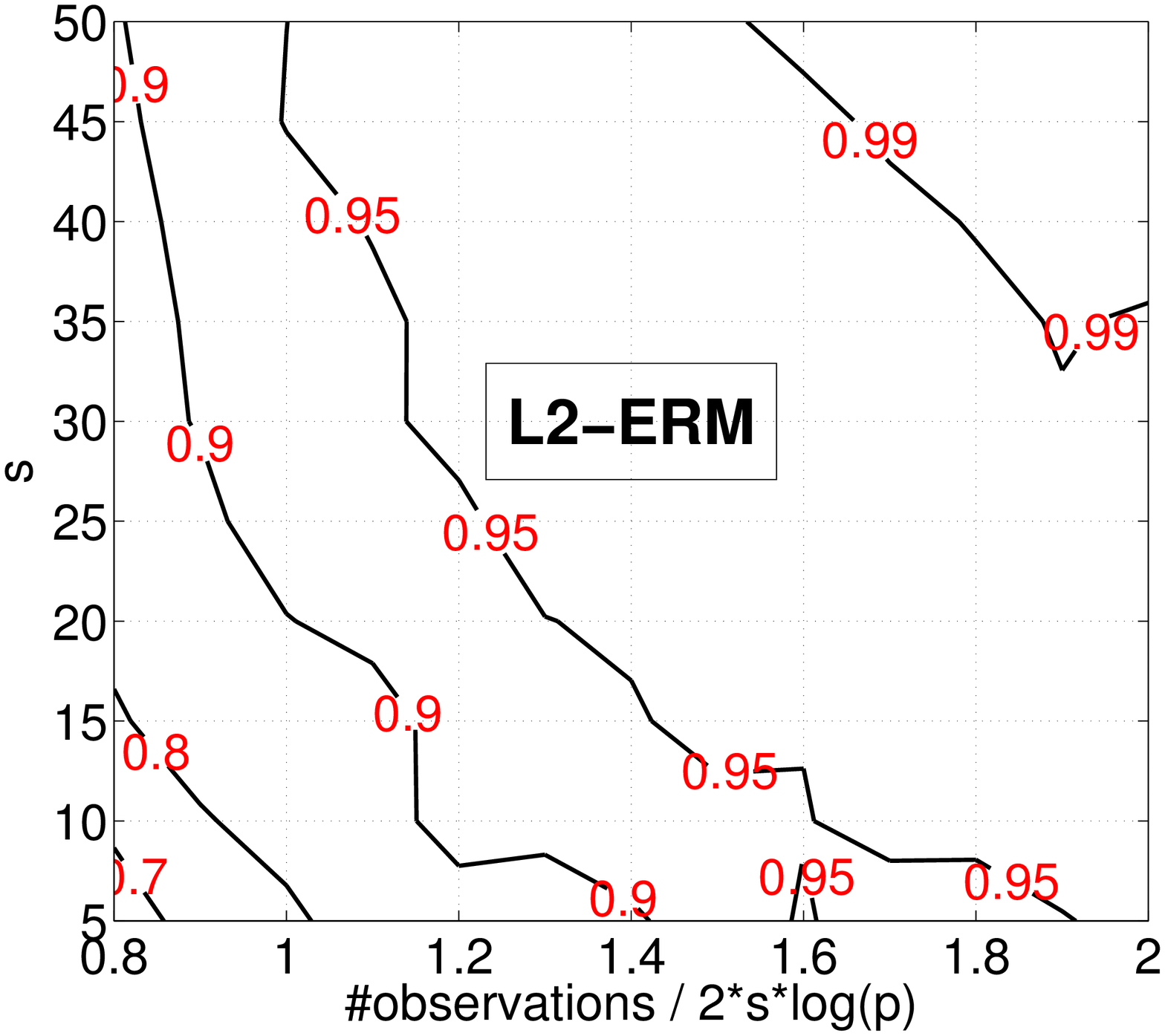}
                 & \hspace*{-0.023\textwidth}   \includegraphics[width=0.333\textwidth]{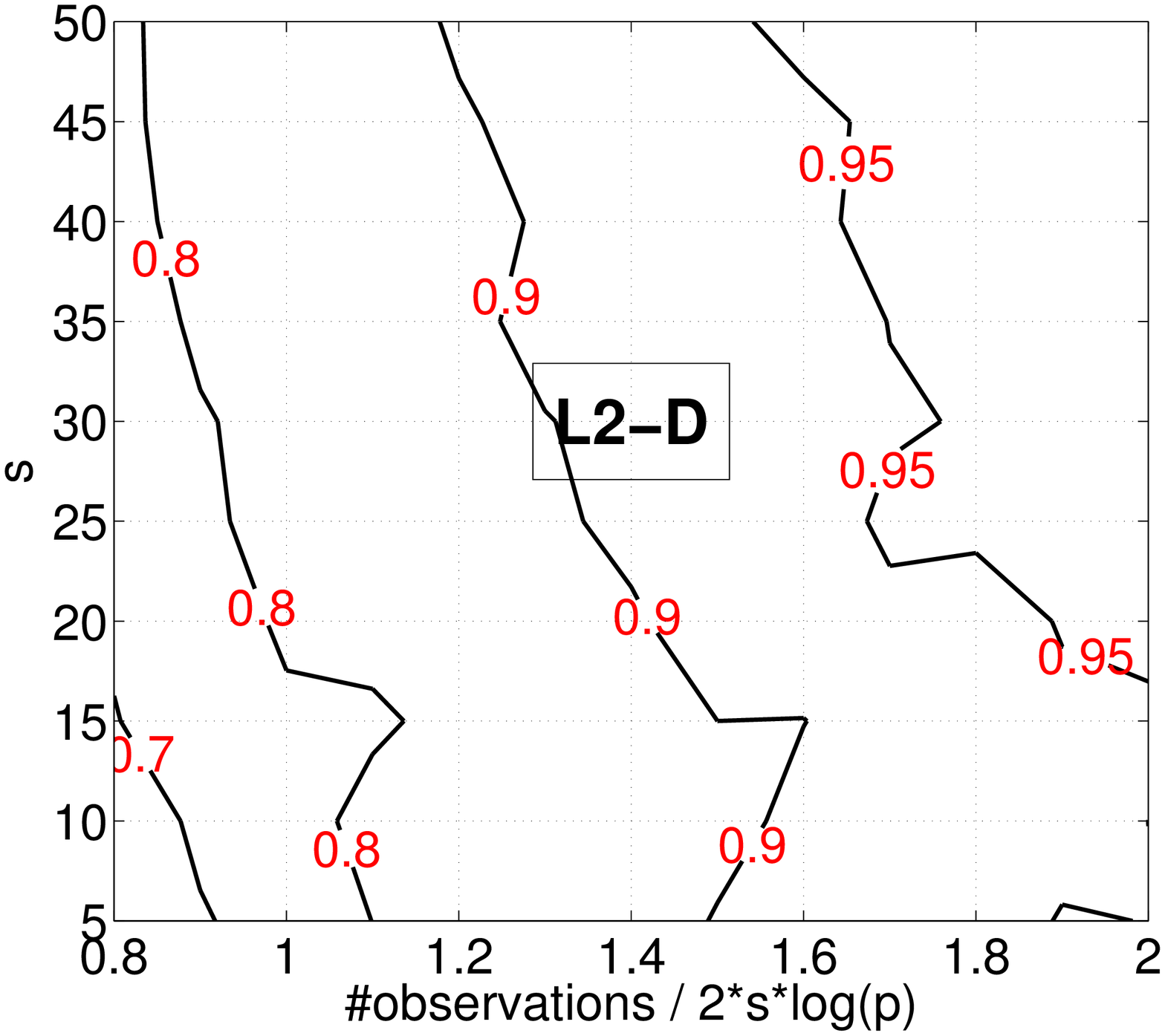}
            \\
\hspace*{-0.033\textwidth} \includegraphics[width=0.333\textwidth]{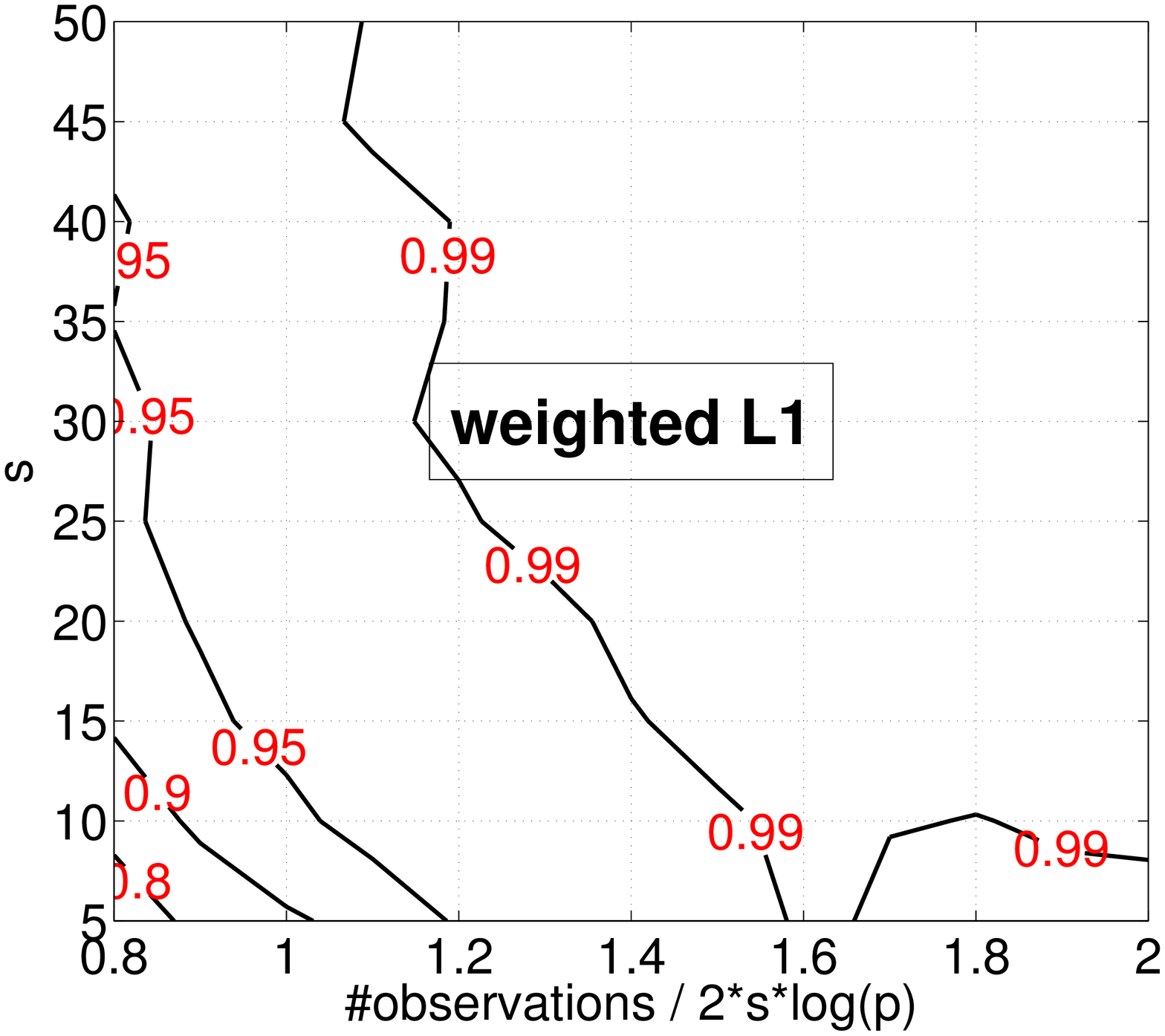}        & \hspace*{-0.028\textwidth}   \includegraphics[width=0.333\textwidth]{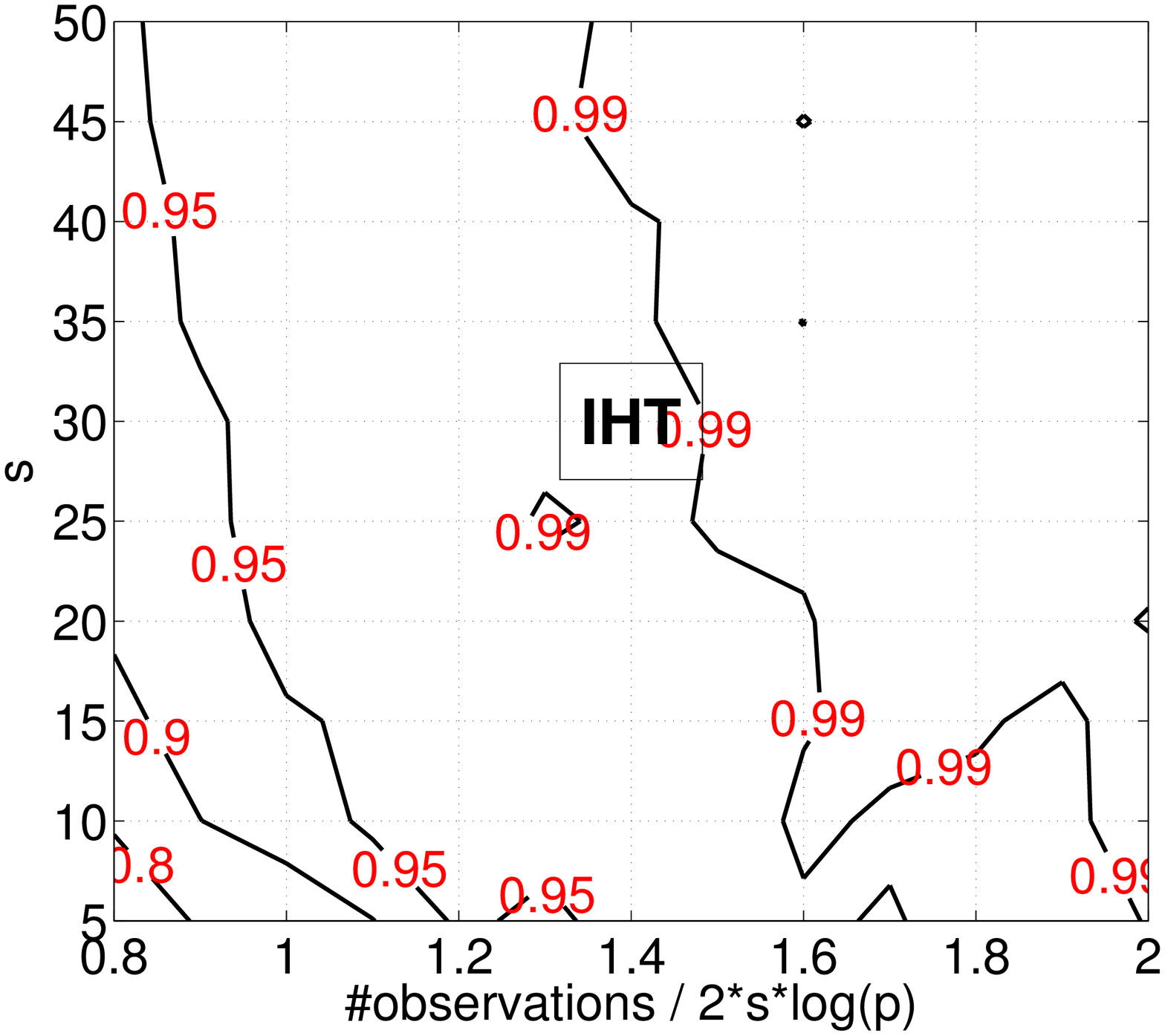}                & \hspace*{-0.023\textwidth}  \includegraphics[width=0.333\textwidth]{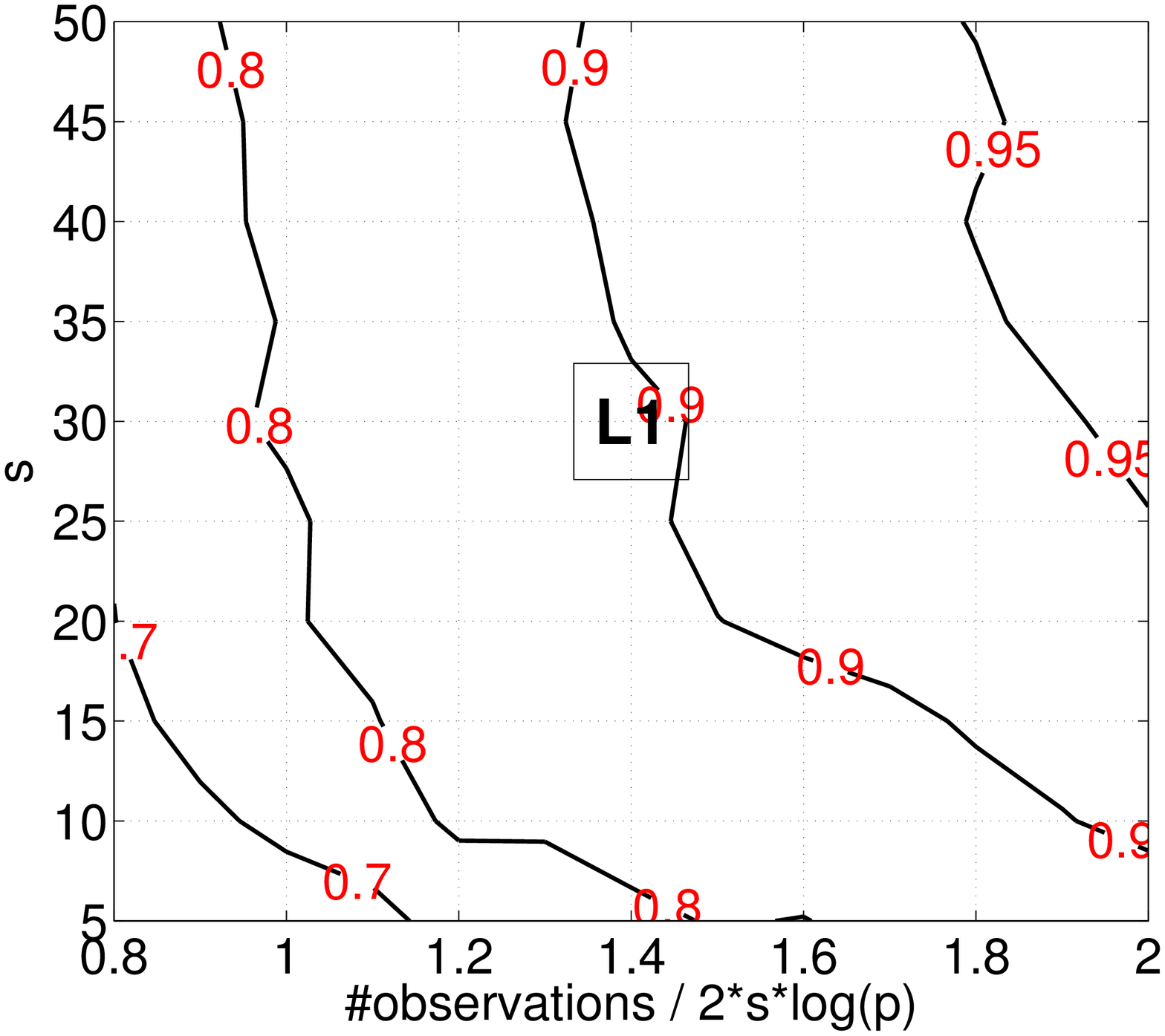}         \\
\end{tabular}
\caption{Upper panel: Average estimation errors $\nnorm{\wh{\theta} - \beta^*}_2$ ($\log_{10}$ scale) in dependence of $n$ over 50 trials for selected values of $s$. Here, $\wh{\theta}$ is a placeholder for any of the the estimators under consideration.  Middle and Lower panel: contour plots of the average Matthew's correlation in dependence of $n$ (horizontal axis) and $s$ (vertical axis) for the contour levels $0.7, 0.8, 0.9, 0.95$. Note that the smaller the area between the lower left corner of the plot and a contour line of a given level, the better the performance.}\label{fig:noisy}\vspace{0.1in}

\end{figure}

\noindent{\bf Results.} The results are summarized in Figures \ref{fig:noisy} and \ref{fig:noisy_IHT}. Turning to the upper
panel of Figure \ref{fig:noisy}, the first observation is that \textsf{'L1'} yields noticeably higher $\ell_2$ estimation errors than \textsf{'ERM'}, which yields a reductions roughly between a factor
of $10^{-.1} \approx 0.79$ and $10^{-.2}\approx 0.63$. A further reduction in error of about the same order is achieved by several of the above methods. Remarkably, the basic two-stage methods, thresholding and weighted $\ell_1$-regularization for the most part outperform the more sophisticated methods. Among the two methods based on negative $\ell_2$-regularization, \textsf{'L2-ERM'} achieves better
performance than \textsf{'L2-D'}. We also investigate sucess in support recovery by comparing $S(\wh{\theta})$ and $S(\beta^*)$, where $\wh{\theta}$ represents
any of the considered estimators, by means of Matthew's correlation coefficient (MCC) defined by
\begin{equation*}
\text{MCC} = (\text{TP} \cdot \text{TN} - \text{FP} \cdot \text{FN})/\left\{(\text{TP} + \text{FP})(\text{TP} + \text{FN})(\text{TN} + \text{FP})(\text{TN} + \text{FN}) \right\}^{1/2},
\end{equation*}
with $\text{TP}$,$\text{FN}$ etc.~denoting true positives, false negatives etc.~The larger the criterion, which takes values in $[0,1]$, the better the performance. The two lower panels
of Figure \ref{fig:noisy} depict the MCCs in the form of contour plots, split by method. The results are consistent with those of the $\ell_2$-errors. The performance of \textsf{'weighted L1'}
and \textsf{'thres'} improves respectively is on par with that of \textsf{'IHT'} which is provided the sparsity level. Figure \ref{fig:noisy_IHT} reveals that this is a key advantage since the
performance drops sharply as the sparsity level is over-specified by an increasing extent.

\newpage\clearpage

\begin{figure}[h!]
\begin{tabular}{lll}
\hspace*{-0.028\textwidth}\includegraphics[width = 0.32\textwidth]{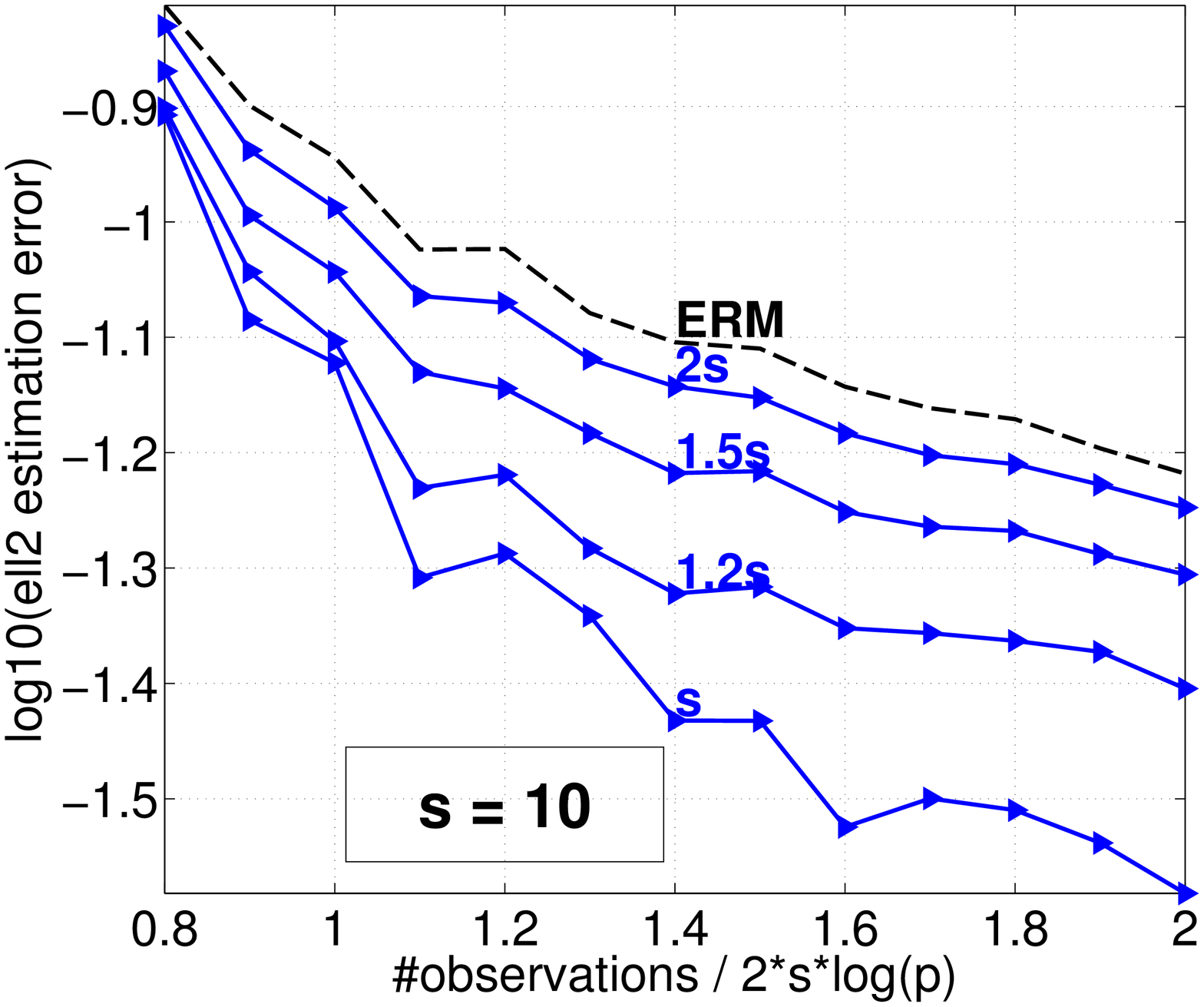} & \hspace*{-0.022\textwidth}\includegraphics[width = 0.32\textwidth]{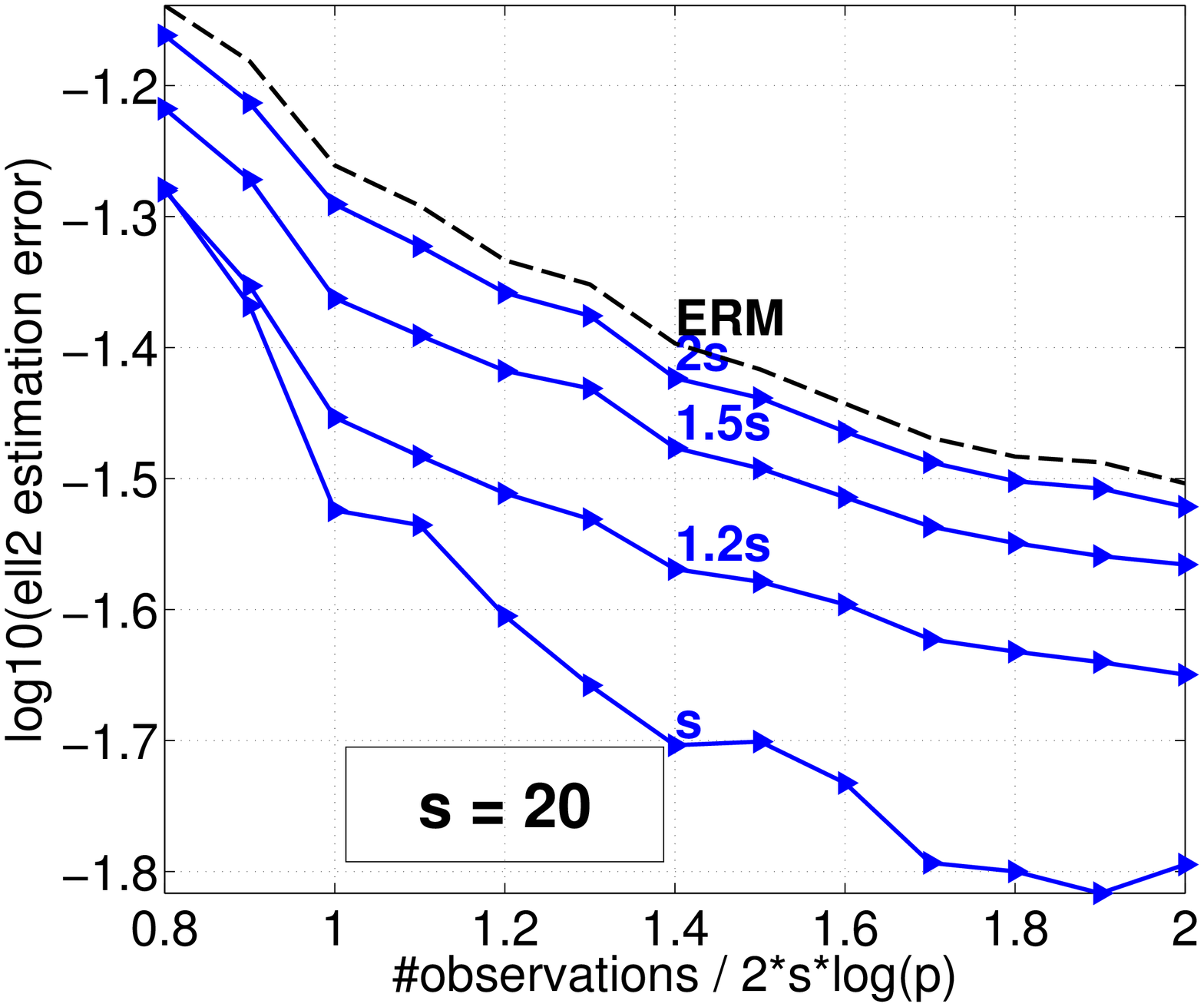} & \hspace*{-0.017\textwidth}\includegraphics[width = 0.32\textwidth]{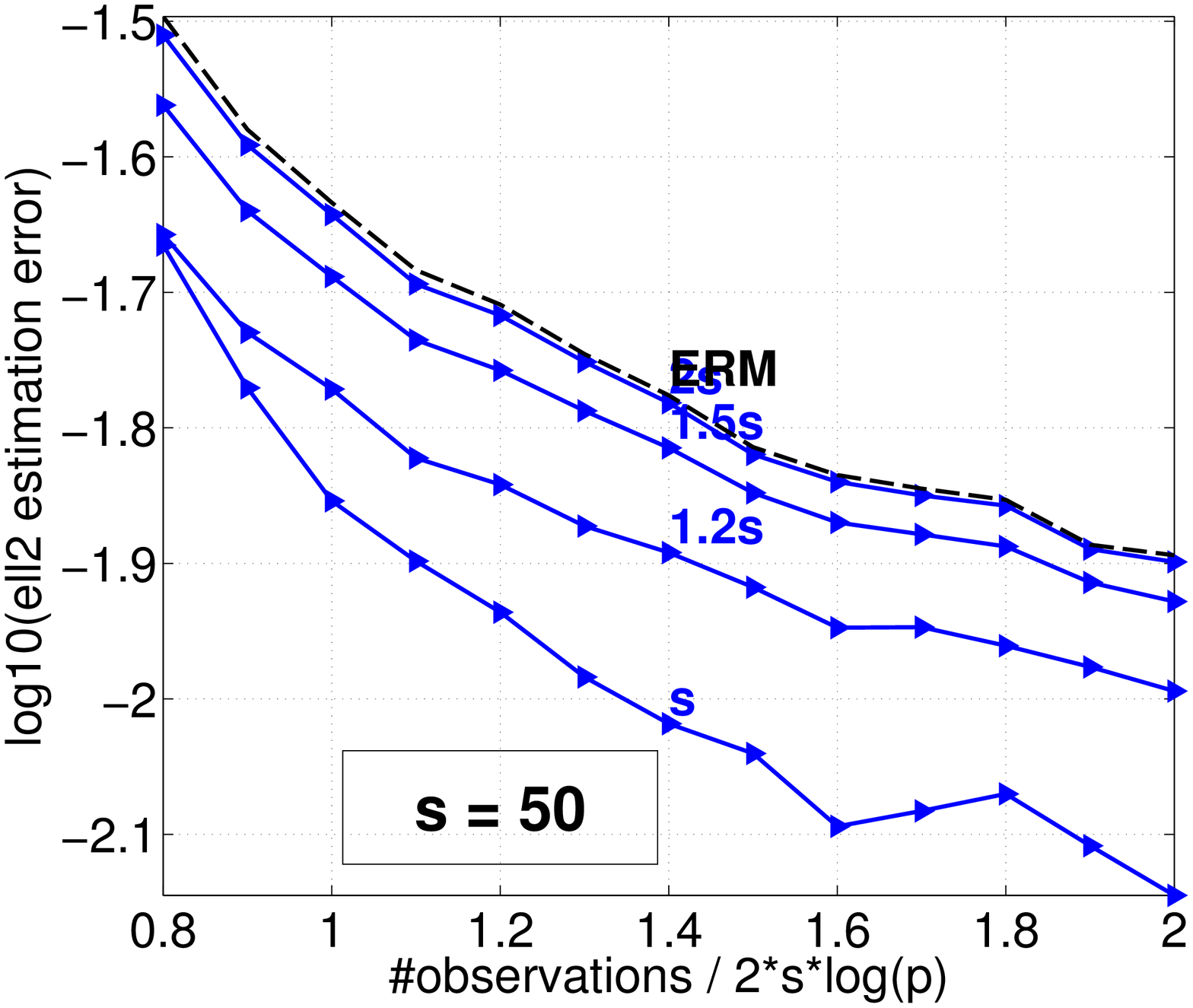} \\
\end{tabular}
\caption{Sensitivity of \textsf{'IHT'} w.r.t.~the choice of $s$. The plots display error curves of IHT run with the correct value of $s$ as appearing in Figure \ref{fig:noisy} as well
with overspecification of $s$ by the factors $1.2, 1.5, 2$. The drop in performance is substantial: for $2s$, the improvement over ERM (here used as a reference) is only minor.}\label{fig:noisy_IHT}
\end{figure}

\subsection{Density estimation}\label{subsec:density}
Let us recall the setup from the corresponding bullet in $\S$\ref{sec:intro}. For simplicity, we here suppose that
the $\{ Z_i \}_{i=1}^n$ are i.i.d.~random variables with density $\phi_{\beta^*}$, where for
$\beta \in \Delta^p$, $\phi_{\beta} = \sum_{j = 1}^p \phi_j \beta_j$ and $\mc{F} = \{\phi_j \}_{j = 1}^p$ is a given
collection of densities. Specifically, we consider univariate Gaussian densities $\phi_j = \phi_{\theta_j}$, where
$\theta_j = (\mu_j, \sigma_j)$ contains mean and standard deviation, $j=1,\ldots,p$. As an example, one might consider
$p_0$ locations and $K$ different standard deviations per location so that $p = p_0 K$, i.e.,~$\theta_{(k-1) p_0 + l} = (\mu_l, \sigma_k)$,
$k=1,\ldots,K$, and $l=1,\ldots,p_0$. This construction provides more flexibility compared to usual kernel density
estimation where the locations equal the data points, a single bandwidth is used, and the coefficients $\beta$ are
all $1/n$. For large $\mc{F}$, sparsity in terms of the coefficients is common as a specific target density can typically
be well approximated by using an appropriate subset of $\mc{F}$ of small cardinality.\\
As in \citet{Bunea2010}, we work with the empirical risk
\begin{equation*}
\vspace{-1ex}
R_n(\beta) = \beta^{\T} Q \beta - 2 c^{\T} \beta,  \quad c = \left(\textstyle \sum\nolimits_{i=1}^n \phi_j(Z_i) / n\right)_{j=1}^p,
\end{equation*}
and $Q=(\scp{\phi_j}{\phi_k})_{j,k=1}^p$, where $\scp{f}{g} = \int_{\R} f g$ for $f$,$g$ such that $\nnorm{f},\nnorm{g} < \infty$ with
$\nnorm{f} = \scp{f}{f}^{1/2}$.\\
In our simulations, we let $p_0 = 100$, $K = 2$, $\sigma_k=k$, $k=1,2$. The locations $\{ \mu_l \}_{l = 1}^{p_0}$ are generated sequentially by selecting
$\mu_1$ randomly from $[0,\delta]$, $\mu_2$ from $[\mu_1+\delta, \mu_1+2\delta]$ etc.~where $\delta$ is chosen such
that the 'correlations' $\scp{\phi_j}{\phi_k}/\nnorm{\phi_j}\nnorm{\phi_k} \leq 0.5$ for all $(j, k)$ corresponding to different locations. An upper bound away
from $1$ is needed to ensure identifiability of $S(\beta^*)$ from finite samples. Data generation, the
methods compared, and the way they are run is almost identical to the previous subsections. Slight changes are made
for $S(\beta^*)$ (still uniformly at random, but it is ruled that any location is selected twice), $b_{\min}^*$ ($\varrho$ set to $2$)
and hyperparameter selection. For the latter, a separate validation data set (also of size $n$) is generated, and hyperparameters
are selected as to minimize the empirical risk from the validation data.\\

\noindent\textbf{Results.} Figure \ref{fig:density} confirms once again that making use of simplex
constraints yields markedly lower error than $\ell_1$-regularization followed by normalization
\citep{Bunea2010}. \textsf{'L2-ERM'} and weighted $\ell_1$ perform best, improving over \textsf{'IHT'}
(which is run with knowledge of $s$).

\begin{figure}[h!]
\begin{tabular}{lll}
\hspace*{-0.023\textheight}
$\begin{array}{c}
\includegraphics[width=0.38\textwidth]{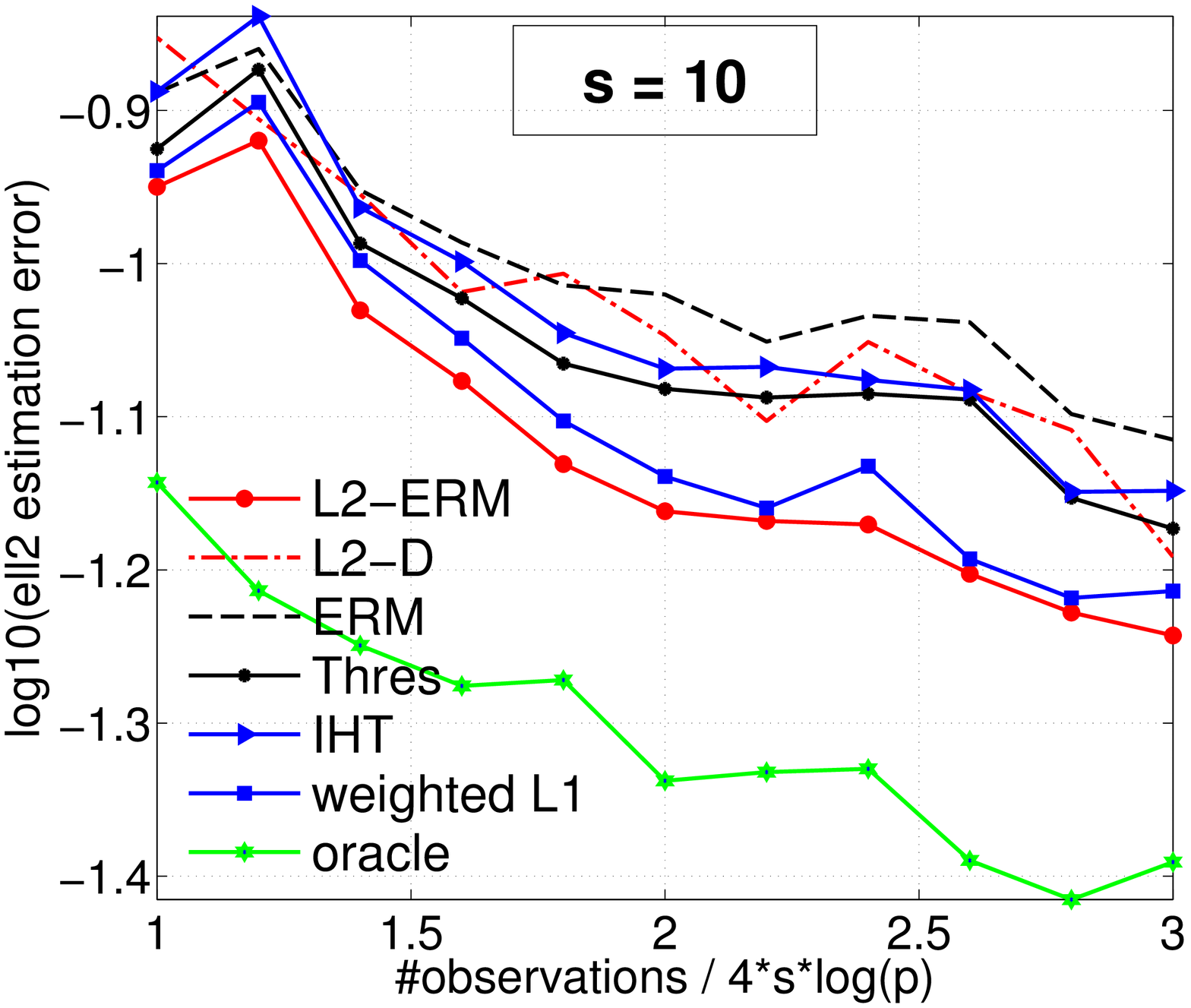} \\
{\tiny   \begin{array}{l} \text{\textbf{Standard errors}(rounded):} \\
\begin{array}{|c|c|c|c|}\hline
\textbf{\textsf{L2-ERM}}  &  \textbf{\textsf{L2-D}} & \textbf{\textsf{ERM}} &  \textbf{\textsf{Thres}} \\
.03  & .03 & .02 & 0.02 \\
\hline
\textbf{\textsf{weightedL1}} & \textbf{\textsf{IHT}} & \textbf{\textsf{L1}} & \textbf{\textsf{oracle}} \\
.03 & .03 & .02 & .02 \\
\hline
 \end{array}
\end{array}
}
\end{array}$
&

\hspace*{-0.026\textheight}
$\begin{array}{c}
\includegraphics[width=0.38\textwidth]{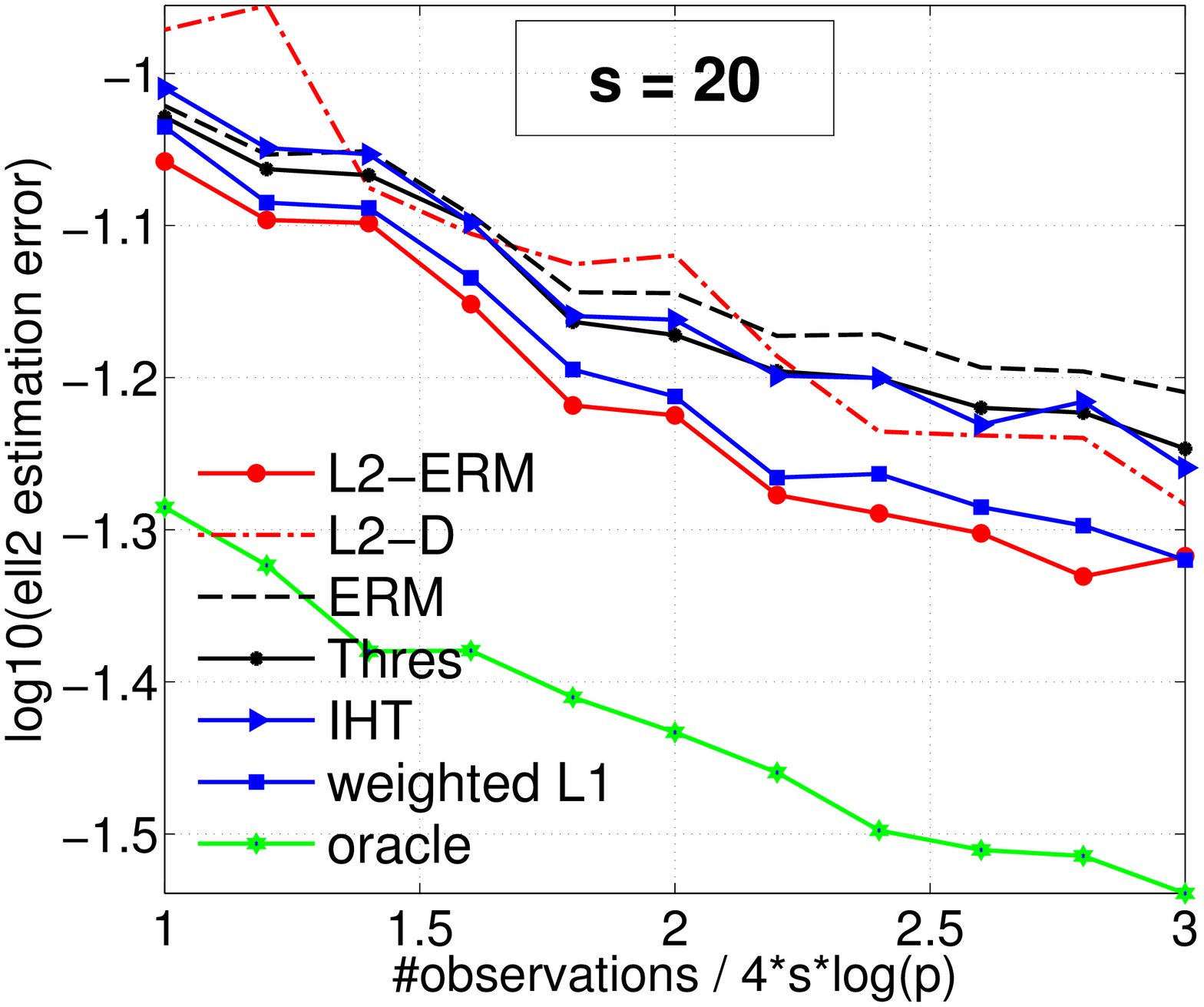} \\
{\tiny   \begin{array}{l} \text{\textbf{Standard errors}(rounded):} \\
\begin{array}{|c|c|c|c|}\hline
\textbf{\textsf{L2-ERM}}  &  \textbf{\textsf{L2-D}} & \textbf{\textsf{ERM}} &  \textbf{\textsf{Thres}} \\
.02  & .02 & .02 & 0.02 \\
\hline
\textbf{\textsf{weightedL1}} & \textbf{\textsf{IHT}} & \textbf{\textsf{L1}} & \textbf{\textsf{oracle}} \\
.02 & .02 & .01 & .01 \\
\hline
 \end{array}
\end{array}
}
\end{array}$
&
\hspace*{-0.0285\textheight}
$\begin{array}{l}
\includegraphics[width=0.225\textwidth]{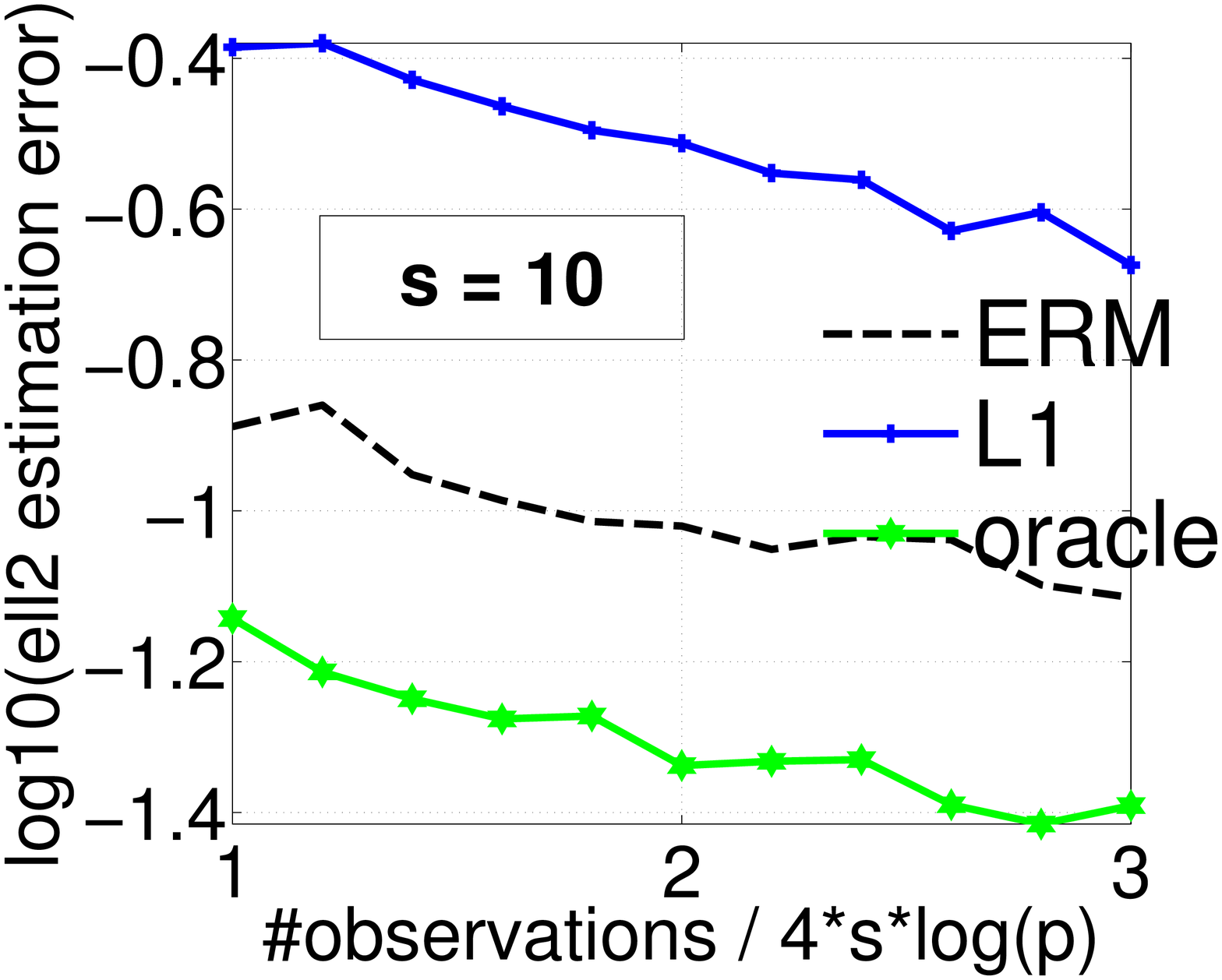}  \\
\includegraphics[width=0.225\textwidth]{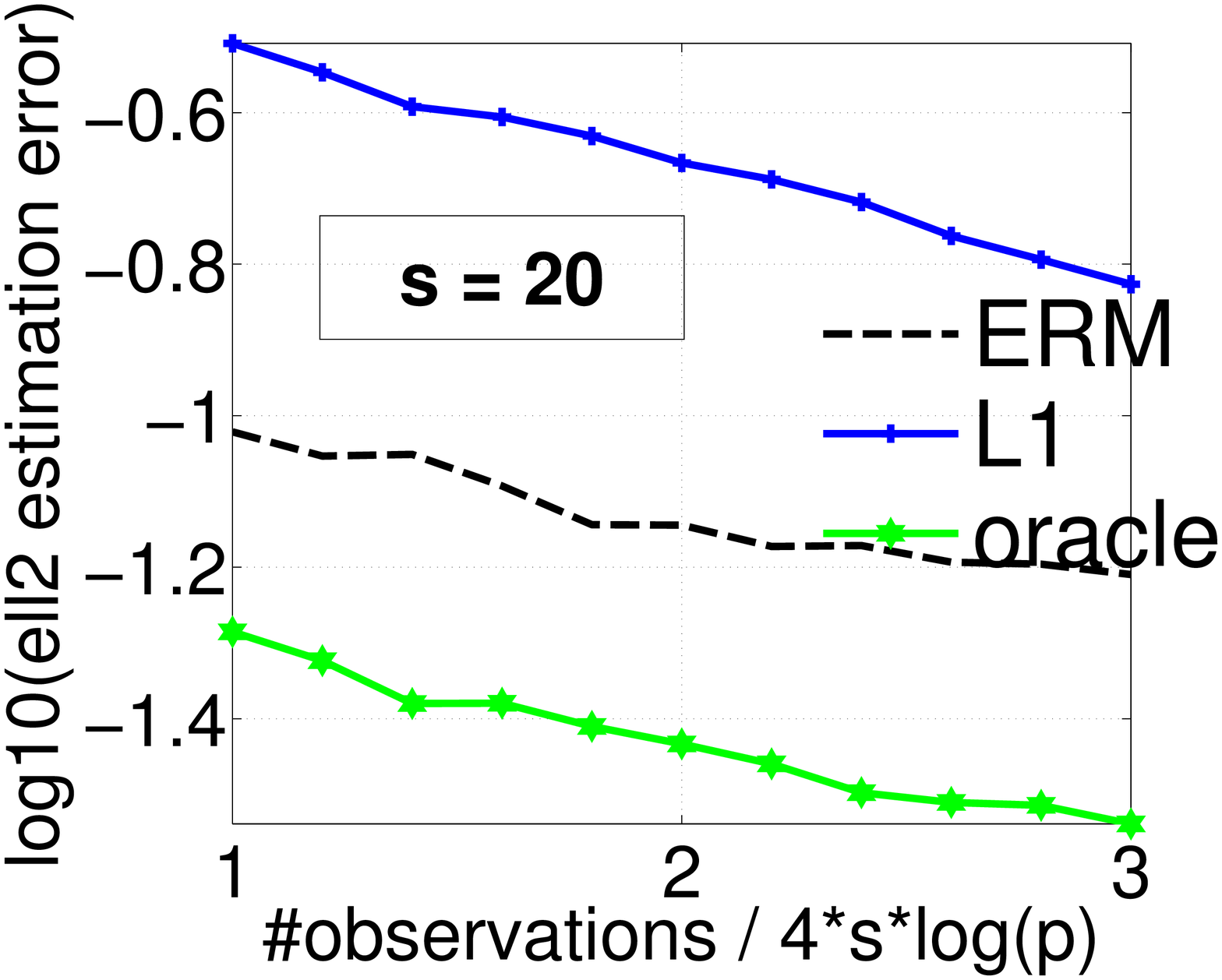}
\end{array}
$

\end{tabular}
\caption{Average estimation errors $\nnorm{\wh{\theta} - \beta^*}_2$ for density estimation over 50 trials. Since the performance of \textsf{'L1'}
falls short of the rest of the competitors, whose differences we would like to focus on, \textsf{'L1'} is compared
to \textsf{'ERM'} and '\textsf{oracle}' in separate plots in the right column. Standard errors are smaller than $0.025$ for
all methods.}
\label{fig:density}
\end{figure}

\subsection{Portfolio Optimization}
We use a data set available from \url{http://host.uniroma3.it/docenti/cesarone/datasetsw3_tardella.html} containing the weekly returns of $p = 2196$ stocks in the NASDAQ index collected during 03/2003 and 04/2008 (264 weeks altogether). For each stock, the expected returns is estimated as the mean return $\wh{\mu}$ from the first four years (208 weeks). Likewise, the covariance of the returns is estimated as the sample covariance $\wh{\Sigma}$ of the returns of the first four years. Given $\wh{\mu}$ and $\wh{\Sigma}$, portfolio selection (without short positions) is based on the optimization problem
\begin{equation}\label{eq:portfolio_basic}
\min_{\beta \in \Delta^p} \beta^{\T} \wh{\Sigma} \beta - \tau \wh{\mu}^{\T} \beta
\end{equation}
where $\tau \in [0, \tau_{\max}]$ is a parameter controlling the trade-off between return and variance of the portfolio. Assuming that $\wh{\mu}$ has a unique maximum entry, $\tau_{\max}$ is defined as the smallest number such that the solution of \eqref{eq:portfolio_basic} has exactly one non-zero entry equal to one at the position of the maximum of $\wh{\mu}$. As observed in \citet{Brodie2009}, the solution of \eqref{eq:portfolio_basic} tends to be sparse already because of the simplex constraint. Sparsity can be further enhanced with the help of the strategies discussed in this paper, treating \eqref{eq:portfolio_basic} as the empirical risk. We here consider \textsf{'L2-ERM'}, \textsf{'weighted L1'}, \textsf{'Thres'} and \textsf{'IHT'} for a grid of values for the regularization parameter (\textsf{'L2-ERM'} and \textsf{'weighted L1'}) respectively sparsity level (\textsf{'L2-ERM'} and \textsf{'Thres'}). The solutions are evaluated by computing the Sharpe ratios (mean return divided by the standard deviation) of the selected portfolios on the return data of the last 56 weeks left out when computing $\wh{\mu}$ and $\wh{\Sigma}$.\\

\begin{figure}[h!]
\begin{center}
\begin{tabular}{cccc}
\hspace*{0.036\textwidth}\includegraphics[width = 0.38\textwidth]{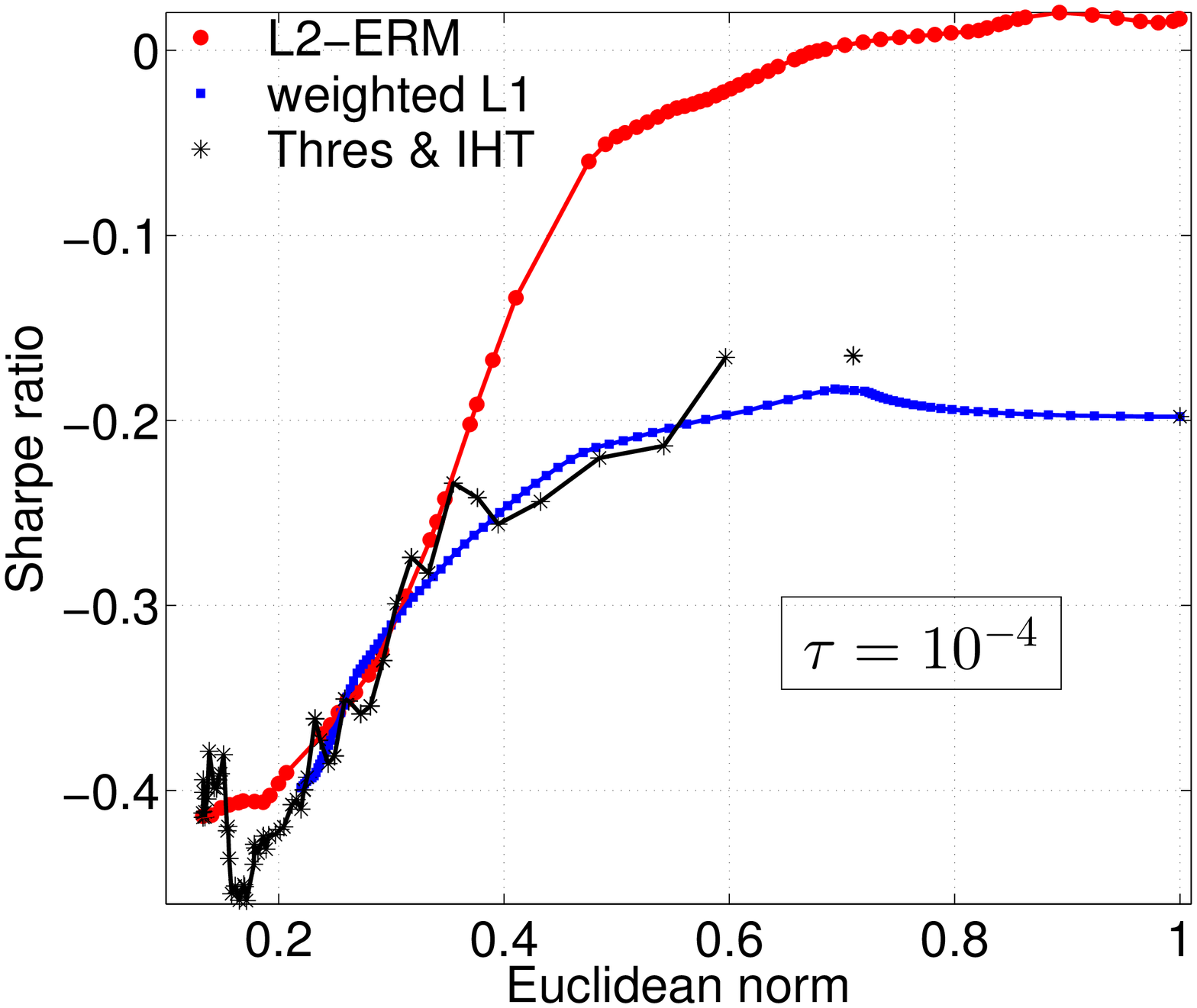} & \hspace*{0.04\textwidth}&
\includegraphics[width = 0.38\textwidth]{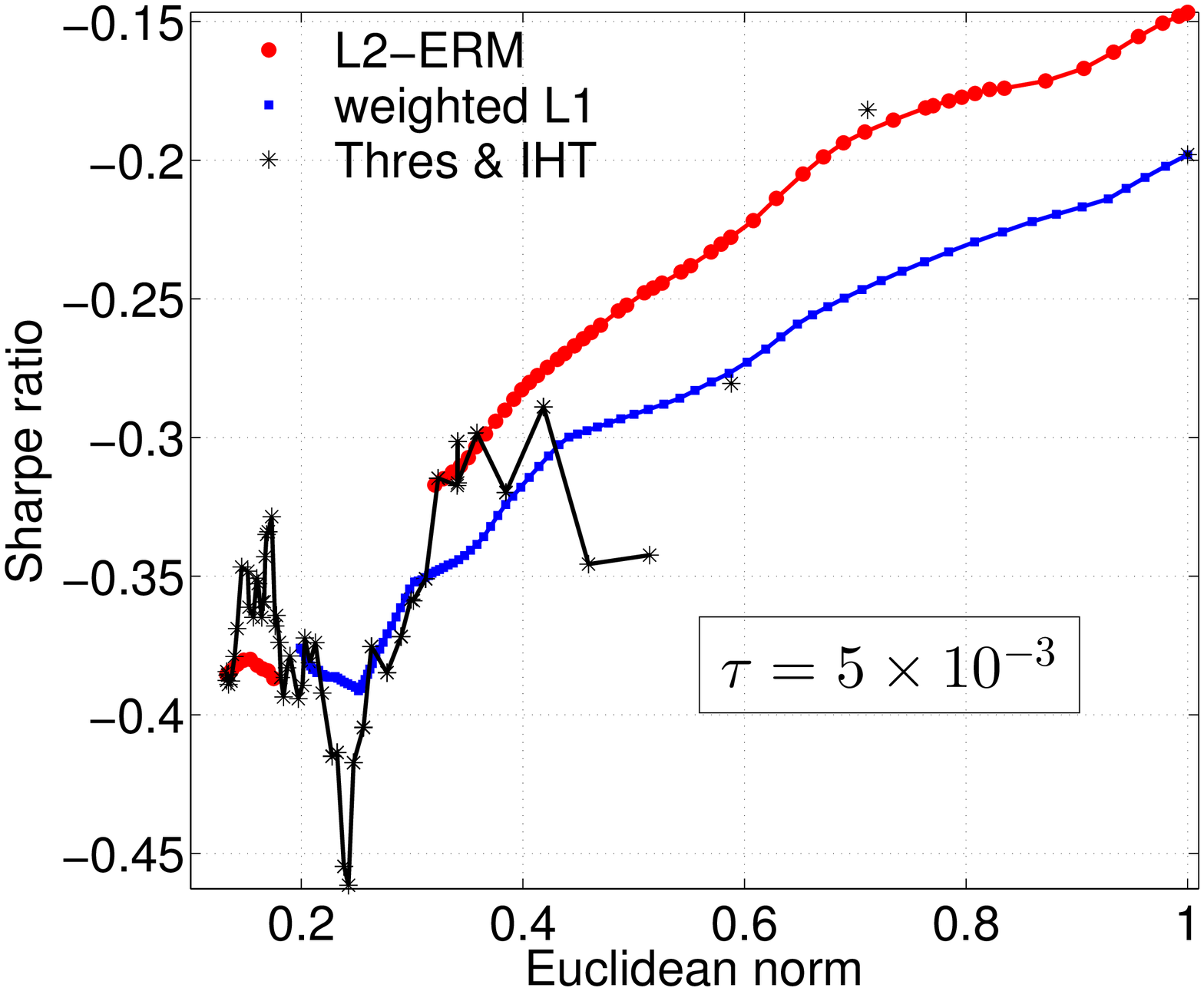} & \hspace*{0.036\textwidth}\\
\end{tabular}
\end{center}
\vspace{-0.3in}
\caption{Sharpe ratios of the portfolios selected by \textsf{'L2-ERM'}, \textsf{'weighted L1'}, \textsf{'Thres'} and \textsf{'IHT'} on the hold-out portion of the NASDAQ data set in dependency of different choices for ther regularization parameter/sparsity level (to allow for joint display, we use the $\ell_2$-norm as measure of sparsity on the horizontal axis). Left panel: $\tau = 10^{-4}$, Right panel: $\tau = 5 \cdot 10^{-3}$, cf.~\eqref{eq:portfolio_basic}. The results of \textsf{'Thres'} and \textsf{'IHT'} are essentially  indistinguishable and are hence not plotted separately for better readability. Note that points that are too far away from each other with respect to the horizontal axis are not connected by lines.}
\label{fig:portfolio_results}
\end{figure}

\noindent\textbf{Results.} Figure \ref{fig:portfolio_results} displays the Sharpe ratios of the portfolios returned by these approaches in dependency of the $\ell_2$-norms of the solutions corresponding to different choices of the regularization parameter respectively sparsity level and two values of $\tau$ in \eqref{eq:portfolio_basic}. One observes that promoting sparsity is beneficial in general. The regularization-based methods \textsf{'L2-ERM'} and \textsf{'weighted L1'} differ from \textsf{'L2-ERM'} and \textsf{'Thres'} (whose results are essentially not distinguishable) in that the former two yield comparatively smooth curves. \textsf{'L2-ERM'} achieves the best Sharpe ratios for a wide range of $\ell_2$-norms for both values of $\tau$.

\subsection{Quantum State Tomography}\label{subsec:Pauli}
We now turn to the matrix case of Section \ref{sec:tomography}. The setup of this subsection
is based on model \eqref{eq:tracereg}, where the measurements $\{ X_i \}_{i = 1}^n$ are chosen
uniformly at random from the (orthogonal) Pauli basis of $\HH^m$ (here, $m = 2^q$ for some integer $q \geq 1$). For $q = 1$,
the Pauli basis of $\HH^2$ is given by the following four matrices:
\begin{equation*}
P_{1,1} = \begin{pmatrix}
           1  &  0   \\
           0  & 1
           \end{pmatrix}, \quad P_{1,2} = \begin{pmatrix}
           0  &  -\sqrt{-1}   \\
           \sqrt{-1}  & 0
           \end{pmatrix},  \quad P_{1,3} = \begin{pmatrix}
           1  &  0   \\
           0  & -1
           \end{pmatrix}, \quad P_{1,4}  = \begin{pmatrix}
           0  &  1   \\
           1  & 0
           \end{pmatrix}.
\end{equation*}
For $q > 1$, the Pauli basis $\{P_{q,1},\ldots,P_{q,m^2} \}$is constructed as the $q$-fold tensor product of $\{P_{1,1},\ P_{1,2},\ P_{1,3},\ P_{1,4} \}$. The set
of measurements is then given by $\{P_{q,i}, i \in \mc{I} \}$, where $\mc{I} \subseteq \{1,\ldots,m^2\}$, $|\mc{I}| = n$, is chosen uniformly at random. Pauli measurements are commonly used in quantum state tomography in order to recover the density matrix of a quantum state (see Section \ref{sec:tomography} above). In \citet{Gross2010}, it is shown that if $B^*$ is of low rank, it can be estimated accurately from few such random measurements by using nuclear norm regularization; the constraint $B^* \in \bm{\Delta}^m$ is not taken advantage of. Proposition \ref{prop:adaptation_m} asserts that this constraint alone is well-suited for recovering matrices of low rank as long as the
measurements satisfy a restricted strong convexity condition (Condition \ref{cond:RSC_m}). It is shown in \cite{Liu11} that Pauli measurements satisfy the matrix
RIP condition of \cite{Recht2010} as long as $n \gtrsim mr \log^6(m)$. Since the matrix RIP condition is stronger than Condition \ref{cond:RSC_m}, Proposition \ref{prop:adaptation_m} applies here. The requirement on $n$ is near-optimal: up to a polylogarithmic factor, it equals
the 'degrees of freedom' of the problem, which is given by $d = mr - r(r-1)/2 \gtrsim m r$, which is the dimension of the space $\TT(B^*) \subset \HH^m$ (cf.~Definition \ref{defn:tangentspace} in the appendix).

\subsubsection{Noiseless measurements}

In the first numerical study, we work with noiseless measurements. We fix $m = 2^7$ and let
$r \in \{1, 2, 5, 10\}$ vary. The target is generated randomly as $B^*=AA^{\T}$, where $A$ is an $m\times r$ matrix, whose entries are drawn i.i.d.~from $N(0,1)$. The number of random Pauli measurements $n$ are varied from $2d$ to $5d$ in steps of $0.5d$, where $d$ equals the 'degrees of freedom' as defined above. For each possible combination of $n$ and $r$, $50$ trials are performed. The following three approaches for recovering $B^*$ are compared.\\

\noindent\textsf{'Feasible set'}: The counterpart to ERM in the noiseless case: finding a point in
\begin{equation}\label{eq:feasibleset_m}
\bm{D}(0) = \{B \in \bm{\Delta}^m:\;\mc{X}^{\star}(\mc{X}(B)-y) = 0 \} = \{B \in \bm{\Delta}^m: \mc{X}(B) = y \},
\end{equation}
where the second identity follows from the orthogonality of the Pauli matrices.\\

\noindent \textsf{'L2}': The counterpart to \eqref{eq:rem_m}/\eqref{eq:dss_m} in the noiseless case, which amounts
to maximizing the Schatten $\ell_2$-norm (i.e.,~Frobenius norm) over \eqref{eq:feasibleset_m}. As initial
iterate for Algorithm \ref{alg:dc}, the output from \textsf{'feasible set'} is used.\\

\noindent \textsf{'IHT'}: The matrix version of iterative hard thresholding under simplex constraints as proposed
by \citet{Kyrilidis2013}. Under the assumption that the rank of the target is known, one tries to
solve directly the rank-constrained optimization problem $\min_{\bm{\Delta}_0^{m}(r)} R_n(B)$ using projected gradient descent. Projections onto $\bm{\Delta}_0^{m}(r)$ can be efficiently computed using partial eigenvalue decompositions. We use a constant step size as in \cite{Kyrilidis2013}. The output of \textsf{'feasible set'} is used as initial iterate.\\

\noindent\textbf{Results.} Figure \ref{fig:matrixRecovery_noiseless} shows a clear benefit of using $\ell_2$-norm maximization
on top of solving the feasibility problem. For \textsf{'L2'}, $2.5d$ measurements suffice to obtain highly accurate solutions, while \textsf{'feasible set'} requires $3.5d$ up to $5d$ measurements. The performance of \textsf{IHT} falls in between the two other approaches even though the knowledge of $r$ provides an extra advantage.

\begin{figure}[h!]
\begin{center}
\subfigure{
\includegraphics[width=0.44\textwidth]{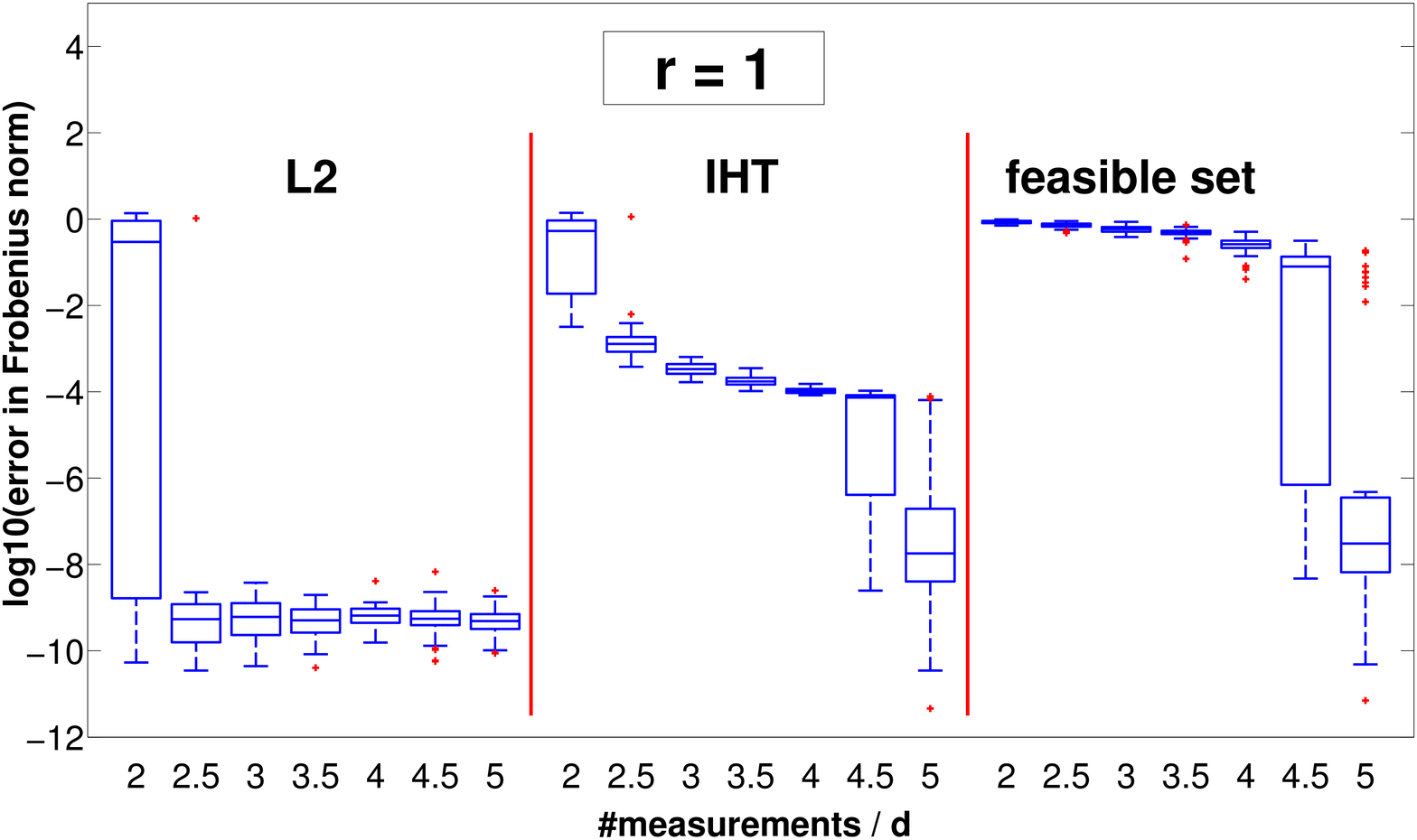}\hspace{0.3in}		
\includegraphics[width=0.44\textwidth]{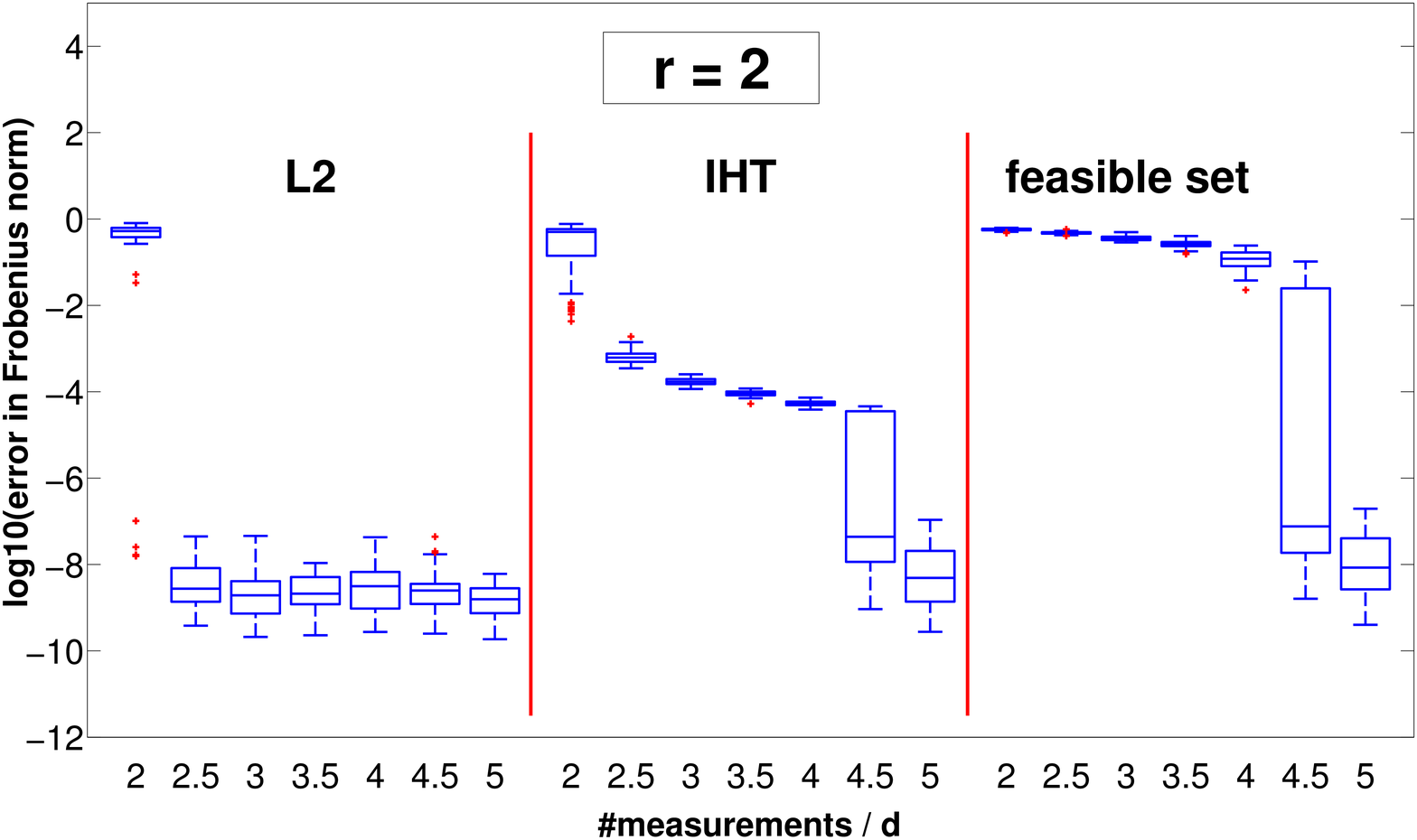}}
\subfigure{
\includegraphics[width=0.44\textwidth]{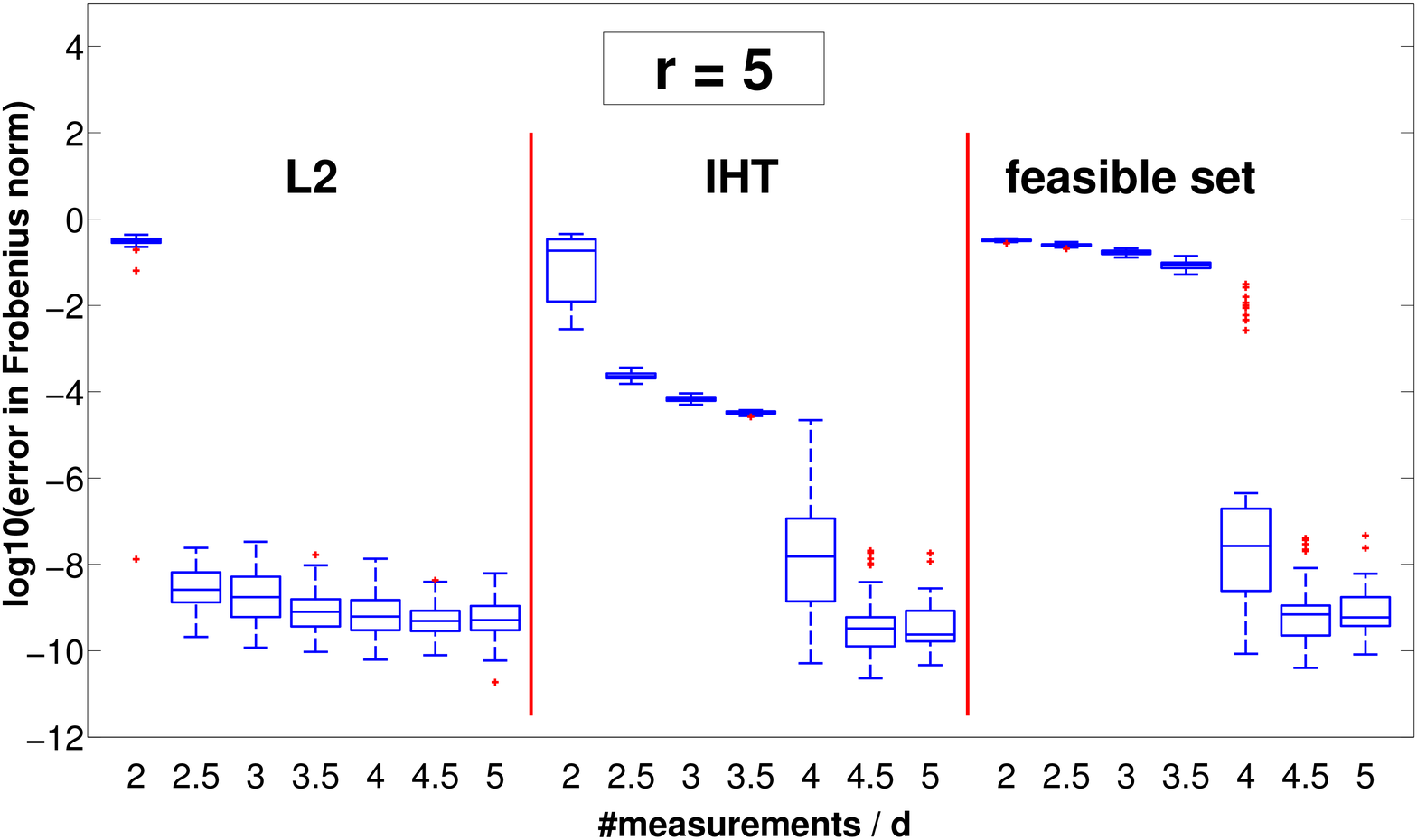}\hspace{0.3in}
\includegraphics[width=0.44\textwidth]{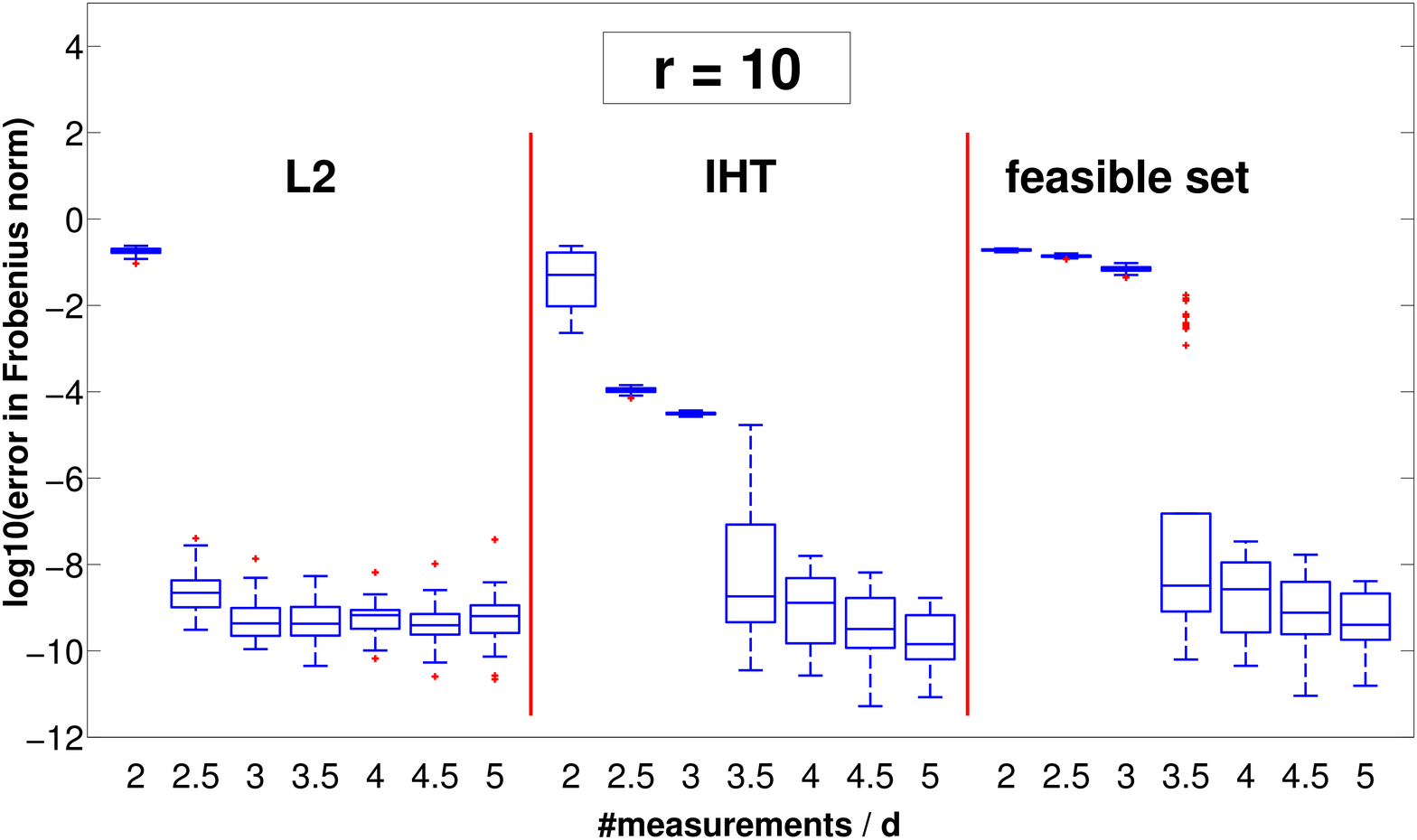}
}
\end{center}
\vspace{-0.2in}
\caption{Boxplots of the errors $\nnorm{\wh{\Theta} - B^*}_2$ (50 trials) in recovering $B^*$ with respect to the Frobenius norm ($\log_{10}$ scale) in dependence of the number of Pauli measurements ($d=$$\,$'degrees of freedom'). Here, $\wh{\Theta}$ is representative for any of the three estimators under consideration.}	
\label{fig:matrixRecovery_noiseless}
\end{figure}

\subsubsection{Noisy measurements}

We maintain the setup of the previous paragraph, but the measurements are now
subject to additive Gaussian noise with standard deviation $\sigma = 0.1$. In order to adjust for the
increased difficulty of the problem, the range for the number of measurements $n$ is multiplied by the factor $\log(m/r)$.
Our comparison covers the following methods.\\[1.25ex]
\textsf{'ERM'}: Empirical risk minimization, the counterpart to \textsf{'Feasible set'} above.\\[1.25ex]
\textsf{'Thres'}: \textsf{'ERM'} followed by eigenvalue thresholding as outlined below Proposition \ref{prop:adaptation_m}.\\[1.25ex]
\textsf{'L2-ERM'}: Regularized ERM with negative $\ell_2$-regularization \eqref{eq:rem}. A grid
search over 20 different values of the regularization parameter $\lambda$ is performed analogously
to the vector case.\\[1.25ex]
\textsf{'weighted L1'}: The approach in \eqref{eq:weightedL1_m}. The grid search for $\lambda$ follows the
vector case.\\[1.25ex]
\textsf{'IHT'}: As in the noiseless case.\\[1.25ex]
\textsf{'L1'}: In analogy to the corresponding approach \eqref{eq:naiveapproach} in the vector case, the unit
trace constraint is dropped, and a nuclear-norm regularized empirical risk is minimized over the positive
semidefinite cone. The result is then divided by its trace. The regularization parameter is fixed to
a single value $\lambda_0 = 2 \sigma \sqrt{\log(m)/n}$ according to the literature \citep{Negahban2011, Koltchinskii2011}.\\

For \textsf{'Thres'}, \textsf{'L2-ERM'} and other methods for which multiple values of a hyperparameter are
considered, hyperparameter selection is done by minimizing a RIC-type criterion. Specifically, for some estimate
$\wh{\Theta}_{\lambda}$ of $B^*$, we use
\begin{equation*}
\text{sel}(\lambda) = R_n(\wh{\Theta}_{\lambda}) + \frac{C \sigma^2 \log(m^2) \nnorm{\wh{\Theta}_{\lambda}}_0}{n}
\end{equation*}
The use of this criterion is justified in light of results in \cite{Klopp2011} on trace regression with rank penalization.
We have experimented with different choices of the constant $C$. Satisfactory results are achieved for $C = 2^6$, which
is the choice underlying the results displayed in Figure \ref{fig:matrixRecovery_noisy}. Once $\lambda$ has been
determined, the matrix of eigenvectors is fixed and the eigenvalues are re-fitted via least squares similar to \eqref{eq:weightedL1_m}.\\

\noindent\textbf{Results.} For space reasons, we only display the results for $r=2,10$ in Figures \ref{fig:matrixRecovery_noisy} and \ref{fig:matrixRecovery_noisy_ihtspecial}. \textsf{'IHT'} achieves best performance even though the error curve of \textsf{'L2'} is essentially
identical for $r = 2$. Figure \ref{fig:matrixRecovery_noisy_ihtspecial} indicates that \textsf{'IHT'} is sensitive to the
choice of $r$: over-specification by a factor of two can lead to a performance that is significantly worse than \textsf{'Thres'}
and only slightly better than \textsf{'ERM'}. Both \textsf{'L2'} and \textsf{'Thres'} are adaptive to the rank which is
correctly recovered in almost all cases. In the matrix case, \textsf{'L2'} improves over \textsf{'Thres'} (as opposed to the vector case), possibly because for \textsf{'Thres'} the eigenvectors remain unchanged compared to \textsf{'ERM'}, only the eigenvalues are modified. The performance of \textsf{'L1'} clearly falls short of all other competitors, which underpins the importance of the unit trace constraint.

\begin{figure}[h!]
\begin{center}
\begin{tabular}{ll}
$\begin{array}{c}
\includegraphics[width = 0.35\textwidth]{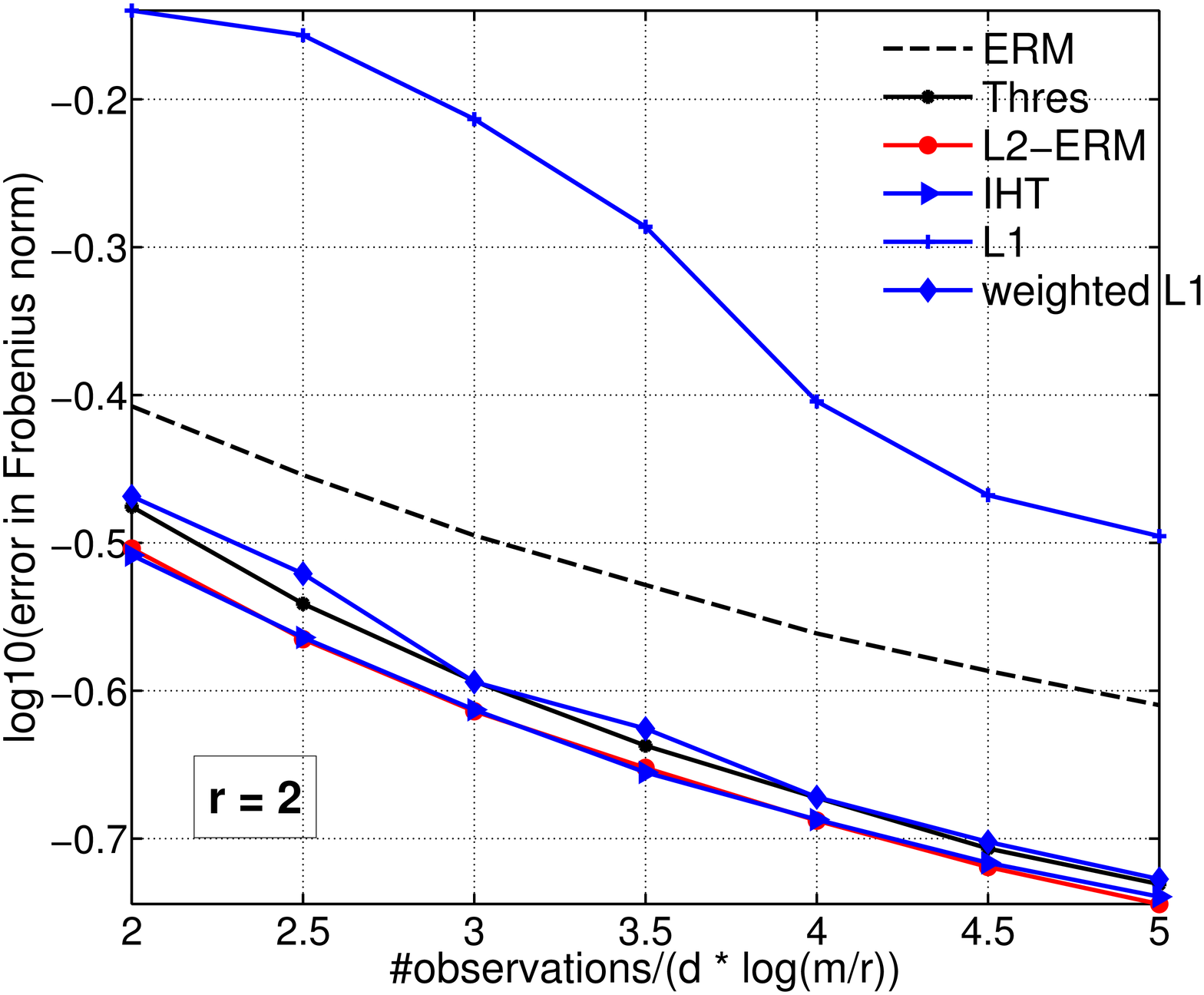} \\
{\scriptsize \begin{array}{l} \text{\textbf{Standard errors}(rounded):} \\
\begin{array}{|c|c|c|}\hline
\textbf{\textsf{L2-ERM}}  &  \textbf{\textsf{ERM}} &  \textbf{\textsf{Thres}} \\
3 \cdot 10^{-3}  & 3 \cdot 10^{-3}  & 3 \cdot 10^{-3}  \\
\hline
\textbf{\textsf{weightedL1}} & \textbf{\textsf{IHT}} & \textbf{\textsf{L1}} \\
1.4 \cdot 10^{-2} &  3 \cdot 10^{-3} & 1.7 \cdot 10^{-2}  \\
\hline
 \end{array}
\end{array}}
\end{array}$
&
$\begin{array}{c}
\includegraphics[width = 0.35\textwidth]{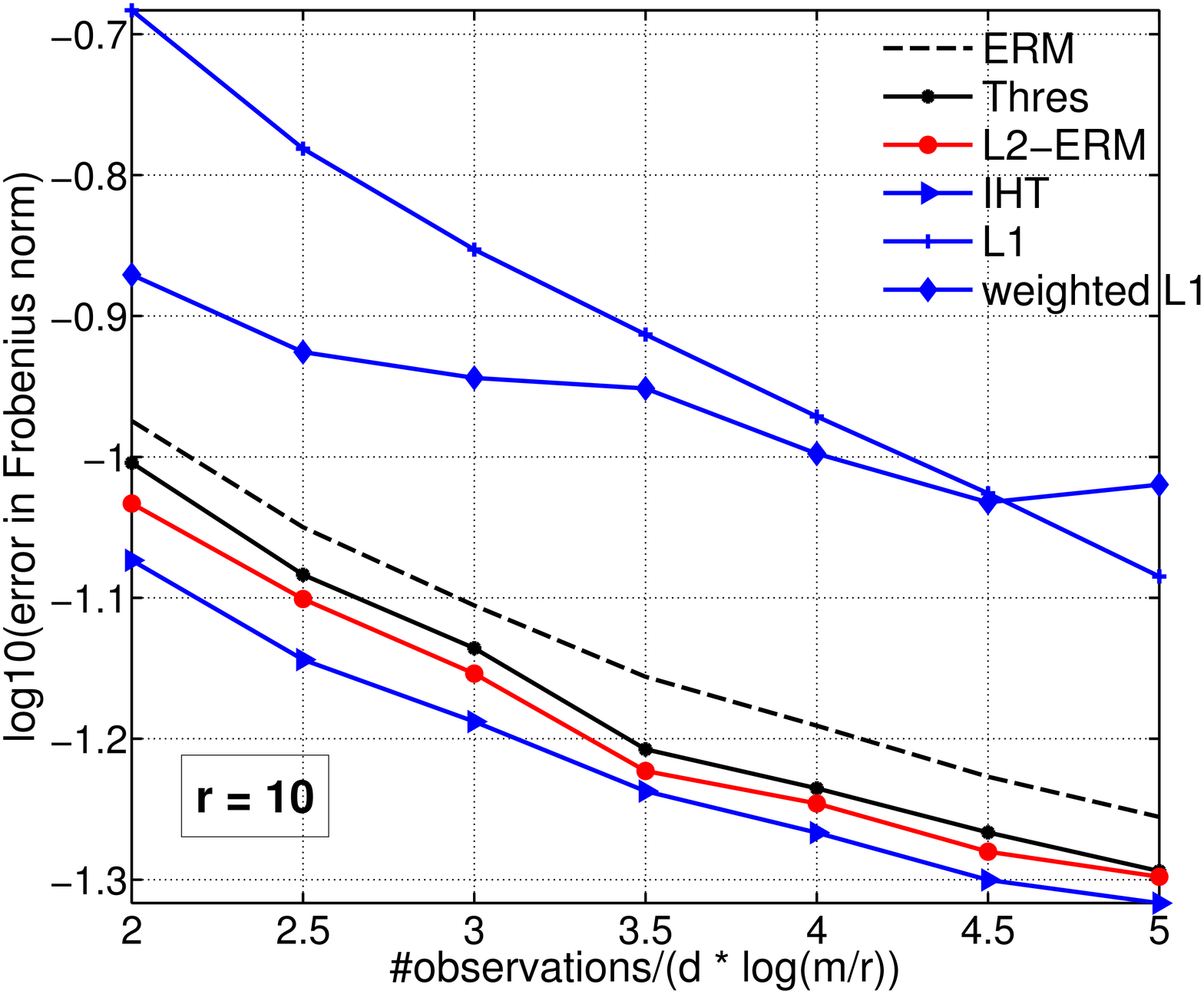}  \\
{ \scriptsize \begin{array}{l} \text{\textbf{Standard errors}(rounded):} \\
 \begin{array}{|c|c|c|}\hline
 \textbf{\textsf{L2-ERM}}  &  \textbf{\textsf{ERM}} &  \textbf{\textsf{Thres}} \\
 2 \cdot 10^{-3}   & 1.2 \cdot 10^{-2} & 1.7 \cdot 10^{-2} \\
 \hline
 \textbf{\textsf{weightedL1}} & \textbf{\textsf{IHT}} & \textbf{\textsf{L1}} \\
 1.2 \cdot 10^{-2} & 1.1 \cdot 10^{-2} &  8 \cdot 10^{-3} \\
 \hline
  \end{array}
\end{array}}
\end{array}$
\\
\\
\includegraphics[width = 0.35\textwidth]{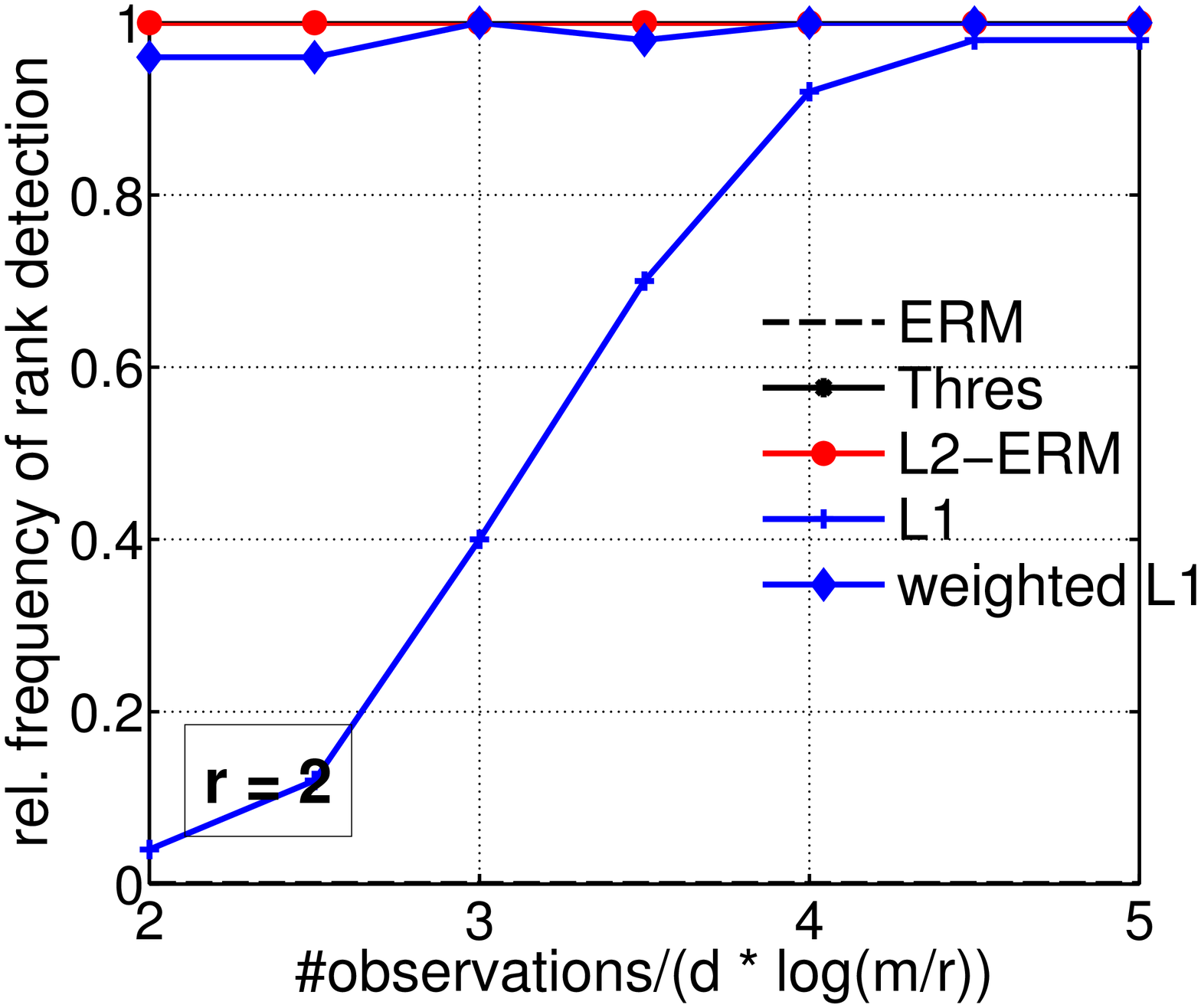} &
 \includegraphics[width = 0.35\textwidth]{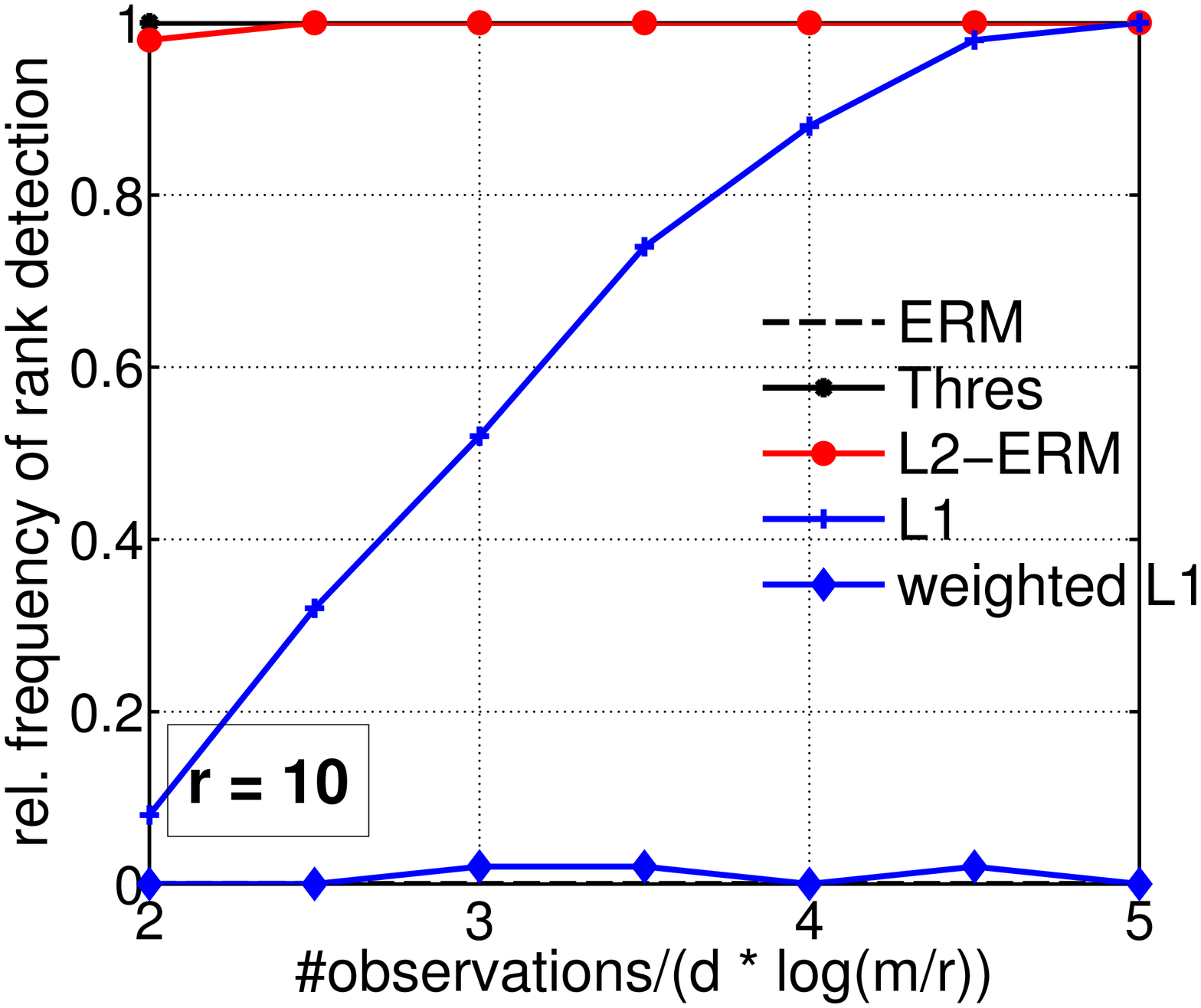}
\end{tabular}
\end{center}
\vspace{-0.2in}
\caption{Bottom: Average estimation errors $\nnorm{\wh{\Theta} -  B^*}_2$ over 50 trials ($\log_{10}$-scale) in dependence
of the number of measurements ($d=$$\,$'degrees of freedom'). Top: Relative frequency of rank detection, i.e., of the event $\{\nnorm{\wh{\Theta}}_0 =  \nnorm{B^*}_0 \}$; for \textsf{'IHT'} this relative frequency is always one, which is not shown in the plots.  Here, $\wh{\Theta}$ is representative for any of the estimators under consideration.}
\label{fig:matrixRecovery_noisy}
\end{figure}

\begin{figure}[h!]
\begin{center}
\includegraphics[width = 0.35\textwidth]{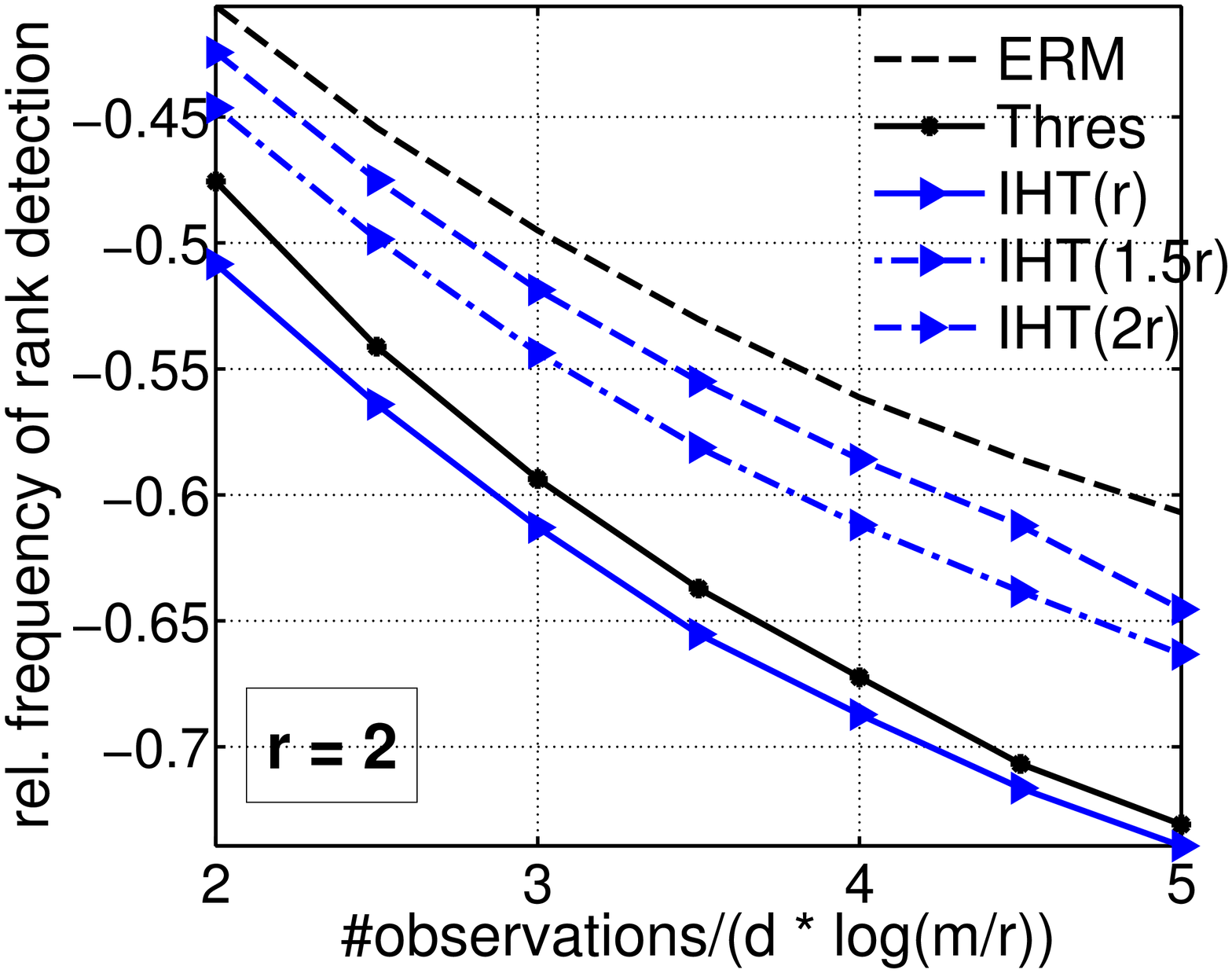}\hspace{0.3in}
\includegraphics[width = 0.35\textwidth]{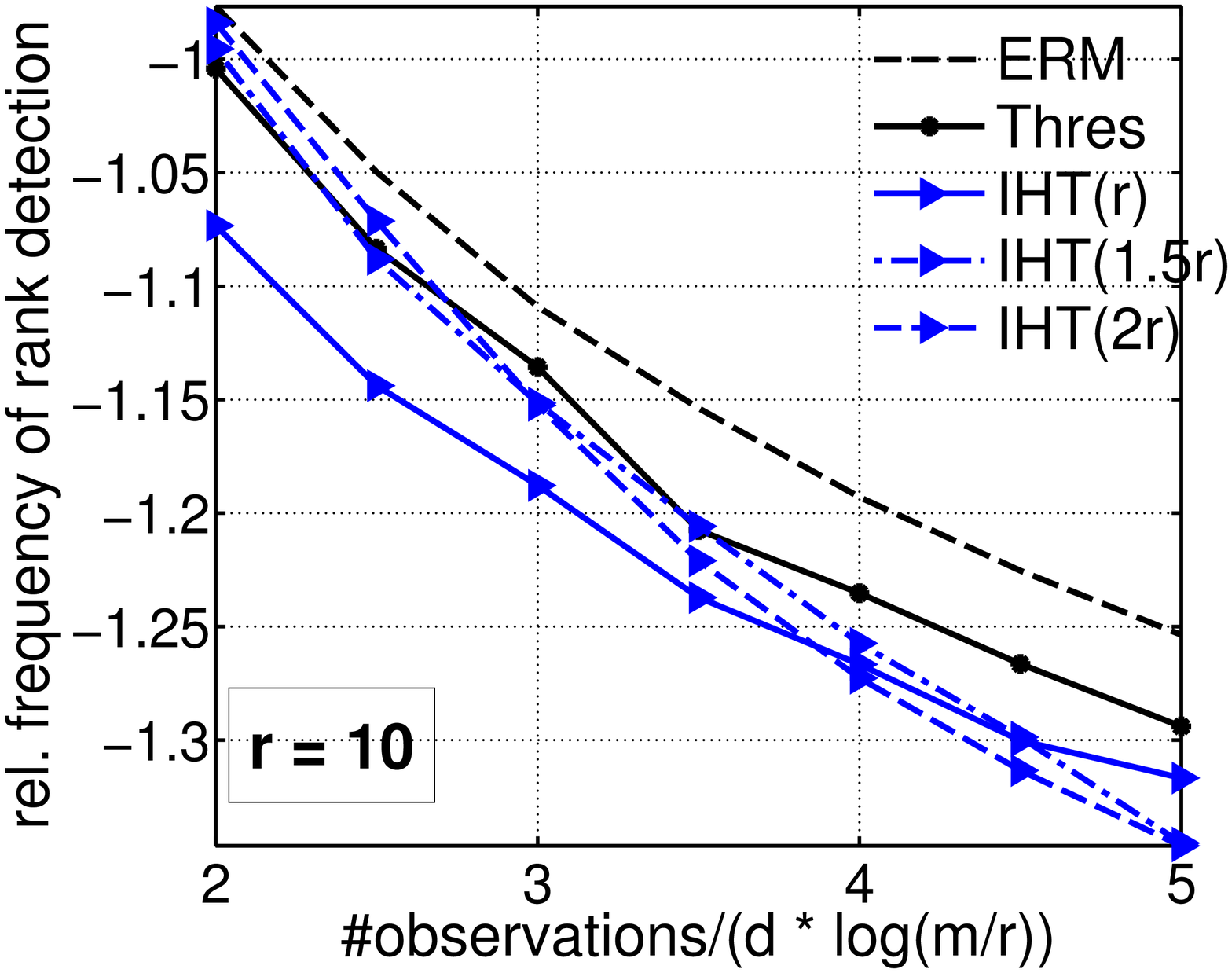}
\end{center}
\vspace{-0.2in}
\caption{Sensitivity of \textsf{'IHT'} w.r.t.~the choice of $r$. The dashed-dotted and dashed lines show the average
estimation errors when \textsf{'IHT'} is run with $1.5r$ and $2r$, respectively. The results of \textsf{'Thres'} and \textsf{'ERM'}
serve as reference.}
\label{fig:matrixRecovery_noisy_ihtspecial}
\end{figure}

\clearpage

\section{Conclusion}\label{sec:conclusion}
Simplex constraints are beneficial in high-dimensional estimation, typically achieving noticeably lower
estimation error than using $\ell_1$-norm regularization in place of the constraint. In order
to enhance sparsity of the solution, simple two-stage methods - thresholding and weighted $\ell_1$-regularization - can
be rather effective. A more principled way to incorporate sparsity is the use of a suitable regularizer.
We have pointed out that under simplex constraints, sparsity cannot be promoted by convex regularizers. We have
therefore considered non-convex alternatives among which regularization by means of the negative $\ell_2$-norm turns out to be
a natural approach lending itself to a straightforward computational strategy. As an attractive feature, there is a direct and practical generalization to the matrix counterpart as opposed to the
two-stage methods.

\section*{Acknowledgments}

The work of Ping Li  and Martin Slawski is partially supported by NSF-DMS-1444124, NSF-III-1360971, and AFOSR-FA9550-13-1-0137. The work of Syama Sundar Rangapuram was partially supported by the ERC starting grant NOLEPRO. The authors would like to thank to Anastasios Kyrillidis for clarifications regarding step size selection for the
iterative hard thresholding method discussed in the present work.

\appendix

\section{Proof of Proposition \ref{prop:excessrisk}}
By definition of $\wh{\beta}$, we have
\begin{align*}
            & R_n(\wh{\beta}) \leq R_n(\beta^*) \\
\Rightarrow \quad & \{R_n(\wh{\beta}) - R(\wh{\beta}) \} + R(\wh{\beta}) \leq  \{R_n(\beta^*) - R(\beta^*)\} + R(\beta^*) \\
\Rightarrow \quad & R(\wh{\beta}) \leq R(\beta^*) + \sup_{\beta \in \mathbb{B}_1^p(\nnorm{\wh{\beta} - \beta^*}_1;\beta^*)} |\underbrace{\{ R_n(\beta) - \{R_n(\beta^*) \} - \{R(\beta) - R(\beta^*)\}}_{\overline{\psi}_n(\beta)}| \\
&\qquad=R(\beta^*) + \overline{\Psi}_n(\nnorm{\wh{\beta} - \beta^*}_1) \\
&\qquad\leq R(\beta^*) + \overline{\Psi}_n(2).
\end{align*}
The last inequality follows from $\wh{\beta} \in \Delta^p$, $\beta^* \in \Delta^p$ and the triangle inequality.\\[1ex]
We now turn to $\ds$. Consider the curve (segment) $\gamma(t) = \beta^* + t (\ds - \beta^*)$ for $t \in [0,1]$
and the function $g(t)=R_n(\beta^* + t (\ds - \beta^*))$. Then $g = R_n \circ \gamma$ is
convex, as it is the composition of an affine and a convex function. Consequently, the derivative
\begin{equation*}
g'(t) = \nabla R_n(\beta^* + t (\ds - \beta^*))^{\T} (\ds - \beta^*)
\end{equation*}
is non-decreasing. As a result, we have
\begin{align*}
R_n(\ds) - R_n(\beta^*) &= \int_0^1  \nabla R_n(\beta^* + t (\ds - \beta^*))^{\T} (\ds - \beta^*) \; dt \\
                              &\leq \nabla R_n(\ds)^{\T} (\ds - \beta^*) \\
                              &\leq \nnorm{\nabla R_n(\ds)}_{\infty}  \nnorm{\ds - \beta^*}_1 \\
                              &\leq \lambda \nnorm{\ds - \beta^*}_1,
\end{align*}
where the first inequality follows from the definition and monotonicity property of $g'$, the second inequality
is H\"older's inequality and the last inequality follows from the definition of $\ds$. Given the above upper
bound on $R_n(\ds) - R_n(\beta^*)$, the proof can be completed by following the scheme used for $\wh{\beta}$.
\section{Proof of Proposition \ref{prop:adaptation}}
We have
\begin{align*}
R_n(\wh{\beta})  - R_n(\beta^*) - \nabla R_n(\beta^*)^{\T} (\wh{\beta} - \beta^*) \geq \kappa \nnorm{\wh{\beta} - \beta^*}_2^2,
\end{align*}
by the $\Delta$-\textsf{RSC}. On the other hand, by the definition of $\wh{\beta}$
\begin{align*}
R_n(\wh{\beta})  - R_n(\beta^*) - \nabla R_n(\beta^*)^{\T} (\wh{\beta} - \beta^*) &\leq -\nabla R_n(\beta^*)^{\T} (\wh{\beta} - \beta^*) \\
&\leq \nnorm{\nabla R_n(\beta^*)}_{\infty} \nnorm{\wh{\beta} - \beta^*}_1.
\end{align*}
Combining these two bounds, we obtain that
\begin{equation*}
\kappa \nnorm{\wh{\beta} - \beta^*}_2^2 \leq \nnorm{\nabla R_n(\beta^*)}_{\infty} \nnorm{\wh{\beta} - \beta^*}_1.
\end{equation*}
This implies that
\begin{align*}
&\nnorm{\wh{\beta} - \beta^*}_2^2 \leq \frac{\nnorm{\nabla R_n(\beta^*)}_{\infty}^2}{\kappa^2} \left(\frac{\nnorm{\wh{\beta} - \beta^*}_1}{\nnorm{\wh{\beta} - \beta^*}_2} \right)^2 \leq \frac{4 s \lambda_*^2}{\kappa^2},\\
&\nnorm{\wh{\beta} - \beta^*}_1 \leq \frac{\nnorm{\nabla R_n(\beta^*)}_{\infty}}{\kappa} \left(\frac{\nnorm{\wh{\beta} - \beta^*}_1}{\nnorm{\wh{\beta} - \beta^*}_2} \right)^2 \leq  \frac{4s \lambda_*}{\kappa},
\end{align*}
where $\lambda_* = \nnorm{\nabla R_n(\beta^*)}_{\infty}$. The rightmost inequalities follow from the fact that $\wh{\beta} - \beta^* \in \mc{C}^{\Delta}(s)$ and hence $\nnorm{\wh{\beta}_{S(\beta^*)^c}}_1 \leq \nnorm{\wh{\beta}_{S(\beta^*)} - \beta_{S(\beta^*)}^*}_1$ so that
\begin{align*}
\nnorm{\wh{\beta} - \beta^*}_1 &= \nnorm{\wh{\beta}_{S(\beta^*)} - \beta_{S(\beta^*)}^*}_1 + \nnorm{\wh{\beta}_{S(\beta^*)^c}}_1 \\
                              &\leq 2 \nnorm{\wh{\beta}_{S(\beta^*)} - \beta_{S(\beta^*)}^*}_1 \leq 2 \sqrt{s} \nnorm{\wh{\beta}_{S(\beta^*)} - \beta_{S(\beta^*)}^*}_2.
\end{align*}
We now turn to $\ds$. Starting from
\begin{align*}
R_n(\ds)  - R_n(\beta^*) - \nabla R_n(\beta^*)^{\T} (\ds - \beta^*) \geq \kappa \nnorm{\ds - \beta^*}_2^2,
\end{align*}
and using the upper bound on $R_n(\ds)  - R_n(\beta^*)$ as derived in the proof of Proposition 1, we obtain
\begin{align*}
\kappa \nnorm{\ds - \beta^*}_2^2 &\leq \nnorm{\nabla R_n(\ds)}_{\infty} \nnorm{\ds - \beta^*}_1 + \nnorm{\nabla R_n(\beta^*)}_{\infty} \nnorm{\ds - \beta^*}_1 \\
&\leq (\lambda + \lambda_*) \nnorm{\ds - \beta^*}_1,
\end{align*}
Arguing similarly as for $\wh{\beta}$, it follows that
\begin{align*}
\nnorm{\ds - \beta^*}_2^2 \leq  \frac{4 s  (\lambda + \lambda_*)^2}{\kappa^2}, \qquad \nnorm{\ds - \beta^*}_1 \leq  \frac{4s (\lambda + \lambda_*)}{\kappa}.
\end{align*}

\section{Proof of Proposition \ref{prop:convergence}}

We prove the statement for problem \eqref{eq:rem}
\begin{align}\label{eq:original}
\min_{\beta \in \Delta^p} R_n(\beta) - \lambda \norm{\beta}^2_2.
\end{align}
The proof for problem \eqref{eq:dss} follows similarly.
The subproblem solved in each iteration in the case of \eqref{eq:original}  is given by
\begin{align}\label{eq:subproblem}
\min_{\beta \in \Delta^p} R_n(\beta) - 2\lambda \inner{\beta^k, \beta - \beta^k}
\end{align}

First note that the constraint sets of both \eqref{eq:original} and \eqref{eq:subproblem} are compact and the objectives are continuous, hence these problems have at least one minimizer by Weierstrass theorem and the minima are finite.

 The current iterate $\beta^k$ is always feasible for \eqref{eq:subproblem}.
Hence the optimal value of \eqref{eq:subproblem} is either $R_n(\beta^k)$ (in which case the algorithm terminates) or strictly smaller than $R_n(\beta^k)$,
\begin{align}\label{eq:descent}
R_n(\beta^{k+1}) -2\lambda \inner{\beta^k,\beta^{k+1}-\beta^k} < R_n(\beta^k).
\end{align}
On the other hand, by convexity of $\lambda\norm{\beta}^2_2 (\lambda \ge 0)$, we have
\begin{align*}
	f(\beta^{k+1}) = R_n(\beta^{k+1}) - \lambda\norm{\beta^{k+1}}_2^2 &\le R_n(\beta^{k+1})-\lambda\norm{\beta^k}_2^2  -2\lambda \inner{\beta^k,\beta^{k+1}-\beta^k}\\
	& \stackrel{\eqref{eq:descent}}{<} R_n(\beta^{k})-\lambda\norm{\beta^k}_2^2\\
	& = f(\beta^{k})
\end{align*}
This establishes the strict monotonicity of the iterates in terms of the objective $f$ of the original problem \eqref{eq:original} until convergence.
It is clear that all the elements of the sequence $\{\beta^k\}$ are feasible for \eqref{eq:original} and satisfy $f^* \le f(\beta^k),\ k \ge 0$, where $f^*$ is the global minimum of \eqref{eq:original}.
Since $\{f(\beta^k)\}$ is a strictly decreasing sequence bounded below by a finite $f^*$, the sequence converges to a limit
\[  \bar{f} = \lim_{k \rightarrow \infty} f(\beta^k).\]

Since all the elements of the sequence $\{\beta^k\}$ are contained in $\Delta^p$, a compact set, there exists a subsequence $\{\beta^{k_i}\}$ converging to an element $\bar{\beta} \in \Delta^p$.
The sequence $\{f(\beta^{k_i})\}$ is a subsequence of $\{f(\beta^k)\}$ that is shown to converge to the limit $\bar{f}$; hence the subsequence $\{f(\beta^{k_i})\}$ also converges to the same limit
\[  \lim_{k \rightarrow \infty} f(\beta^{k_i}) = \bar{f}.\]

Let us define $\phi_{\bar{\beta}}(\beta) = R_n(\beta) - 2\lambda \inner{\bar{\beta}, \beta - \bar{\beta}}$.
We now argue that $ \bar{\beta} \in \argmin_{\beta \in \Delta^p} \phi_{\bar{\beta}}(\beta)$.
To see this note that $\bar{\beta}$ is feasible for this problem and hence $\min_{\beta \in \Delta^p} \phi_{\bar{\beta}}(\beta) \le f(\bar{\beta}) = \bar{f}$.
Assume for the sake of contradiction that a minimizer $\check{\beta}$ of this problem has a strictly smaller objective,
\[ \phi_{\bar{\beta}}(\check{\beta}) =  R_n(\check{ \beta}) - 2\lambda \inner{\bar{\beta}, \check{\beta} - \bar{\beta}} < \bar{f}.\]
Similar to the argument above regarding strict descent, we can show that
\[ f(\check{\beta})  < \bar{f},\]
which contradicts the fact that the sequence $\{f(\beta^k)\}$ has converged to the limit $\bar{f}$.
Thus, we must have,
\[\bar{\beta} \in \argmin_{\beta \in \Delta^p} R_n(\beta) - 2\lambda \inner{\bar{\beta}, \beta - \bar{\beta}}.\]
The first order optimality condition for $\bar{\beta}$ then implies
\[ -\nabla R_n(\bar{\beta}) + 2\lambda \bar{\beta} \in N_{\Delta^p}(\bar{\beta}),\]
where $N_{\Delta^p}(\bar{\beta})$ is the normal cone of $\Delta^p$ at $\bar{\beta}$ (see e.g.~\citet{Rockafellar2004} for a definition).
Note that this is exactly the first-order optimality condition for the original problem \eqref{eq:original}.
Finally note that the argument is true for any subsequence $\{\beta^{k_i}\}$ and hence each of such subsequences and consequently the original sequence $\{\beta^k\}$ converge to the same limit $\bar{\beta}$, which has been shown to satisfy the required optimality condition.

\section{Proof of Proposition \ref{prop:denoising}}

The optimization problem under consideration is equivalent to the following one:
\begin{equation}\label{eq:denoising_suppw}
\min_{\beta \in \Delta^n} \left(\frac{1}{n} - \lambda \right) \nnorm{\beta}_2^2 - \frac{2}{n} \mathbf{Z}^{\T} \beta.
\end{equation}
For $\lambda \geq 1/n$, the objective becomes concave. If $\lambda > 1/n$, the objective
is strictly concave and the unique minimum is attained at one of the vertices $\{e_i \}_{i = 1}^n$
of $\Delta^n$. It must be any unit vector $e_i$ for which $z_i = \max_{1 \leq k \leq n} z_k$.
Since we have assumed that $z_{(1)} > \ldots > z_{(n)}$, such vector is unique. If
$\lambda = 1/n$, we have
\begin{equation*}
\rer \in \conv \left\{e_i:\;  z_i = \max_{1 \leq k \leq n} z_k \right\}.
\end{equation*}
By the same argument as above, that convex hull equals the unique unit vector $e_i$ for which $z_i = \max_{1 \leq k \leq n} z_k$.\\
For $0 \leq \lambda < 1/n$, the problem becomes strictly convex. Setting $\gamma = 1 - n\lambda$,
problem \eqref{eq:denoising_suppw} is equivalent to
\begin{equation*}
\min_{\beta \in \Delta^n} \gamma \nnorm{\beta}_2^2 - 2 \mathbf{Z}^{\T} \beta.
\end{equation*}
Re-arranging terms, this can be seen to be equivalent to
\begin{equation*}
\min_{\beta \in \Delta^n} \norm{\beta - \frac{\mathbf{Z}}{\gamma}}_2^2,
\end{equation*}
i.e.~$\rer = \Pi_{\Delta^n}(\mathbf{Z}/\gamma)$, with $\Pi_{\Delta^n}$ denoting the
Euclidean projection onto $\Delta^n$.\\
Suppose that the realizations $\mathbf{z} = (z_i)_{i = 1}^n$ are arranged such that
\begin{equation*}
z_1 = \beta_1^* + \eps_1 > z_2 = \beta_2^* + \eps_2 > \ldots > z_s = \beta_s^* + \eps_{s} > z_{s + 1} = \eps_{s+1} > \ldots > z_{p} = \eps_{p}.
\end{equation*}
Under the event $\{ b_{\min}^* = \min_{i \in S(\beta^*)} |\beta_i^*| \geq 2 \max_{1 \leq i \leq n} |\eps_i|\}$, this
can be assumed without loss of generality. The projection of $\mathbf{z}/\gamma$ onto $\Delta^n$ can
then can be expressed as (cf.~\cite{Kyrilidis2013})
\begin{equation*}
\left( \Pi_{\Delta^n}(\mathbf{z}/\gamma) \right)_i = \max\{z_i/\gamma - \tau, 0 \}, \quad \text{where} \; \tau
= \frac{1}{q} \left(\sum_{i = 1}^q (z_i/\gamma) -  1 \right),
\end{equation*}
and
\begin{equation*}
q = \max \left\{k:\; (z_k/\gamma) > \frac{1}{k} \left(\sum_{i = 1}^k (z_i/\gamma) - 1 \right) \right \}.
\end{equation*}
In order to establish that $S(\rer) = S(\beta^*)$, it remains to be shown that under the given conditions on $b_{\min}^*$ and $\lambda$ respectively $\gamma$, it holds that
\begin{align*}
(a)\quad& \frac{\beta_s^* + \eps_s}{\gamma} > \frac{1}{\gamma}\frac{\beta_1^* + \ldots + \beta_s^* - \gamma}{s} + \frac{1}{\gamma} \frac{\eps_1 + \ldots + \eps_s}{s} =  \frac{1}{s} \, \frac{1 - \gamma}{\gamma} + \frac{1}{\gamma} \frac{\eps_1 + \ldots + \eps_s}{s} \\
\Longleftrightarrow &\; \beta_s^* > \frac{1}{s}(\{ 1 - \gamma \} - \left\{\eps_1 + \ldots + \eps_s - s\eps_s \right\}).
\end{align*}
and
\begin{equation*}
(b) \quad \frac{\eps_{s+1}}{\gamma} < \frac{1}{\gamma}\frac{1 - \gamma}{s + 1} +  \frac{1}{\gamma} \frac{\eps_1 + \ldots + \eps_s + \eps_{s+1}}{s+1}.
\end{equation*}
Re-arranging (b), we find that
\begin{equation*}
n \lambda = (1 - \gamma)  > s \eps_{s+1} - (\eps_1 + \ldots + \eps_s),
\end{equation*}
which is implied by
\begin{equation*}
n \lambda > 2 s \max_{1 \leq i \leq n} |\eps_i|.
\end{equation*}
Likewise, the inequality in (a) holds as long as
\begin{equation*}
\beta_s^* > \frac{n \lambda}{s} + 2 \max_i |\eps_i|. 
\end{equation*}
This concludes the proof.

\section{Proof of Proposition \ref{prop:adaptation_m}}

Before providing a proof of Proposition 5, we first provide a precise definition of the linear spaces $\TT(B)$, $B \in \bm{B}_0^m(r)  \subset \HH^m$.
\begin{defnApp}\label{defn:tangentspace} Let $B \in \bm{B}_0^m(r)$ have eigenvalue decomposition $B = U \Lambda U^{H}$, where
\begin{equation*}
U = \left[ \begin{array}{cc}
                             U_{\pa} & U_{\perp} \\
                             \mbox{{\footnotesize $m \times r$}} & \mbox{{\footnotesize $m \times (m - r)$}}
                             \end{array} \right] \begin{bmatrix}
                             \Lambda_{r}   &  0_{r \times (m-r)}  \\
                                 0_{(m-r) \times r}        &  0_{(m - r) \times (m-r)}
                           \end{bmatrix}
                           \end{equation*}
for $\Lambda_r$ real and diagonal. We then define
\begin{equation*}
\TT(B) = \{M \in \mathbb{H}^m:\; M = U_{\pa} \Gamma + \Gamma^{H} U_{\pa}^{H}, \quad \Gamma \in \C^{r \times m} \}.
\end{equation*}
\end{defnApp}
It is immediate from the definition of $\TT(B)$ that its orthogonal complement is given by
\begin{equation*}
\mathbb{T}(B)^{\perp} = \{M \in \mathbb{H}^m:\; M = U_{\perp} A U_{\perp}^{H}, \quad A \in \mathbb{H}^{m-r} \}.
\end{equation*}

\begin{bew} We first show that $\wh{\Phi} = \wh{B} - B^* \in \mc{K}^{\bm{\Delta}}(r)$, where we recall that
\begin{align*}
&\mc{K}^{\bm{\Delta}}(r) = \{\Phi \in  \mathbb{H}^m: \,\exists B \in \bm{B}_0^m(r) \, \text{s.t.} \\
&\tr(\Pi_{\mathbb{T}(B)^{\perp}}(\Phi)) = -\tr( \Pi_{\mathbb{T}(B)} \Phi) \; \text{and} \;  \Pi_{\mathbb{T}(B)^{\perp}}(\Phi) \gec 0 \}.
\end{align*}
Define the shortcuts $\wh{\Phi}_{\TT} = \Pi_{\mathbb{T}(B^*)} \wh{\Phi}$ and $\wh{\Phi}_{\TT^{\pe}} = \Pi_{\mathbb{T}(B^*)^{\perp}} \wh{\Phi}$.
Since $\wh{B}$ is feasible, it must hold that $\tr(\wh{\Phi}) = 0$ and thus $\tr(\wh{\Phi}_{\TT^{\pe}}) = -\tr(\wh{\Phi}_{\TT})$. Since
$\wh{B}$ must also be positive definite, it must hold that $\tr(\wh{B}{W}) \geq 0$ for all $W \in \TT(B^*)^{\pe}$, $W \gec 0$. We have
\begin{equation*}
\tr(\wh{B}{W}) = \tr((B^* + \wh{\Phi})W) = \tr(\wh{\Phi}_{\TT^{\pe}} W) \; \; \forall W \in \TT(B^*)^{\pe},
\end{equation*}
since $B^* \in  \TT(B^*)$. We conclude that $\tr(\wh{\Phi}_{\TT^{\pe}} W) \geq 0$ for all $W \in \TT(B^*)^{\pe}$, $W \gec 0$,
and thus $\wh{\Phi}_{\TT^{\pe}} \gec 0$. Altogether, we have shown that $\wh{\Phi} \in \mc{K}^{\bm{\Delta}}(r)$.\\
Since $\wh{B}$ is a minimizer, we have
\begin{equation*}
\frac{1}{n} \nnorm{\mathbf{Y} - \mc{X}(\wh{B})}_2^2 \leq \frac{1}{n} \nnorm{\mathbf{Y} - \mc{X}(B^*)}_2^2 \\
\end{equation*}
After re-arranging terms, we obtain
\begin{align*}
\frac{1}{n} \nnorm{\mc{X}(B^* - \wh{B})}_2^2
&\leq
\frac{2}{n} \scp{\eps}{\mc{X}(\wh{B} - B^*)} \\
&= \frac{2}{n} \scp{\mc{X}^{\star}(\eps)}{\wh{B} - B^*} \\
&\leq 2 \nnorm{\mc{X}^{\star}(\eps)/n}_{\infty} \nnorm{\wh{B} - B^*}_1 \\
&=\lambda_* \nnorm{\wh{B} - B^*}_1.
\end{align*}
where $\mc{X}^{\star}$ is the adjoint of $\mc{X}$. By $\bm{\Delta}$-\textsf{RSC}, we now have
\begin{equation*}
\frac{1}{n} \nnorm{\mc{X}(B^* - \wh{B})}_2^2 \geq \kappa \nnorm{B^* - \wh{B}}_2^2.
\end{equation*}
Combining this with the preceding upper bound, we hence obtain
\begin{align*}
&\nnorm{\wh{B} - B^*}_2^2 \leq \frac{\lambda_*^2}{\kappa^2} \left(\frac{\nnorm{\wh{B} - B^*}_1}{\nnorm{\wh{B} - B^*}_2} \right)^2 \leq \frac{8 r \lambda_*^2}{\kappa^2},\\
&\nnorm{\wh{B} - B^*}_1 \leq \frac{\lambda_*}{\kappa} \left(\frac{\nnorm{\wh{B} - B^*}_1}{\nnorm{\wh{B} - B^*}_2} \right)^2 \leq  \frac{8r \lambda_*}{\kappa},
\end{align*}
The rightmost inequalities follow from the fact that $\wh{B} - B^* = \wh{\Phi} \in \mc{K}^{\bm{\Delta}}(r)$ and hence
$\nnorm{\wh{\Phi}_{\TT^{\pe}}}_1 \leq \nnorm{\wh{\Phi}_{\TT}}_1$ so that
\begin{align*}
\nnorm{\wh{B} - B^*}_1 = \nnorm{\wh{\Phi}}_1 &= \nnorm{\wh{\Phi}_{\TT}}_1 + \nnorm{\wh{\Phi}_{\TT^{\pe}}}_1 \\
                      &\leq 2 \nnorm{\wh{\Phi}_{\TT}}_1 \\
&\leq 2 \sqrt{2r} \nnorm{\wh{\Phi}_{\TT}}_2 \leq  2 \sqrt{2r} \nnorm{\wh{B} - B^*}_2,
\end{align*}
where for the third inequality, we have used that $\nnorm{M}_0 \leq 2r$ for all $M \in \TT(B^*)$.\\
The bound for $\wt{B}_{\lambda}$ can be established by combining the proof scheme used for $ \wt{\beta}_{\lambda}$
with the scheme used for $\wh{B}$ and is thus omitted.
\end{bew}

\newpage

\section{Proof of Proposition \ref{prop:denoising_matrix}}
We start by expanding the objective function of the optimization problem under consideration. Define
$\sym^m:\HH^{m} \cap \R^{m \times m}$ which is a subspace of $\HH^m$ which is isometrically isomorphic (w.r.t.~to the standard inner
product) to $\R^{\delta_m}$, $\delta_m = m(m+1)/2$ under the isometry $\mc{X}$ \eqref{eq:Xi_svec}. Therefore,
\begin{align}
\frac{1}{n}\nnorm{\mathbf{Y} - \mc{X}(B)}_2^2 &= \frac{1}{n} \nnorm{\mc{X}^{\star}(\mathbf{Y}) - B}_2^2 \notag \\
                                              &= \frac{1}{n} \nnorm{B^* + E - B}_2^2, \quad E \coloneq \mc{X}^{\star}(\eps), \notag \\
                                              &= \frac{1}{n} \nnorm{\Upsilon - B}_2^2, \quad \Upsilon \coloneq B^* + E. \label{eq:matrixdenoising}
\end{align}
It follows directly from the definition of $\mc{X}^{\star}$ that the symmetric random matrix $E = (\eps_{jk})_{1 \leq j,k \leq m}$
is distributed according to the Gaussian orthogonal ensemble (GOE, see e.g.~\cite{Tao2012}), i.e.,~$E \sim \text{GOE}(m)$, where
\begin{align*}
\vspace*{-1ex}
\text{GOE}(m) = \{ X = (x_{jk})_{1 \leq j,k \leq m}, \; &\{ x_{jj} \}_{j = 1}^m \overset{\text{i.i.d.}}{\sim} N(0,1/m),\\
 &\{ x_{jk}=x_{kj} \}_{1 \leq j < k \leq m} \overset{\text{i.i.d.}}{\sim} N(0,1/2m)  \}.
\end{align*}
In virtue of \eqref{eq:matrixdenoising}, we have
\begin{align*}
\min_{B \in \bm{\Delta}^m} \frac{1}{n} \nnorm{\mathbf{Y} - \mc{X}(B)}_2^2
&= \min_{B \in \bm{\Delta}^m} \left\{(1/n - \lambda) \nnorm{B}_2^2 - \frac{2}{n} \scp{\Upsilon}{B} \right\} + \frac{1}{n} \nnorm{\Upsilon}_2^2.
\end{align*}
At this point, the proof parallels the proof of Proposition \ref{prop:denoising}. We see that for $\lambda \geq 1/n$, $\wh{B}_{\lambda}^{\ell_2} = u_1 u_1^{\T}$, where $u_1$ is the eigenvector of $\Upsilon$ corresponding to its largest eigenvalue. This follows from the duality of the Schatten $\ell_1$/$\ell_{\infty}$ norms and the fact
that for all feasible $B$, it holds that $\nnorm{B}_2^2 \leq \nnorm{B}_1^2 = 1$ with equality if and only if $B$ has rank one. Conversely, if $0 \leq \lambda < 1/n$, we define $\gamma \coloneq 1 - n \lambda > 0$ and deduce that the optimization problem in the previous display
is equivalent to $\min_{B \in \bm{\Delta}^m} \nnorm{\Upsilon/\gamma - B}_2^2$ with minimizer $\wh{B}_{\lambda}^{\ell_2} = U
\text{diag}(\{\wh{\phi}_j \}_{j=1}^m) U^{\T}$, where $\wh{\phi} = \Pi_{\Delta^m}(\upsilon/\gamma)$ with $\upsilon = (\upsilon_j)_{j = 1}^m$ denoting
the eigenvalues of $\Upsilon$ (in decreasing order) corresponding to the eigenvectors in $U$. We now prove the last claim of the proposition, combining the proof of Proposition \ref{prop:denoising} for the vector case with concentration results by \cite{Peng2012} for the spectrum of the random matrix $\Upsilon = B^* + E$, which are here rephrased as follows. Define
\begin{equation*}
\wt{\phi}_{j}^* = \begin{cases}
                   \phi_j^* + \frac{\sigma^2}{\phi_j^*} \quad &\text{if} \; \sigma < \phi_j^* \leq 1 \\
                   2 \sigma \quad &\text{if} \; 0 \leq \phi_j^* \leq \sigma, \quad j=1,\ldots,m,
                   \end{cases}
\end{equation*}
where we recall that the $\{ \phi_j^* \}_{j = 1}^m$ denote the ordered eigenvalues of $B^*$ and $\sigma^2$ is the variance of the noise (up to a scaling factor of $1/m$). We then have
\begin{align*}
\p(\upsilon_j  \geq  \wt{\phi}_{j}^* + t) &\leq C_1 \exp(-c_1 m t^2 / \sigma^2), \quad j=1,\ldots,m.
\end{align*}
Furthermore, let $r_0$ denote the number of eigenvalues of $B^*$ that are larger than $\sigma$. Then, there is a constant $c_0 > 0$ so that
if $r \leq c_0 m$, it holds that
\begin{align*}
\p(\upsilon_j \leq \wt{\phi}_{j}^* - t - 2 \sigma) &\leq \exp(-c_2 m/\sigma^2) + C_2' \exp(-c_2' m t^2 / \sigma^2), \quad j=1,\ldots,r_0,
\end{align*}
where $c_1, c_2, C_1, C_2, C_2'$ are positive constants.\\
It needs to be shown that for a suitable choice of $\lambda$ and for $\phi_r^*$ large enough, it holds that
$\nnorm{\wh{B}_{\lambda}^{\ell_2}}_0 = \nnorm{B^*}_0 = r$ with high probability as specified in the proposition. This is the case if and only if $\wh{\phi} = \Pi_{\Delta^m}(\upsilon/\gamma)$ has precisely $r$ non-zero entries.\\

\noindent a) $\nnorm{\wh{\phi}}_0 \geq r$:\\[0.5ex]
It follows from the proof in the vector case that a) is satisfied if
\begin{align*}
\frac{\upsilon_r}{\gamma} > \frac{\upsilon_1 + \ldots + \upsilon_r - \gamma}{r \gamma}
\end{align*}
Write $\xi_j = \upsilon_j - \wt{\phi}_j^{*}$, $b_j = \wt{\phi}_j^{*} - \phi_j^*$, $j = 1,\ldots,m$, and $\overline{\xi} = \max_{1 \leq j \leq m} \xi_j$, $\underline{\xi} = \min_{1 \leq j \leq r_0} \xi_j$. Then the above condition can equivalently be expressed as
\begin{align*}
\upsilon_r >& \frac{1}{r} \left\{\sum_{j = 1}^r ( \wt{\phi}_j^{*} + \xi_j)   - \gamma \right\} \\
           =& \frac{1}{r} \left\{\sum_{j = 1}^r (b_j + \xi_j) + (1 - \gamma) \right\}, \quad \text{since} \sum_{j = 1}^r \phi_j^* = 1\\
           =& \frac{1}{r} \sum_{j = 1}^r (b_j + \xi_j) + \frac{n \lambda}{r}
\end{align*}
As $\phi_j^* \geq 5 \sigma$ for $j=1,\ldots,r$ by assumption, we have $r = r_0$ and
\begin{equation*}
\frac{1}{r} \sum_{j = 1}^r (b_j + \xi_j) \leq   \sigma + \overline{\xi}.
\end{equation*}
Since $\upsilon_r \geq \phi_r^* + \underline{\xi}$, we obtain the sufficient condition
\begin{equation*}
(A)\qquad \phi_r^* > -\underline{\xi} +  \sigma + \overline{\xi} + \frac{n \lambda}{r}.
\end{equation*}
b) $\nnorm{\wh{\phi}}_0 \leq r$\\[1ex]
In analogy to a), we start with the condition
\begin{equation*}
\frac{\upsilon_{r+ 1}}{\gamma} < \frac{\upsilon_1 + \ldots + \upsilon_r + \upsilon_{r + 1} - \gamma}{(r + 1) \gamma}
\end{equation*}
After cancelling $\gamma$ on both sides, we lower bound the right hand side as follows:
\begin{align*}
\frac{\upsilon_1 + \ldots + \upsilon_r + \upsilon_{r + 1} - \gamma}{r+1} \geq \frac{(1 - \gamma) + \upsilon_{r + 1} + r \underline{\xi}}{r+1}.
\end{align*}
Back-subtituting this lower bound, we obtain the following sufficient condition
\begin{equation*}
(B)\qquad \lambda > \frac{r}{n} (\upsilon_{r + 1} - \underline{\xi}).
\end{equation*}
Consider the following two events:
\begin{align*}
E_1:\;\,\{\overline{\xi} >  \sigma \}, \qquad E_2:\;\,\{\underline{\xi} < -3 \sigma   \}
\end{align*}
The concentration results stated above yield that $\p(E_1 \cup E_2) \leq C \exp(-c m)$ for constants
$c, C > 0$. Note that conditional on the complement of $E_1 \cup E_2$, $\upsilon_{r+1} \leq 3 \sigma$ so that condition $(B)$ is fulfilled
as long as $\lambda > 6 \sigma r/n$. Likewise, condition $(A)$ is fulfilled as long as
$\phi_r^* > 5 \sigma + n \lambda/r$.


\bibliographystyle{plainnat}
\bibliography{../paper/references_sparsesimplex_long}

\end{document}